\pdfoutput=1
\documentclass[useAMS,usenatbib,nofootinbib]{mn2e}

\usepackage[pdftex,pdfpagemode={UseOutlines},bookmarks,bookmarksopen,colorlinks,linkcolor={blue},citecolor={green},urlcolor={red}]{hyperref}

\usepackage[pdftex]{graphicx}

\usepackage{amsmath}
\usepackage{url}
\usepackage{natbib}
\usepackage{rotating}

\newcommand{\snr}{S/N}
\newcommand{\scuba}{SCUBA-2}
\newcommand{\rms}{rms}

\def\lsim{\mathrel{\lower2.5pt\vbox{\lineskip=0pt\baselineskip=0pt
          \hbox{$<$}\hbox{$\sim$}}}}

\def\gsim{\mathrel{\lower2.5pt\vbox{\lineskip=0pt\baselineskip=0pt
          \hbox{$>$}\hbox{$\sim$}}}}

\def\sinc{\mathrm{sinc}}

\newcommand{\model}[1]{\texttt{#1}}

\title[\scuba: iterative map-making with SMURF]{\scuba: iterative map-making
with the Sub-Millimetre User Reduction Facility}

\author[Edward~L.~Chapin~et~al.]{
  \parbox[t]{\textwidth}{
    Edward~L.~Chapin$^{1,2}$\thanks{E-mail:~echapin@sciops.esa.int,
      Present address: XMM SOC, ESAC, Apartado 78, 28691 Villanueva de
      la Ca\~nada, Madrid, Spain},
    David~S.~Berry$^{2}$,
    Andrew~G.~Gibb$^{1}$,
    Tim~Jenness$^{2}$,
    Douglas~Scott$^{1}$,
    Remo~P.~J.~Tilanus$^{2,3}$,
    Frossie~Economou$^2$\thanks{Present address: National Optical
      Astronomy Observatory, 950 N.\ Cherry Avenue, Tucson, AZ 85719,
      USA},
    Wayne ~S.~Holland$^{4,5}$
  }
  \\
  \\
  $^{1}$Dept. of Physics \& Astronomy, University of British Columbia,
  6224 Agricultural Road, Vancouver, B.C. V6T 1Z1, Canada\\
  $^{2}$JointAstronomy Centre, 660 N. A`oh\={o}k\={u} Place, University
  Park, Hilo, Hawaii 96720, USA\\
  $^{3}$Netherlands Organisation for Scientific Research,
  Laan van Nieuw Oost-Indie 300, NL-2509 AC The Hague, The Netherlands\\
  $^{4}$UK Astronomy Technology Centre, Royal Observatory, Blackford
  Hill, Edinburgh EH9 3HJ\\
  $^{5}$Institute for Astronomy, University of Edinburgh, Royal
  Observatory, Blackford Hill Edinburgh, EH9 3HJ}
\begin{document}

\label{firstpage}

\maketitle

\begin{abstract}
  The Submillimetre Common User Bolometer Array 2 (\scuba) is an
  instrument operating on the 15-m James Clerk Maxwell Telescope,
  nominally consisting of 5120 bolometers in each of two simultaneous
  imaging bands centred over 450 and 850\,\micron. The camera is
  operated by scanning across the sky and recording data at a rate of
  200\,Hz. As the largest of a new generation of multiplexed kilopixel
  bolometer cameras operating in the (sub)millimetre, \scuba\ data
  analysis represents a significant challenge.  We describe the
  production of maps using the Sub-Millimetre User Reduction Facility
  (SMURF) in which we have adopted a fast, iterative approach to
  map-making that enables data reduction on single, modern, high-end
  desktop computers, with execution times that are typically shorter
  than the observing times.  SMURF is used in an automated setting,
  both at the telescope for real-time feedback to observers, as well
  as for the production of science products for the JCMT Science
  Archive at the Canadian Astronomy Data Centre. Three detailed case
  studies are used to: (i) explore convergence properties of the
  map-maker using simple prior constraints (Uranus -- a point source);
  (ii) achieve the white-noise limit for faint point-source studies
  (extragalactic blank-field survey of the Lockman Hole); and (iii)
  demonstrate that our strategy is capable of recovering angular
  scales comparable to the size of the array footprint (approximately
  5\,arcmin) for bright extended sources (star-forming region M17).
\end{abstract}

\begin{keywords}
methods: data analysis, techniques: image processing, submillimetre:
general, methods: observational
\end{keywords}

\section{Introduction}
\label{sec:intro}

The Submillimetre Common User Bolometer Array 2
\citep[\scuba,][]{holland2013} is a new instrument for the 15-m James
Clerk Maxwell Telescope (JCMT) on Mauna Kea, Hawai'i. The camera
simultaneously images the sky in two broad bands centred over 450 and
850\,\micron, with approximately 7.5 and 14.5\,arcsec full-width at
half-maximum (FWHM) point spread functions (PSFs). The focal planes at
each wavelength are populated with 4 rectangular subarrays, consisting
of $40 \times 32$ bolometers each, and together subtend a nearly
7\,arcmin $\times$ 7\,arcmin footprint on the sky (excluding gaps, the
continuous solid angle is about 43\,arcmin$^{2}$ per focal plane).
This paper describes the properties of \scuba\ data that are relevant
for producing maps of the imaging data, and the Submillimetre User
Reduction Facility, SMURF, a software package for performing the
reduction written using the Starlink Software Environment
\citep{1993ASPC...52..229W,2009ASPC..411..418J}. The details of the
instrument design, performance, and calibration are given in two
companion papers: \citet{holland2013} and \citet{dempsey2013}.

Over the last twenty years, observations in the submillimetre band
(defined here to be 200--1200\,\micron) have helped revolutionise
several important areas of astrophysics, including: discovering
through blind surveys a class of massive dusty star-forming galaxies
in the early ($z>2$) Universe, now referred to as submillimetre
galaxies, or SMGs; the characterisation of the early stages of
star-formation by identifying the dense, cold regions in molecular
clouds where stars may eventually form; and identifying debris disks
around nearby stars, helping us understand the early stages of planet
formation.  With $10,240$ nominal detectors \citep[of which $\sim$70\%
work and are typically useful,][]{holland2013}, \scuba\ is presently
the largest of a new generation of multiplexed kilopixel
(sub)millimetre bolometer cameras, which also includes the cameras for
the South Pole Telescope \citep[SPT,][]{carlstrom2011} and the Atacama
Cosmology Telescope \citep[ACT,][]{swetz2011}, both dedicated
experiments for studying anisotropies in the Cosmic Microwave
Background (CMB) using similar technology, the latter of which uses the
same time-domain multiplexed readout electronics as \scuba\
\citep{battistelli2008}.

Submillimetre imaging cameras generally maximise sensitivity using
bolometers, rather than coherent detectors, which are limited by white
photon and phonon noise from the instrument and ambient
backgrounds. The low-frequency noise, however, is typically dominated
by sources which produce slow variations in the background (e.g.,
thermal variations within the cryostat, and the atmosphere for
ground-based cameras), and drifts in the readout electronics. Such
noise has a power spectrum $\propto 1/f^\alpha$ ($\alpha>0$), and the
frequency at which it is comparable to the white noise level is called
the ``$1/f$ knee''. Since the low-frequency noise is largely
correlated between all, or subsets, of the bolometers in time, it can
be suppressed during map-making, since astronomical signals have the
distinct property that they are fixed in a sky reference frame
(assuming they are not time-varying). If successful, the noise in the
resulting map is said to be ``white noise limited'', meaning that it
is uncorrelated spatially, and has an amplitude that scales as the
$\mathrm{NEFD}/\sqrt{t}$, where the NEFD is the noise-equivalent
flux-density (the white noise level of a bolometer in 1\,s of
integration), and $t$ is the amount of integration time in a map
pixel.

There are numerous ways to attack this map-making problem, both in
terms of the data-collection method, and processing. The two most
important principles to follow in terms of scan strategy are: (i) to
modulate the astronomical signals of interest in such a way that they
appear in the lowest-noise regions of the bolometer noise power
spectrum, i.e., above the $1/f$ knee; and (ii) to provide good
``cross-linking'', in which each portion of the map is scanned at a
range of position angles, again, to help distinguish time-varying
noise features from fixed astronomical sources. In the case of \scuba,
(i) is achieved through fast-scanning of the entire telescope (up to
600\,arcsec\,sec$^{-1}$), such that significant drift in the
bolometers due to low-frequency noise occurs more slowly than the
crossing times for the astronomical scales of interest; and (ii) by
offering scan patterns that cross the sky at a wide range of position
angles. Such methods are now used by virtually all existing
ground-based submillimetre cameras
\citep[e.g.,][]{glenn1998,weferling2002,wilson2008,kovacs2008b}, in
preference to ``chopping'' methods (where the secondary is moved
quickly to modulate the signal) that were more appropriate for older
instruments that had poorer low-frequency noise performance, and were
only sensitive to modest angular scales. We note that under some
circumstances cross-linking may not be essential. For example, SPT,
which was designed to measure CMB anisotropies, scans entirely in
azimuth \citep{schaffer2011}. While this strategy limits its ability
to recover large angular scales transverse to the scan direction, the
anisotropic filtering in its maps can be accounted for during analysis
to achieve SPT's focussed objectives \citep[in contrast, the similar
ACT experiment uses cross-linking to improve its response to
large-angular scales,][]{das2011}. This approach is less practical as
a general solution for \scuba, however, which must serve a broader
range of scientific interests.

There are three general styles of map-making that are relevant to
reducing bolometer data in the literature. The simplest ``direct
methods'' involve some basic processing of the data to remove as much
noise contamination as possible, (e.g., using baseline removal and
other simple filters), and then re-gridding these cleaned data into a
map. Such was the basic recipe for the reduction of chopped data from
SCUBA-2's predecessor SCUBA
\citep{1998ASPC..145..216J,2000ASPC..216..559J}, and MAMBO
\citep[e.g.,][]{omont2001}, another camera from the same generation. A
more recent example is the analysis of SPT data
\citep{schaffer2011}. Generally speaking, such methods are fast,
although depending on the science goals and noise properties of the
data, they may not achieve the best noise performance on the angular
scales of interest. A method for reducing data in fields of faint
point sources with Bolocam \citep[e.g.,][]{laurent2005}, and its
younger sibling the Aztronomical Thermal Emission Camera \citep[AzTEC,
e.g.,][]{scott2008}, is principal component analysis (PCA) cleaning. A
statistical ``black-box'' removes the most correlated components of
the bolometer signals, enabling the detection of point-sources close
to the theoretical white-noise limits of the detectors, with
reasonable computation times when small numbers of bolometers are
involved (hundreds rather than thousands of detectors). However, PCA
cleaning is not a good solution for producing maps of extended
structures, since such sources produce correlated signals amongst many
detectors, and are removed by this procedure \citep[an exception is
the iterative PCA approach of][]{aguirre2011}. Furthermore, in the
case of \scuba, performing PCA on even a single subarray (typically
$\sim$900 detectors) can be prohibitively slow.

The best existing map-making strategies for recovering information on
all angular scales are maximum likelihood techniques, in which the
time-series data are expressed as a sampling of the ``true'' map of
the sky plus noise, and then the equation is inverted to estimate the
map as some weighted linear combination of the data that minimizes the
variance.  The first good description of this method appears in
\citet{janssen1992} for the production of maps from the COsmic
Background Explorer (COBE -- the description is relevant despite the
fact that it used a differential radiometer instead of
bolometers). Other descriptions in the experimental CMB literature
include \citet{tegmark1997} and \citet{stompor2002}, and an
application to data from SCUBA is described in \citet{borys2004}.  The
downside to such methods is that they can be both computationally
expensive and have large memory requirements. While for some
experimental designs fast iterative methods for the inversion do exist
without requiring excessive amounts of memory, such as that of
\citet{wright1996} \citep[which was later implemented for SCUBA
by][]{johnstone2000}, for the more general map-making problem,
involving many detectors, matters are significantly complicated both
by the need to measure the cross spectra for all unique pairs of
bolometers, as well as performing the inversion itself. A
maximum-likelihood method was successfully used to make maps from ACT
data using approximately the same number of detectors as a single
\scuba\ subarray at a single wavelength. However, the calculation is
extremely resource intensive and requires a large cluster in order to
run; a single map can take 10\,CPU\,years \citep{fowler2010}. Perhaps
the most promising maximum-likelihood method that may one day be
applied to \scuba\ data is ``SANEPIC'', which produced maps from
Balloon-borne Large-Aperture Submillimeter Telescope data, involving
hundreds of bolometers, while correctly incorporating inter-bolometer
noise correlations \citep[][]{patanchon2008}.

The third approach adopted here for \scuba\ is a compromise between
the previous two methods. Under the assumption that a significant
portion of the (predominantly low-frequency) non-white noise sources
can be modelled, an iterative solution is obtained for both the
astronomical image and the parameters of the noise model. Since the
remaining (non-modelled) noise sources are assumed to be white, a
single scalar \rms\ may be calculated for all of the data points from
a given bolometer to characterise its noise distribution (since the
noise at any instant in time is uncorrelated with others, and with the
data from other bolometers), greatly simplifying the inversion step
that is so complicated in the maximum-likelihood methods. In the
iterative approach memory usage scales linearly with $N$, the product
of the number of bolometers, $N_\mathrm{b}$, with the number of
samples in time, $N_\mathrm{t}$. The computation time is the product
of the number of iterations, $N_\mathrm{iter}$, with the time per
iteration, $t_\mathrm{iter}$. For the algorithm described in this
paper, $N_\mathrm{iter}$ is typically in the range $\sim$5--100
(depending on the \snr\ of large-scale structures, and the size of the
map), and $t_\mathrm{iter}$ scales as $N_\mathrm{t} \log
N_\mathrm{t}$, since Fast Fourier Transforms (FFTs) of the time-series
are typically performed. By comparison, the fastest maximum-likelihood
methods that account for inter-bolometer correlations have an
execution time with an $N_\mathrm{b}^2$ dependence, although it is
possible to limit memory usage to be linear in $N_\mathrm{t}$
\citep[e.g.,][]{patanchon2008}.

Iterative approaches to map-making have a long history in the
submillimetre, such as the pipeline recipe for fitting and removing
baseline drifts in SCUBA scan-map data as the astronomical image
estimate improved \citep{1999ASPC..172..171J}. The closest modern
relatives are the Comprehensive Reduction Utility
\citep[CRUSH,][]{kovacs2008} for the Submillimeter High-Angular
Resolution Camera 2 (SHARC-2), the pipeline developed for the Bolocam
Galactic Plane Survey \citep{aguirre2011}, and the Bolometer array
data Analysis software (BoA) for the Large APEX BOlometer CAmera
\citep[LABOCA,][]{schuller2012}.

A reasonable model for correlated noise in \scuba\ is a single
``common-mode'' signal seen by all of the bolometers. Iterative
estimation and removal of this signal significantly lowers the $1/f$
knee, without compromising structures on angular scales smaller than
the array footprint. Residual independent drifts at lower frequencies
are removed with an iterative high-pass filter. This strategy enables
SMURF to reduce data on time scales commensurate with the observing
times (or significantly faster), on single high-end desktop
computers. For reference, all of the data analysis in this paper was
performed on a machine with a 64-bit 8-core central processing unit
operating at 2.7\,GHz, and 48\,Gb of memory. SMURF has been
successfully used as part of a real-time pipeline \citep[based on the
Observatory Reduction and Acquisition Control - Data Reduction system,
or ORAC-DR,][]{gibb2005,2008AN....329..295C} offering feedback to
observers at the telescope. The pipeline is also used to generate
products for the JCMT Science Archive \citep{2011ASPC..442..203E}
hosted by the Canadian Astronomy Data Centre (CADC).

This paper is organized as follows. We first describe the properties
of \scuba\ data, including a principal component analysis to reveal
correlated noise features, in Section~\ref{sec:data}. Next, the
details of the SMURF algorithm (pre-processing steps and the iterative
solution) are given in Section~\ref{sec:algorithm}. The paper is
concluded in Section~\ref{sec:examples} with three detailed test cases
that span the majority of observation types likely to be undertaken
with \scuba, with an emphasis on the mitigation of divergence
problems, and characterising the output maps: (i) Uranus, a bright,
compact source (Section~\ref{sec:point}); (ii) the Lockman Hole, a
blind survey of faint point-like sources (Section~\ref{sec:cosmo});
and (iii) the star-forming region M17, including bright, extended
emission (Section~\ref{sec:extended}). All of the data analysed in
this paper are publicly available through the CADC \scuba\ raw-data
queries page for the dates and observation numbers given in the
text\footnote{\url{http://www.cadc-ccda.hia-iha.nrc-cnrc.gc.ca/jcmt/search/scuba2}}.
All of the analysis was performed using the Starlink \textsc{kapuahi}
release from 2012.

\section{\scuba\ data properties}
\label{sec:data}

In this section we summarise how \scuba\ works,
(Section~\ref{sec:bolos}), give examples of typical bolometers signals
(Sections~\ref{sec:bolosignal}), discuss the impacts of magnetic field
pickup (Section~\ref{sec:magpickup}) and sky noise
(Section~\ref{sec:skynoise}), and finally use of principal component
analysis to explore the correlated noise properties of \scuba\ data
(Section~\ref{sec:pca}).

\subsection{Description of \scuba}
\label{sec:bolos}

While the details of how \scuba\ works are described in
\citet{holland2013}, and its calibration in \citet{dempsey2013}, we
summarise the basic concepts relevant to map-making here.

Incoming light passes through a beam-splitter, and then bandpass
filters, providing simultaneous illumination and wavelength definition
of both the 450 and 850\,\micron\ focal planes. Each focal plane is
populated by four ``subarrays'' (labelled s4a--s4d at 450\,\micron,
and s8a--s8d at 850\,\micron), each consisting of 32 columns and 40
rows of bolometers. The bolometers are thermal absorbers (with a
response time constant of $\tau \sim 1$\,ms) coupled to
superconducting transition-edge sensors (TESs) for
thermometry. Temperature variations in the TESs produce changing
currents, and therefore varying magnetic fields, which are detected
and amplified using chains of superconducting quantum interference
devices (SQUIDs), before the larger output currents are
digitised. Each detector has its own SQUID for the first-stage of the
amplification, but the remaining stages occur within a common chain of
SQUIDs for each of the 32 columns. All 40 rows are read out in
sequence at a row-visit frequency of about 12\,kHz. Such a high sample
rate is unnecessary to produce maps, so the data are re-sampled to
approximately 200\,Hz before writing to disk. This rate provides a
sample every 3.0\,arcsec, or approximately one third of the
450\,\micron\ diffraction-limited full-width at half-maximum (FWHM) --
a typical rule-of-thumb for adequately sampling a Gaussian point
spread function (PSF) -- at the maximum scanning speed of
600\,arcsec\,s$^{-1}$. There is an additional 41st row of SQUID
readouts that are not connected to TESs. These ``dark SQUIDs'' track
non-thermal noise sources that are common to each column's amplifier
chain. The relationship between the output digitised current, $I$, and
the input power, $P$, or $dI/dP$ is established using flatfield
observations immediately prior to science observations, in which the
output signal is measured throughout a ramp of the pixel heaters
(which provide a known input power); see Section~2.1 in
\citet{dempsey2013} and Section~5.3 in
\citet{holland2013} for more details. Finally, the conversion to
astronomical flux units from power involves a correction for
atmospheric extinction \citep[primarily using the 183\,GHz (1.6\,mm)
JCMT Water Vapour Monitor to track line-of-sight opacity variations,
see Section~3 in][]{dempsey2013}, and the application of a flux
conversion factor (FCF) which is established from regular measurements
of calibrators such as Uranus \citep[Section~5 in][]{dempsey2013}.

In addition to uncorrelated white noise, and low-frequency (and often
correlated from bolometer to bolometer) $1/f$ drifts, bolometer power
spectra exhibit a roll-off at frequencies approaching Nyquist, which
is due to the anti-aliasing filter that forms part of the 200\,Hz
re-sampling stage. This anti-aliasing filter also introduces an
effective lag of $\sim$6\,ms. Adding this lag to the $\sim$1\,ms
thermal time constant, as well as a negative shift of about $-$2.5\,ms
caused by a half-sample offset due to the synchronization between the
pointing and bolometer data, yields a net lag of approximately
4.5\,ms. There are also line features in the spectra that are thought
to be produced primarily at frequencies far above the final 200\,Hz
sample rate, which are aliased to lower frequencies during the
multiplexed readout stage before they can be removed by the
anti-aliasing filter.

We note that the noise performance of \scuba\ has evolved over
time. An initial ``\scuba\ shared-risk observing'' period (S2SRO) took
place during February and March 2010, during which each of the 450 and
850\,\micron\ focal planes were populated with single subarrays, s4a
and s8d, respectively. In addition to having significantly fewer
available bolometers than the current fully-commissioned instrument
(and therefore a mapping speed reduced by a factor of $\sim$4),
instabilities in the fridge led to a large oscillating signal with a
period of about 25\,s that was correlated amongst the detectors. After
upgrading and commissioning the instrument, a servo using newly added
thermometers in the focal planes has effectively mitigated this
problem \citep[Section~2.5 in][]{holland2013}. In addition, there were
improvements to the magnetic shielding (reducing magnetic field
pickup, as described in Section~\ref{sec:magpickup}), as well as more
effective removal of aliased noise sources (which has reduced both the
presence of line features and the mean white noise level).

In order to reduce the impact of low-frequency noise on the final
maps, \scuba\ scan strategies have been designed to provide: (i) good
cross-linking (visiting every point of the mapped area at different
scan angles); and (ii) minimal accelerations to reduce turn-around
overheads (which could be quite large given the 600\,arcsec\,s$^{-1}$
maximum scan speed). For areas larger than the array footprint a
rectangular ``PONG'' pattern is used, in which the boresight travels
in approximately straight lines and ``bounces'' off the edges at
45\,degree angles until the area is uniformly filled in. It is also
usually combined with a rotation through a number of fixed position
angles to create a ``rotating PONG'' with even better
cross-linking. For smaller areas (of order the array footprint, or
point sources), in which the PONG turn-around overheads would be
large, a ``constant velocity daisy'' (CV Daisy) is used.  Here the
telescope moves in a circle, whose centre also slowly traces out a
small circle. For a more complete description of the \scuba\ observing
modes, see Section~5 in \citet{holland2013} and also
\citet{2010SPIE.7740E..66K}.

\begin{figure*}
\centering

\begin{minipage}[h]{0.495\linewidth}
\textbf{(a) S2SRO} \\

\includegraphics[width=\linewidth]{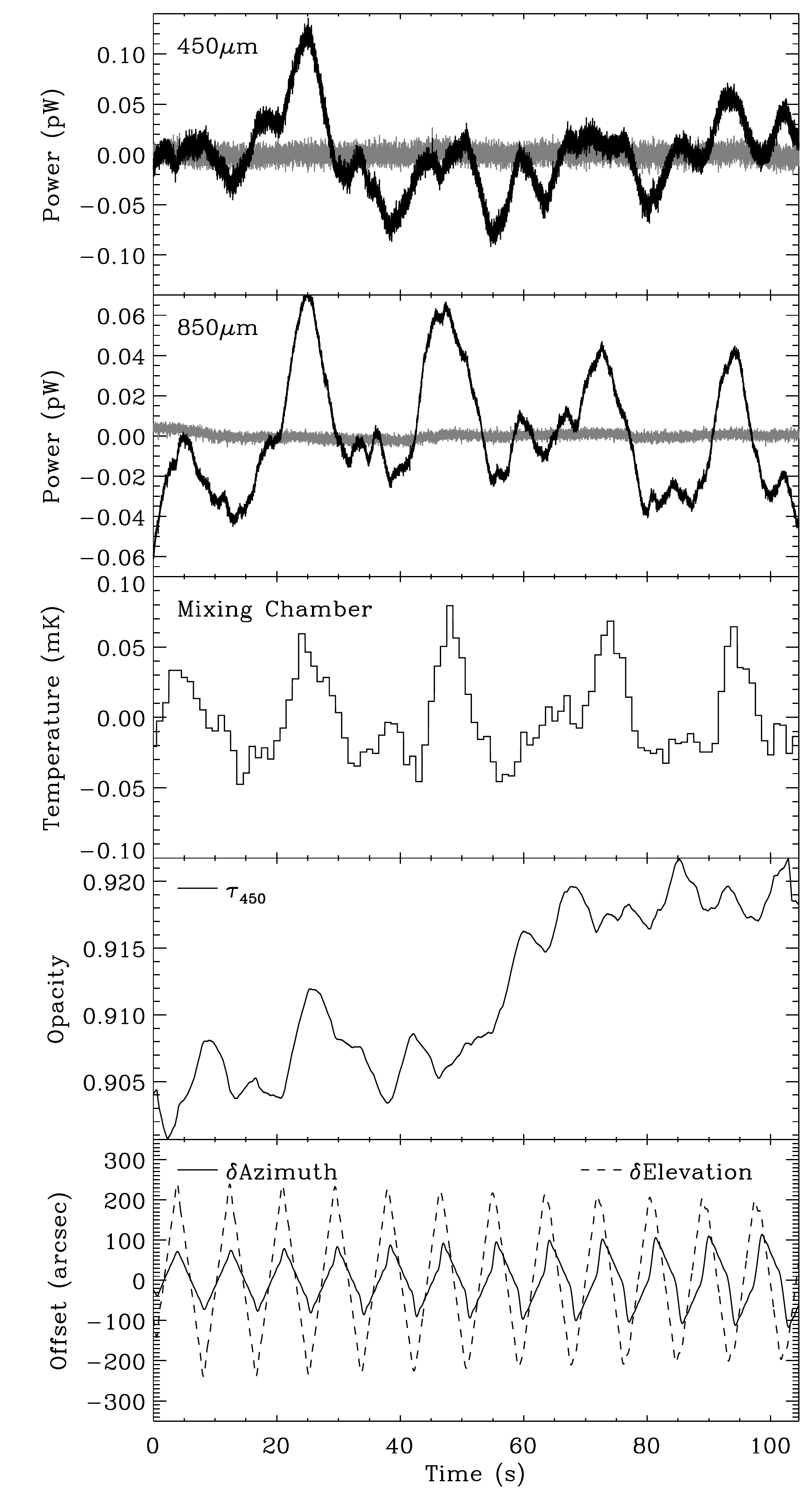}
\end{minipage}
\begin{minipage}[h]{0.495\linewidth}
\textbf{(b) fully-commissioned} \\

\includegraphics[width=\linewidth]{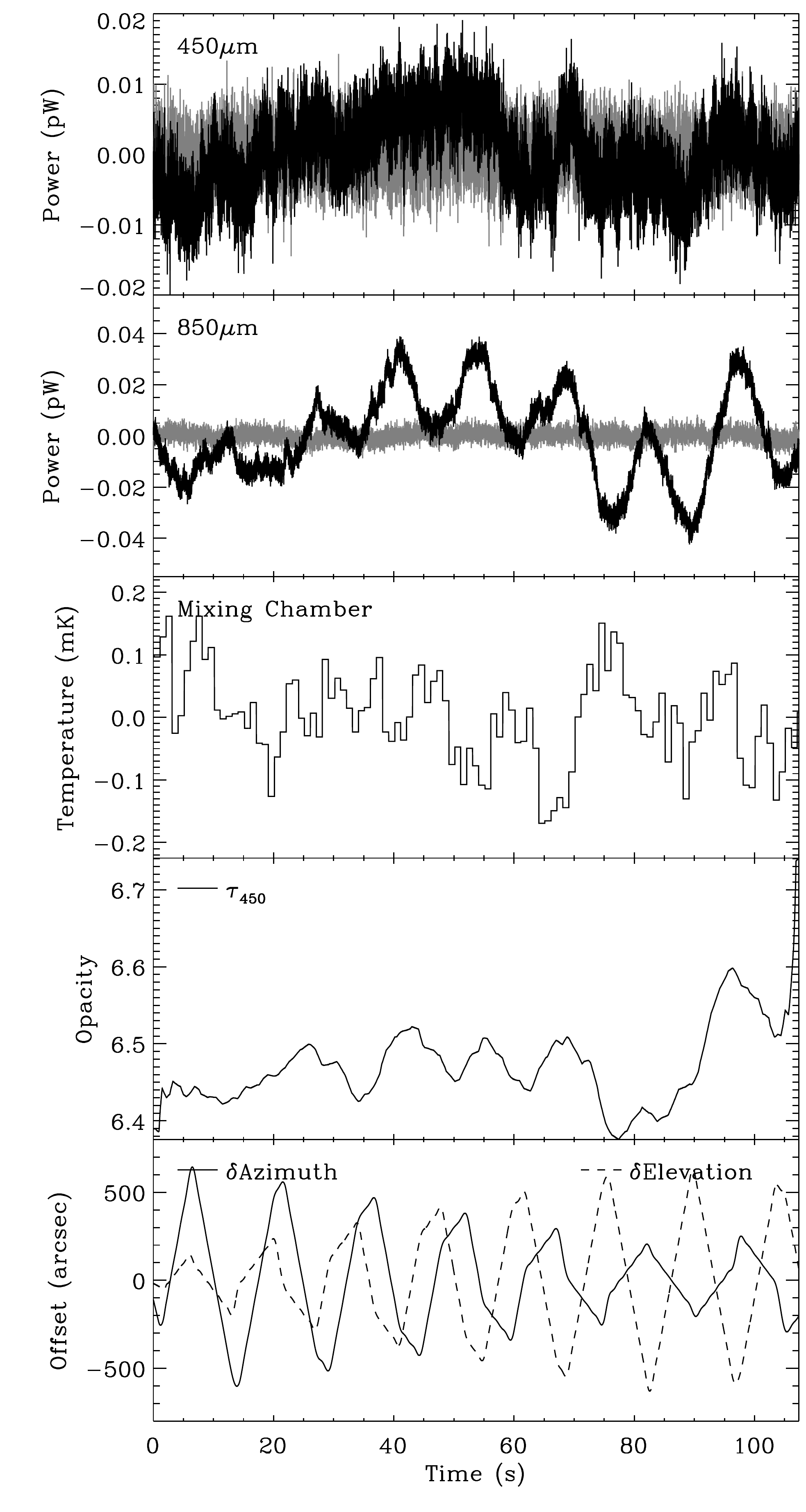}
\end{minipage}

\caption{A comparison between single bolometer time-series in each of
  the 450 and 850\,\micron\ bands with the mixing chamber temperature,
  the line-of-sight 450\,\micron\ opacity derived from the JCMT water
  vapour monitor, and azimuth/elevation pointing offsets, before and
  after upgrading the instrument. The grey signals over-plotted in the
  top two panels show the residual time-series after removing the
  common-mode signals.  (a) Data taken during the S2SRO period,
  observation 29 on 2010 March 13. There is a strong correlation
  between the bolometers and the roughly $\sim$25\,s oscillation in
  the fridge, but only a minor correlation with the opacity and
  telescope motion. The nearly flat common-mode subtracted signals
  show: (i) that most of the low-frequency signal is common to all of
  the bolometers; and (ii) the non-correlated, and predominantly white
  noise at 450\,\micron\ is significantly larger than at
  850\,\micron. (b) Data taken with the fully-commissioned instrument,
  observation 38 on 2011 November 12. Unlike the S2SRO data, there is
  no strong signal produced by variations in the fridge
  temperature. The 450\,\micron\ data have minimal correlated
  low-frequency noise (as evidenced by the lack of a difference
  between the raw and common-mode subtracted signals), although the
  850\,\micron\ data show a common signal correlated with the opacity
  and telescope motion (note that this scan has a significantly larger
  amplitude than the S2SRO data set), mostly likely caused by a
  combination of changing airmass (anti-correlated with opacity), and
  magnetic field pickup (see Section~\ref{sec:magpickup} and
  Fig.~\ref{fig:magpickup}).}
\label{fig:bolos_mix}
\end{figure*}

\begin{figure*}
\centering

\begin{minipage}[h]{0.495\linewidth}
\textbf{(a) S2SRO} \\

\includegraphics[width=\linewidth]{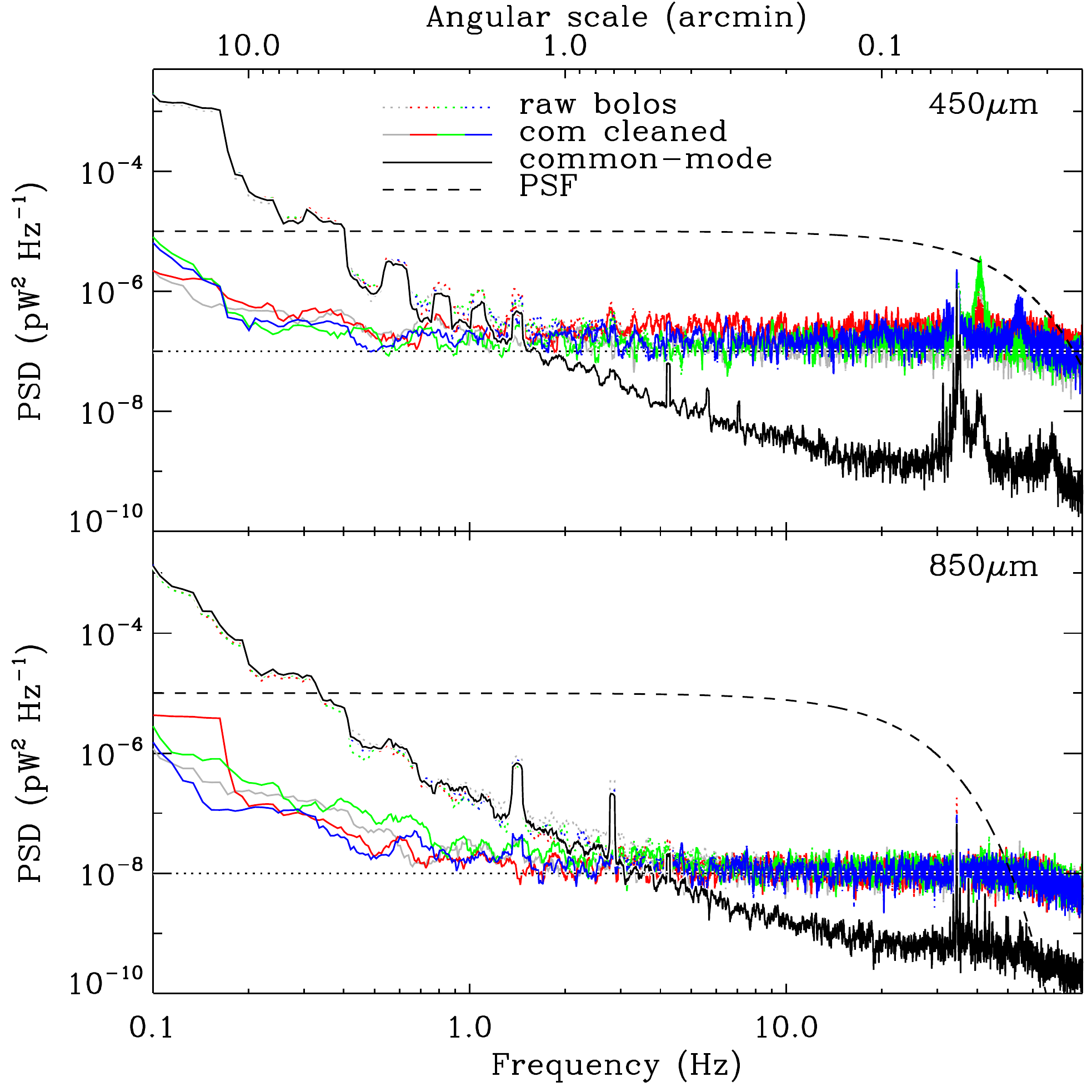}
\end{minipage}
\begin{minipage}[h]{0.495\linewidth}
\textbf{(b) fully-commissioned} \\

\includegraphics[width=\linewidth]{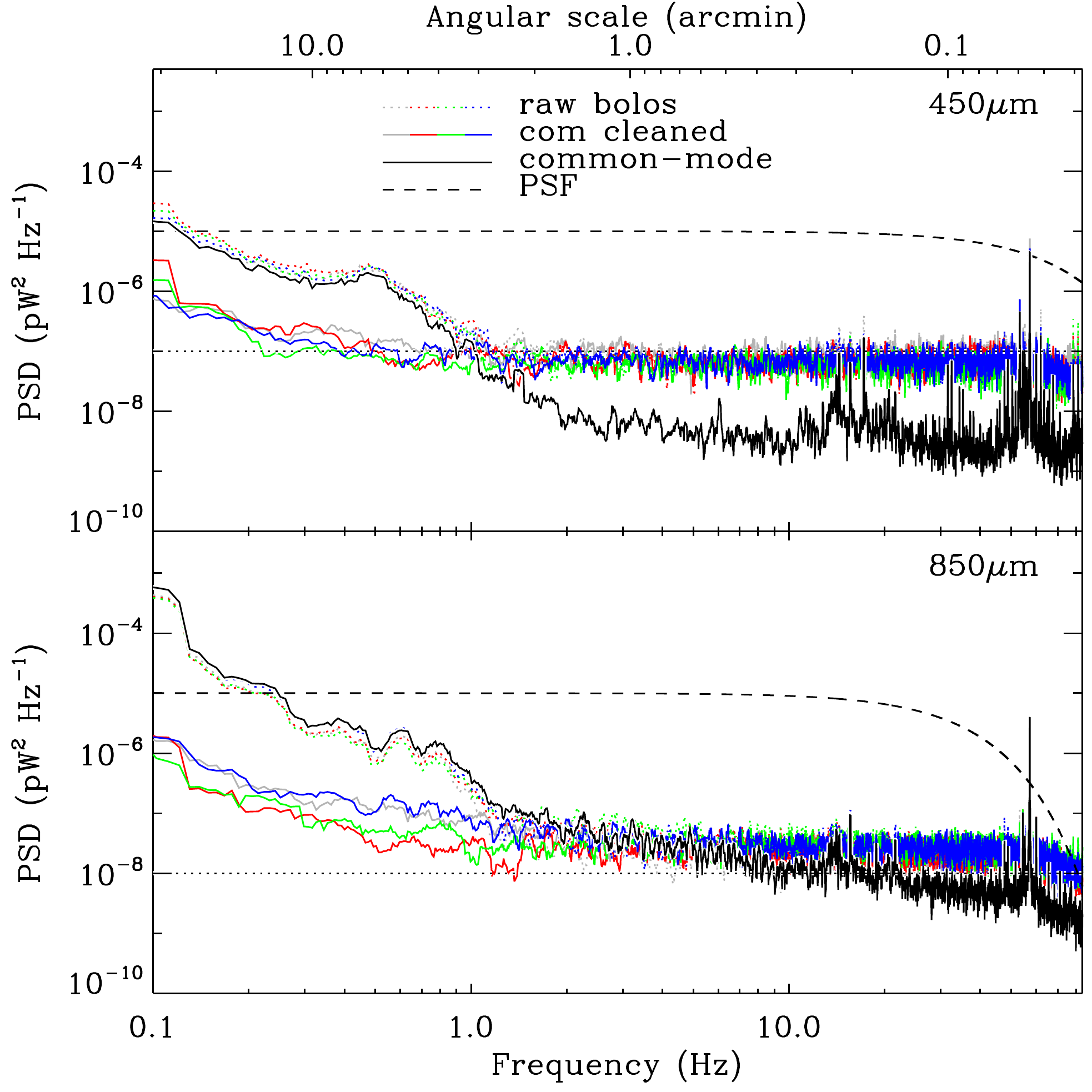}
\end{minipage}

\caption{Bolometer power spectral densities (PSDs) for the same two
  data sets (before and after upgrades) used in
  Fig.~\ref{fig:bolos_mix}. The PSDs have been boxcar smoothed with a
  width of 0.1\,Hz to reduce the noise slightly and clarify some
  features. Four of the most sensitive bolometers have been selected
  at each wavelength, and the flat-fielded and step-corrected (but
  otherwise raw) PSDs are shown as coloured dotted lines (the blue
  signals are for the same time series as those shown in
  Fig.~\ref{fig:bolos_mix}).  The solid black lines are the PSDs of
  the common-mode signals at each wavelength, and the solid coloured
  lines show the PSDs of the bolometers once the common-mode is
  removed.  Finally, the dashed black lines show the spectral shape
  produced by a point source given the scan speeds for the two
  observations (120\,arcsec\,s$^{-1}$ in (a), and
  190\,arcsec\,s$^{-1}$ in (b); for reference, the top horizontal axes
  shows the conversion from frequency to angular scale. Horizontal
  dotted lines at $10^{-7}$ and $10^{-8}$\,pW\,Hz$^{-1}$ at 450 and
  850\,\micron, respectively, are provided as a visual reference for
  the white-noise levels. In addition to $1/f$ and white noise
  components, and line features, all of the PSDs exhibit the gradual
  roll-off of the anti-aliasing filter just below the Nyquist
  frequency.  (a) For the S2SRO data, particularly at 450\,\micron,
  there are broad line features in the PSDs at both wavelengths above
  $\sim$35\,Hz.  At lower frequencies, the bolometer signals exhibit
  clear $1/f$ knees at approximately 2 and 3\,Hz at 450 and
  850\,\micron. Common-mode subtraction removes most of the correlated
  fridge oscillation signal, lowering the $1/f$ knees to approximately
  0.6 and 1.0\,Hz at 450 and 850\,\micron, respectively. (b) Data from
  the fully-commissioned instrument tend to have lower $1/f$ knees,
  and fewer line features. For this example, the improvement is most
  striking at 450\,\micron\ where the PSD is approximately two
  orders-of-magnitude lower in the fully-commissioned data at
  0.1\,Hz. The white noise performance, however, is similar for the
  two subarrays (s4a and s8b) that were used both before and after the
  upgrades.}
\label{fig:pspec}
\end{figure*}

\subsection{Typical bolometer signals}
\label{sec:bolosignal}

In Fig.~\ref{fig:bolos_mix} we show sample time-series from single
bolometers in each of the 450 and 850\,\micron\ focal planes, as well
as variations in the mixing chamber temperature (though not located in
the focal plane itself, it is certainly correlated with the
temperature of the detectors), the line-of-sight atmospheric opacity
as derived from the JCMT water vapour monitor (WVM), and the telescope
pointing, for two data sets, before and after the upgrades that
followed S2SRO.

In the S2SRO data (Fig.~\ref{fig:bolos_mix}a), observation 29 on 2010
March 13, both bolometers share significant long-timescale structure
($\gsim10$\,s) that appears to be related to variations in the fridge
base temperature, although the similarity is clearly greater at
850\,\micron. Note that the total power in the fluctuations at
450\,\micron\ are comparable to those at 850\,\micron\ as one might
expect if there is a comparable varying thermal load from the fridge
at each wavelength that dominates. We also note that there is no
obvious strong correlation between the low-frequency signal structure,
at either wavelength, with the telescope motion. However, there is a
suggestion that the shorter-timescale behaviour of the opacity is
anti-correlated with the elevation, as expected.

The low-frequency signal component of the S2SRO bolometer output is
also highly correlated amongst bolometers in the same subarray. We
have calculated a common-mode signal, $\mathbf{c}(t)$, as the average
time-series of all the working bolometers. We then fit the amplitude
of $\mathbf{c}(t)$ at each wavelength to the signals shown in
Fig.~\ref{fig:bolos_mix} and remove it, yielding the grey residual
signals. These residuals are nearly white, although still with
noticeable long-timescale variations. The white noise is greater at
450\,\micron\ as one would expect from the larger backgrounds compared
to 850\,\micron.

Data from the fully commissioned instrument, observation 38 on 2011
November 12, are shown in Fig.~\ref{fig:bolos_mix}b. Having solved the
fridge oscillation problem, these data no longer exhibit a correlation
with the mixing chamber (however, note that variations are still seen
in the mixing chamber signal; such variations do not necessarily
reflect changes in the focal plane temperature). There is minimal
correlated low-frequency noise in these 450\,\micron\ data, resulting
in little difference between the raw and common-mode subtracted
data. The 850\,\micron\ channel, however, exhibits a more significant
signal that is obviously correlated with the telescope motion. It is
also correlated amongst many of the detectors, and common-mode removal
corrects it to a large extent. Part of the reason that this is seen
here, and not in the S2SRO data set shown, is that the amplitude of
the scan pattern is larger. This ``scan-synchronous'' noise is
attributed to a combination of magnetic field pickup, as described in
Section~\ref{sec:magpickup}, and sky brightness variations due to
changes in elevation (not to be confused with underlying changes in
the atmosphere; a homogeneous atmosphere will appear brighter with
increasing airmass). Also similar to the S2SRO data, an
anti-correlation between the opacity and elevation is apparent.

Next, in Fig.~\ref{fig:pspec} we produce power spectral density (PSD)
plots for four of the most sensitive bolometers from both focal
planes, using the same two data sets. To produce this figure, we
follow the convention that the PSD as a function of frequency,
$\mathbf{P}(f)$, is normalised such that the integral over frequency
gives the same variance as the time-series variance across the full
time-series. In other words, given a bolometer signal $\mathbf{b}(t)$,
\begin{equation}
\label{eq:psd}
\langle\mathbf{b}^2(t)\rangle = 2 \int_0^{f_\mathrm{N}} \mathbf{P}(f)
df ,
\end{equation}
where we only integrate over the positive frequencies up to
$f_\mathrm{N}$, the Nyquist frequency, and the factor of 2 accounts
for the (equal) power that appears at negative frequencies in the
discrete Fourier Transform. The units of the PSD written in this form
are pW$^2$\,Hz$^{-1}$.

The dotted coloured lines in Fig.~\ref{fig:pspec} show the PSDs for
raw, though flat-fielded and step-corrected (Section~\ref{sec:steps})
data. At each wavelength, and in both data sets, there are clear $1/f$
knees at frequencies ranging from roughly 1 to 2\,Hz, followed by a
predominantly white spectrum punctuated by line features at higher
frequencies, and finally a roll-off caused by the anti-aliasing filter
above $\gsim 70$\,Hz. The correlation between the low-frequency
components of the different bolometer signals is large. The solid
black lines in Fig.~\ref{fig:pspec} indicate the PSDs of the
common-modes $\mathbf{c}(t)$ at each wavelength, which reproduce much
of the low-frequency structure, as well as some of the
higher-frequency line features. The $\mathbf{c}(t)$ otherwise drop
substantially below the individual bolometer PSDs at high frequency,
as expected if the bolometers are dominated by uncorrelated white
noise in that part of the spectrum. The common-mode signals are fit to
each bolometer time series and removed as in Fig.~\ref{fig:bolos_mix},
and the resulting PSDs are shown as solid coloured lines.  For
reference, the top horizontal axes have been converted to angular
scale assuming the scan speeds of each observation
(120\,arcsec\,s$^{-1}$ in Fig.~\ref{fig:pspec}a, and
190\,arcsec\,s$^{-1}$ in Fig.~\ref{fig:pspec}b). The power spectra of
unresolved point-sources are also shown as dashed lines in each band
(arbitrarily normalised), showing that the smallest features
resolvable by the telescope are only minimally affected by the excess
noise in the line features at these scan speeds (this may not be the
case at higher scan speeds).

In the S2SRO data (Fig.~\ref{fig:pspec}a), the low-frequency noise is
very correlated between the detectors (note the tight scatter in the
dotted coloured lines, and their resemblance to the common-mode), due
to it being dominated by the fridge oscillations. Common-mode
subtraction is very effective, reducing the $1/f$ knees in both
wavelengths by about a factor of 5.

Raw data from the fully-commissioned instrument
(Fig.~\ref{fig:pspec}b) generally have less significant $1/f$ noise as
compared with the S2SRO data, and it is considerably less correlated
amongst detectors (larger spread in the dotted coloured lines,
particularly noticeable in these 450\,\micron\ data), leading to a
less drastic improvement upon common-mode removal. However, since the
data are less dominated by low-frequency noise to begin with, the
signals generally require less aggressive high-pass filtering to
produce maps, and therefore retain larger-scale structures than with
the S2SRO data. Furthermore, there are generally fewer line features
in the PSDs of bolometers in the fully-commissioned
instrument. Finally, these two data sets illustrate that the white
noise performance (NEFD) is in fact similar before and after the
upgrades for these two subarrays (s4a at 450\,\micron, and s8b at
850\,\micron). The main improvements are a reduction in correlated and
line noise features mentioned above, the larger number of working
bolometers and field-of-view (approximately a factor of 4).

\subsection{Magnetic field pickup}
\label{sec:magpickup}

\begin{figure}
\centering
\includegraphics[width=\linewidth]{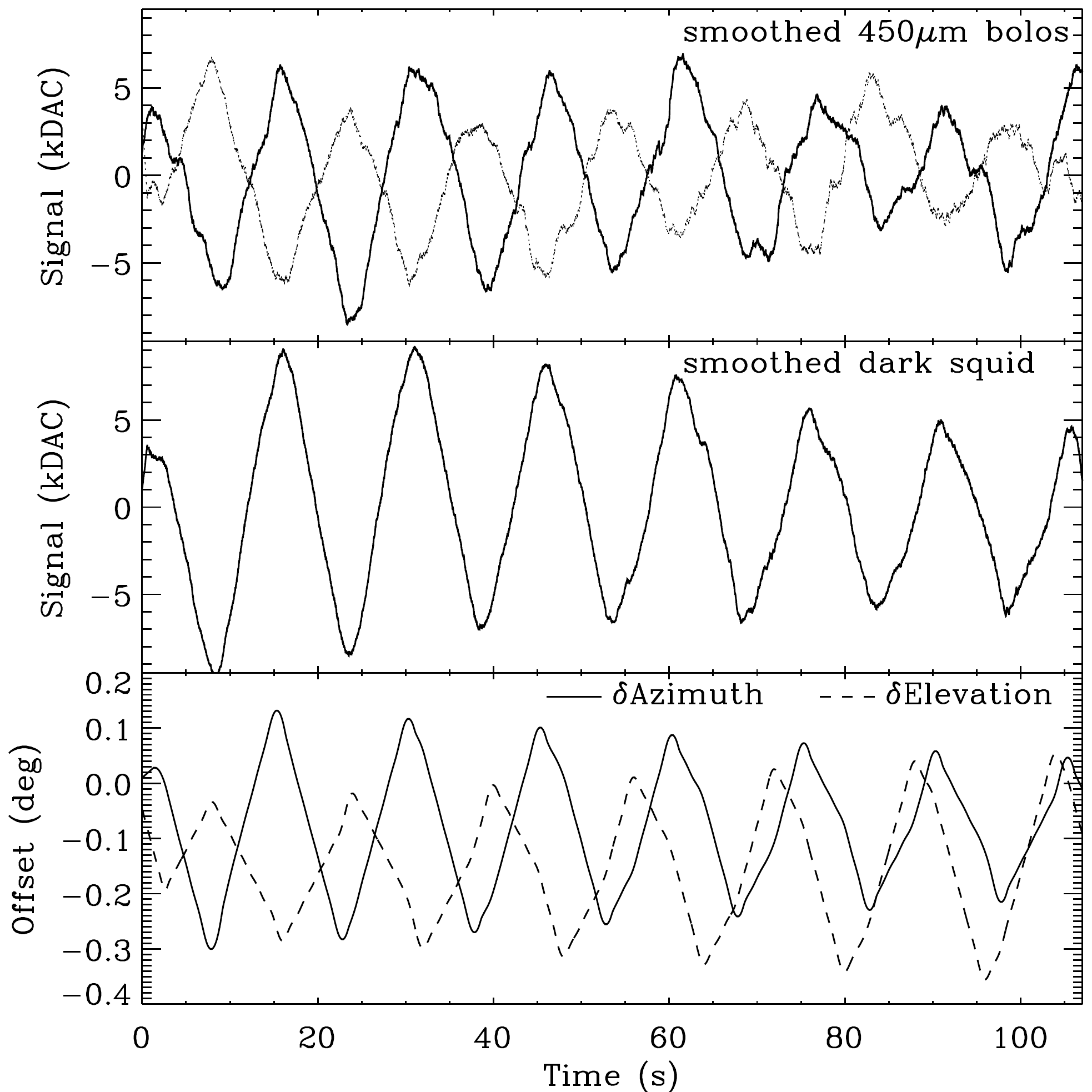}
\caption{Evidence for significant magnetic field pickup for
  observation 16 on 2010 February 28 at 450\,\micron\ (s4a subarray).
  The top panel shows two un-flatfielded (but mean-subtracted and step
  corrected) bolometer time-series from the same column, with a 200
  sample boxcar smoothing (approximately 1\,s), illustrating that they
  are dominated by a similar signal with opposite signs. The second
  panel shows the dark squid signal for the column, also
  mean-subtracted and with the same boxcar smoothing. The bottom panel
  shows the azimuthal and elevation offsets from the map centre (mean
  azimuth and elevation 171.9$^\circ$ and 68.0$^\circ$,
  respectively). Only the azimuthal signal is obviously correlated
  with the dark squids and bolometer signals, which suggests a
  magnetic field stationary with respect to the telescope dome as the
  source, since its direction with respect to the cryostat only
  changes with azimuthal motion.}
\label{fig:magpickup}
\end{figure}

An additional noise source that is significant primarily for wide-area
scans (and more so during the S2SRO period) is magnetic field
pickup. Since the bolometer signals are detected through the
amplification of magnetic fields, any additional changing fields
within the instrument will add to the noise.

Example data from the 450\,\micron\ subarray s4a where pickup appears
to be significant (observation 16 on 2010 February 28) are shown in
Fig.~\ref{fig:magpickup}. The time-series for two bolometers in the
same column (not flatfielded) show that there is a strong signal with
a similar shape, but opposite signs. This behaviour is seen across the
entire array. The dark squid signal for the same columns exhibits a
similar shape and amplitude. Since the dark squid has no thermal
absorber or TES attached to it, this observed signal is not likely to
be optical or thermal in nature (although there can be some cross-talk
with the bolometers). Due to the fact that the sign of the gain in
each stage of SQUID amplification is random (although the combined
gain is constrained to be negative), and since magnetic field pickup
is only seen at the input to the second stage, the pickup can appear
with random signs for bolometers along a column, giving it a distinct
signature from other common signals that always appear with the same
sign.

The telescope pointing offsets for this approximately 0.5\,degree
diameter scan are also shown in Fig.~\ref{fig:magpickup}. Since the
phase of the azimuth offsets from the map centre in this scan pattern
slowly drifts with respect to the elevation offsets, it is clear that
the bolometer and dark squid signals are detecting a noise source that
is correlated only with the azimuthal motion and not the
elevation. This behaviour would be expected if if there were a strong
magnetic field fixed with respect to the telescope dome (i.e., the
earth's magnetic field). Since \scuba\ is mounted on a Nasmyth
platform, only azimuthal motion will change the direction of such a
field with respect to the cryostat. Tests have shown that, as in this
example, large scans in azimuth generically produce pickup. In
contrast, and as expected, changes in elevation results in pickup that
is approximately three orders-of-magnitude smaller.

\subsection{Sky noise}
\label{sec:skynoise}

\begin{figure}
\centering
\includegraphics[width=\linewidth]{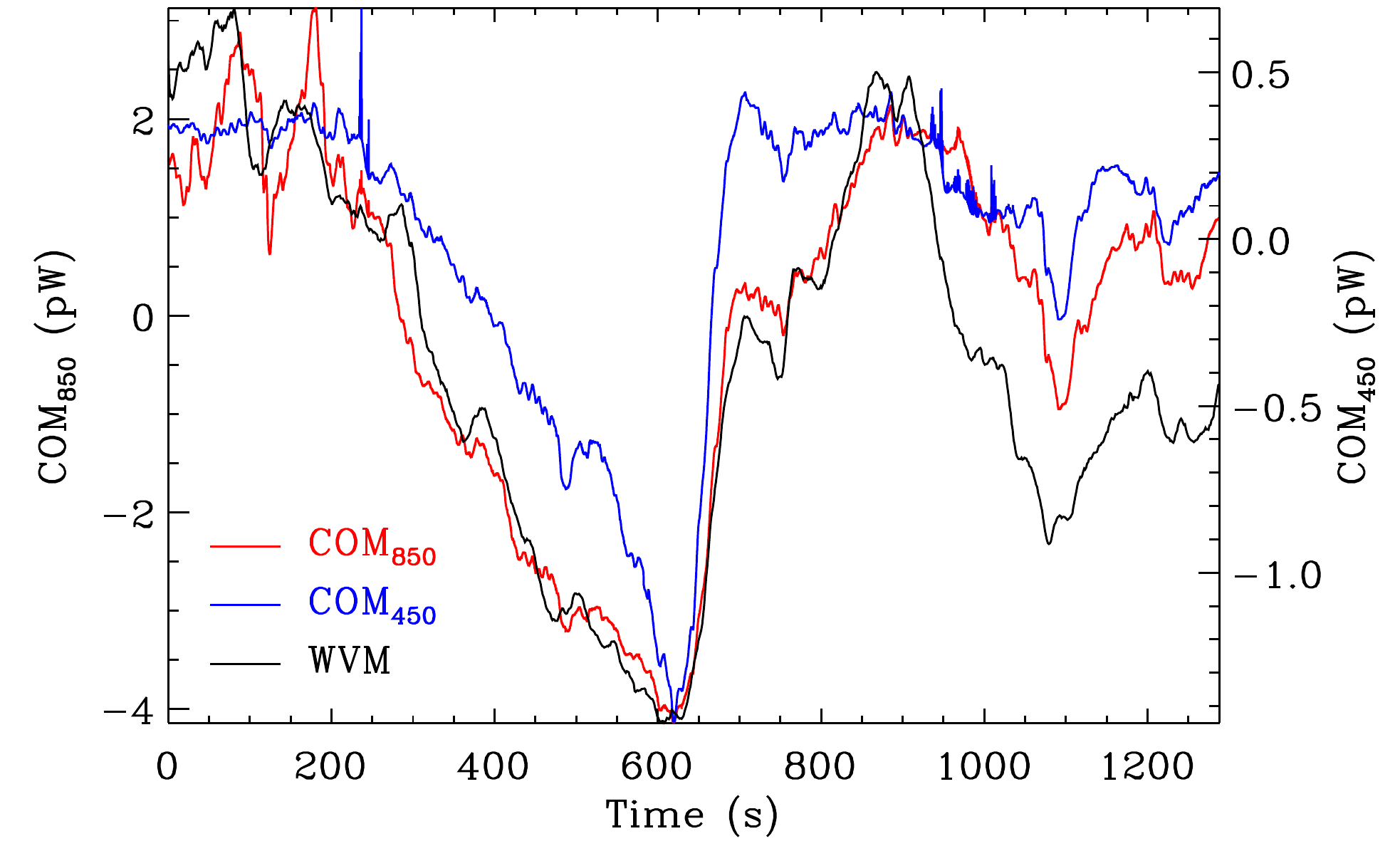}
\caption{Illustration of the dominance of atmospheric variations on
  long-timescales using 35\,min. of data taken in highly-variable
  weather conditions (line-of-sight 450\,\micron\ opacity ranging from
  3.6 to 7.7) from the fully-commissioned instrument, observation 18
  on 2011 March 24. Red and blue lines show the 850 and 450\,\micron\
  common-mode signals, corresponding to the left- and right- vertical
  axes, respectively. The black line shows the line-of-sight opacity
  from the WVM as a proxy for atmospheric emission (arbitrary
  scale). Bolometer baseline drifts on timescales $\gsim$100\,s are
  clearly correlated with the atmosphere in this case.}
\label{fig:skynoise}
\end{figure}

The data shown in Figs.~\ref{fig:bolos_mix} and \ref{fig:pspec} were
taken during fairly stable weather conditions (although with different
450\,\micron\ line-of-sight opacities $\sim$0.9 and 6.5 in the S2SRO
and post-upgrade data sets, respectively), and are dominated at low
frequencies ($\lsim$2\,Hz) by scan-synchronous noise, rather than
underlying variations in the atmosphere, or ``sky noise''. However,
only $\sim$100\,s of data were analysed. In Fig.~\ref{fig:skynoise},
an order-of-magnitude longer (fully-commissioned) data set is plotted
from observation 18 on 2011 March 24, during highly variable weather
conditions (450\,\micron\ line-of-sight opacities ranging from 3.6 to
7.7, scaled from the WVM data which are shown as a black line). Unlike
the earlier plots, it is clear that the bolometer baseline drifts on
timescales $\gsim$100\,s are tightly correlated with the atmospheric
opacity, which serves as a proxy for atmospheric emission. The weaker
correlation between the WVM and the 450\,\micron\ bolometers is
probably due to the atmosphere being more transmissive at 850\micron,
therefore the emission detected at 450\,\micron\ comes from water
vapour that is closer to the telescope than that observed at the
longer wavelengths.

While the relative importance of sky noise at low frequencies is
highly time- and scan pattern-dependent, in general it does not have a
major impact on map-making. Upon common-mode removal (which is an
integral part of our map-making strategy) uncorrelated noise remains
(e.g., solid lines in Fig.~\ref{fig:pspec}). As we will see in
Section~\ref{sec:pca} and \ref{sec:experimental}, this remaining noise
does not have a smoothly-varying correlation pattern across the focal
plane as one might expect if it were due to resolved ``clouds'' of
atmospheric emission. A high-pass filter with an edge frequency in the
range $\sim$0.1--1.0\,Hz is usually employed to remove this residual
noise (Section~\ref{sec:components}), and will also remove any sky
noise that is not already subtracted by the common-mode.

\subsection{Principal component analysis}
\label{sec:pca}

A method that has been used to remove correlated noise as part of the
map-making procedure for Bolocam and AzTEC is Principal Component
Analysis \citep[PCA,][]{laurent2005,scott2008,perera2008}. Here we use
PCA to further explore correlated signals in \scuba\ data.

The basic method is as follows: (i) a covariance matrix is built up
for all pairs $(i,j)$ of the $N$ bolometer time-series,
$\langle\mathbf{b}_i(t),\mathbf{b}_j(t)\rangle$; and (ii) a singular
value decomposition identifies a new set of statistically uncorrelated
basis vectors, $\mathbf{\xi}_i(t)$ (i.e., whose covariance matrix is
diagonal), such that each of the bolometer time-series is a linear
combination of the basis vectors, or components i.e., $\mathbf{b}_i(t)
= \bar{\mathbf{\xi}} \mathbf{\lambda}_i^\mathrm{T}$, where each row of
the matrix $\bar{\mathbf{\xi}}$ is a basis vector, and
$\mathbf{\lambda}_i^\mathrm{T}$ is a column vector containing the
corresponding amplitudes. The $\mathbf{\xi}_i(t)$ are normalised by
$[\sum_t \mathbf{\xi}_i^2(t)]^{1/2}$, such that the amplitude of each
component is proportional to its \rms. In the earlier analyses
mentioned, the low-frequency noise is assumed to be encapsulated in
those components with the largest eigenvalues. Removing the projection
of the time series along those components then significantly reduces
$1/f$ noise while retaining most of the (higher-frequency) signal in
point-like sources

\begin{figure*}
\centering
\includegraphics[width=0.85\linewidth]{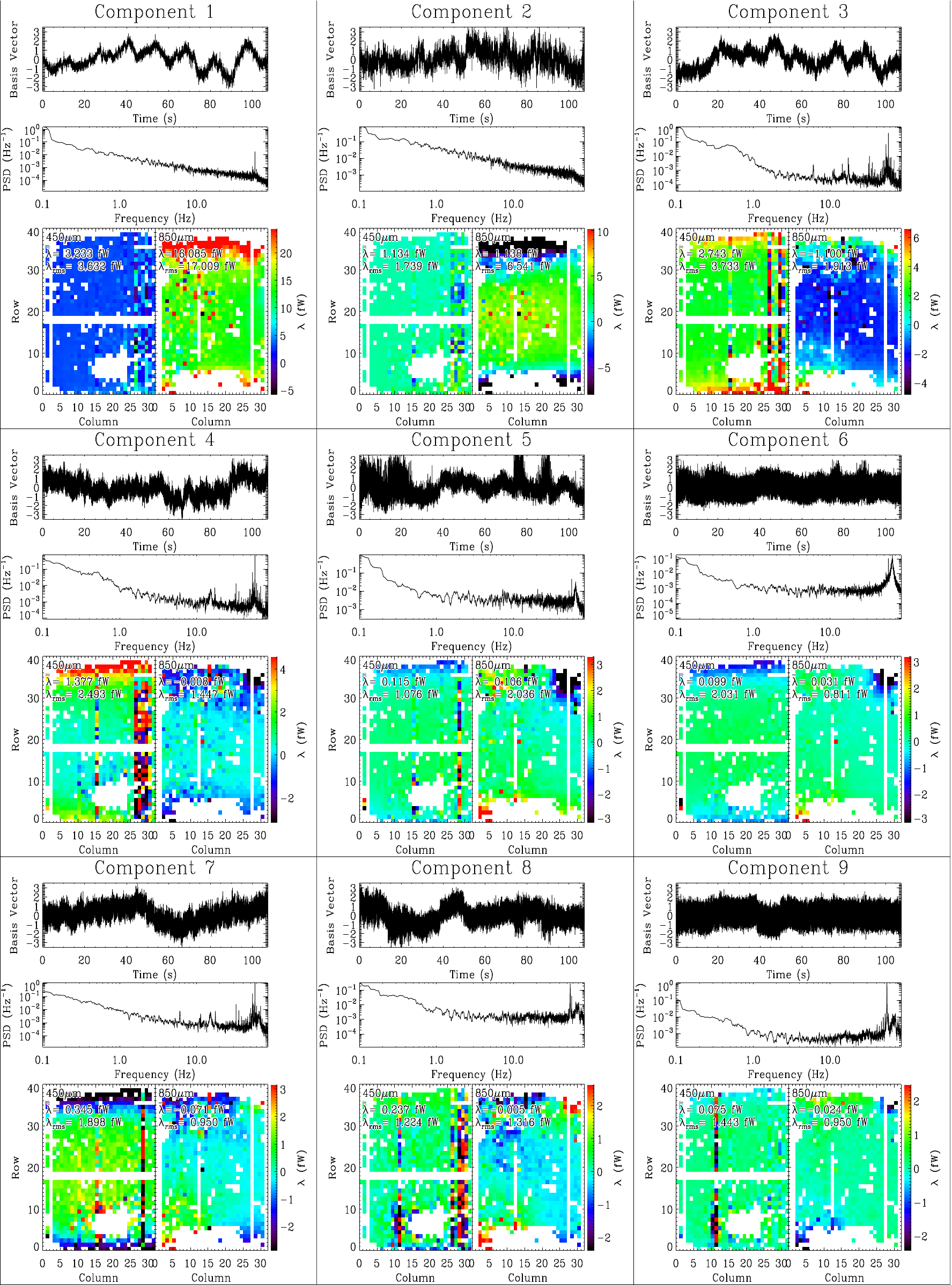}
\caption{The first nine components from a principal component
  analysis, ranked by the mean amplitudes of combined 450 (s4a
  subarray) and 850\,\micron\ (s8b subarray) time-series bolometer
  data, for the same observation as that used in
  Figs.~\ref{fig:bolos_mix}b and \ref{fig:pspec}b. For each component
  the top plot shows the time-series of the basis vectors normalised
  by their \rms, the middle plot its PSD, and the bottom coloured
  panels the ampliudes for the bolometers at each wavelength (the
  contribution of the basis vectors to the time-series). For
  reference, both the mean, $\bar{\lambda}$, and \rms,
  $\lambda_\mathrm{rms}$, amplitudes for the bolometers in each
  subarray are also shown. Most of the correlated $1/f$ signal is
  encapsulated in Component~1, and appears to be dominated by
  scan-synchronous noise (compare with the scan pattern in
  Fig.~\ref{fig:bolos_mix}b). Some of this signal also appears in
  Component~3. Component~2 is another $1/f$ signal that is dominant in
  the 850\,\micron\ subarray. Most of the remaining components consist
  of weaker $1/f$ noise and line features predominantly above
  $\sim$30\,Hz. The amplitude maps show correlated patterns along
  columns suggesting pickup in their common amplifier chains.}
\label{fig:pca}
\end{figure*}

A novel feature of our \scuba\ analysis is that we can perform PCA
with 450 and 850\,\micron\ data simultaneously, potentially helping us
to differentiate thermal and optical noise signals (e.g., from the
atmosphere) that might appear in both wavelengths, from other noise
sources that are restricted to single subarrays (such as readout
noise). In Fig.~\ref{fig:pca} we show the results for a combined
analysis of the s4a and s8b subarrays for the first $\sim$100\,s of
the same observation that was used in Figs.~\ref{fig:bolos_mix}b and
\ref{fig:pspec}b. The nine most significant components are shown
(ranked by the mean amplitudes across all working bolometers in both
bands). The top and middle panels for each component show the
time-series and PSDs of the normalised basis vectors. The bottom
panels are maps indicating the significance (amplitudes) of the
component across the focal plane (with the mean $\bar{\lambda}$ and
and \rms, $\lambda_\mathrm{rms}$, of the amplitudes for all of the
bolometers calculated separately at each wavelength also shown).

The majority of the correlated signal at both wavelengths in
Fig.~\ref{fig:pca} is produced by Component~1, in which the basis
vector time-series exhibits a roughly periodic signal that resembles
the scan pattern in Fig.~\ref{fig:bolos_mix}b. While this correlated
signal appears at both wavelengths, referring to the maps of
amplitudes, they are considerably stronger at 850\,\micron\
(consistent with the visual appearance of the bolometer signals in
Fig.~\ref{fig:pca}). Also note that the PSD for this basis vector
exhibits nearly a pure $1/f$ signature with minimal line features, and
no white-noise plateau (suggesting that these are high-\snr\
measurements of a purely low-frequency drift). Component~3 is weaker,
yet also shows evidence of the scan-synchronous noise, combined with
high-frequency line features that appear primarily in columns 26, 29
and 30 of the 450\,\micron\ subarray. Unlike Component~1, it is
generally more dominant at 450\,\micron.

Why are there two components that seem to be dominated by
scan-synchronous noise (which is presumably a mixture of magnetic
field pickup and brightness variations in the atmosphere)?  First, the
smoothly-varying gradients in the amplitude maps (primarily at
850\,\micron) suggests that this noise has a different response across
the focal plane; in this case the PCA has identified two orthogonal
shapes that, when mixed in different quantities, can reproduce most of
the signal in each bolometer. It is also likely that the atmospheric
contribution is stronger at 450\,\micron\ (i.e., predominantly
Component~3).

Like Component~1, Component~2 is a purely $1/f$ signal, with an
amplitude comparable to Component~3, that appears almost exclusively
at 850\,\micron. Since it is not obviously correlated with the
telescope motion, yet has an amplitude that varies smoothly across the
850\,\micron\ subarray, it may represent the PCA's attempt to reproduce
some non-linear response to magnetic field and/or atmospheric pickup

The remaining, weaker, Components~4--9 share some common
features. They all have some degree of $1/f$ noise, suggesting further
residual noise related to the telescope motion and/or other slow
baseline drifts. All of them also exhibit line features in their
spectra, which in many cases appear at random amplitudes along columns
(e.g. columns 15, and 26--30 at 450\,\micron\ in Component~4), or in a
sampling of isolated bolometers (e.g. column 13, row 19 at
850\,\micron\ in Component~6). The signals that appear along columns
are probably related to magnetic field pickup. Other sources of lines
are thought to be aliased high-frequency noise, and also the 60\,Hz
mains which appear in some bolometers.

This example illustrates some of the types of correlated signals and
patterns that PCA can identify in \scuba\ data. While there are
typically one or two strong signal components detected, in general,
the details can vary significantly from data set to data set, and the
lengths of the time-series analysed. The reason for this is that many
of the noise sources are not stationary in time (e.g.,
scan-synchronous noise which obviously depends on the scan pattern,
and the tuning of the subarrays). It should also be clear from this
example that while PCA offers some helpful insight into the various
sources of noise, it does not necessarily identify clean patterns that
can easily be modelled. For example, while there are clearly
correlation patterns along columns as expected from the common
amplification chain, the intensities of these high-frequency signals
appear random in the eigenvalue maps, and a simple common-mode signal
estimated from the data for each column would not remove
it. Regardless, the dominant noise in excess of the fundamental white
noise limit, in all cases, is at low-frequencies. This conclusion is
central to our data reduction strategy described in
Section~\ref{sec:algorithm}.

\section{Production of maps}
\label{sec:algorithm}

The approach taken by SMURF to reducing \scuba\ data is to model (and
remove) predominantly low-frequency noise sources that are correlated
amongst detectors, and to iterate this process along with estimates of
the map. Significant experimentation with different models for noise
sources during \scuba\ commissioning required a highly flexible
software framework, and a configurable user interface. To achieve this
goal, while minimizing development time, it was decided to build SMURF
as a Starlink package \citep{2009ASPC..411..418J}, which provides
access to a large suite of libraries (including a commanding and
messaging interface, astrometric coordinate conversions and generation
of standard WCS information, file formats etc.). Furthermore,
Starlink\footnote{\url{http://www.starlink.ac.uk}} is
open-source\footnote{\url{https://github.com/Starlink}} (distributed
under the GNU General Public Licence v3), and already used extensively
at the Joint Astronomy Centre (host of the JCMT) for many other
systems \citep{jenness2011}, which helps with interoperability. Though
originally written in FORTRAN, many of the core Starlink libraries are
now ported to native C, or at least have a C interface. It was
therefore decided to develop SMURF in C as well (rather than C++, for
example, which would have required adding further dependencies to
Starlink), although we have taken an object-oriented approach. For
example, all of the data for a given signal component model are
encapsulated in a C structure, and there is a standard interface for
all functions that handle models (member data and functions for the
class, respectively). In this way it is easy to extend SMURF with new
models. Parallelisation is incorporated in the most time-consuming
low-level routines using threads (e.g., performing Fast Fourier
Transforms, re-gridding the data), usually handling either independent
blocks of bolometers, or blocks of time, in each thread, depending on
the nature of the calculation. We have found that for typical data
sets the processing time scales well with the number of central
processing unit (CPU) cores, although beyond 8 the returns are
diminished due to tasks that cannot be parallelised (e.g., reading the
data from disk).

\begin{figure}
\centering
\includegraphics[width=\linewidth]{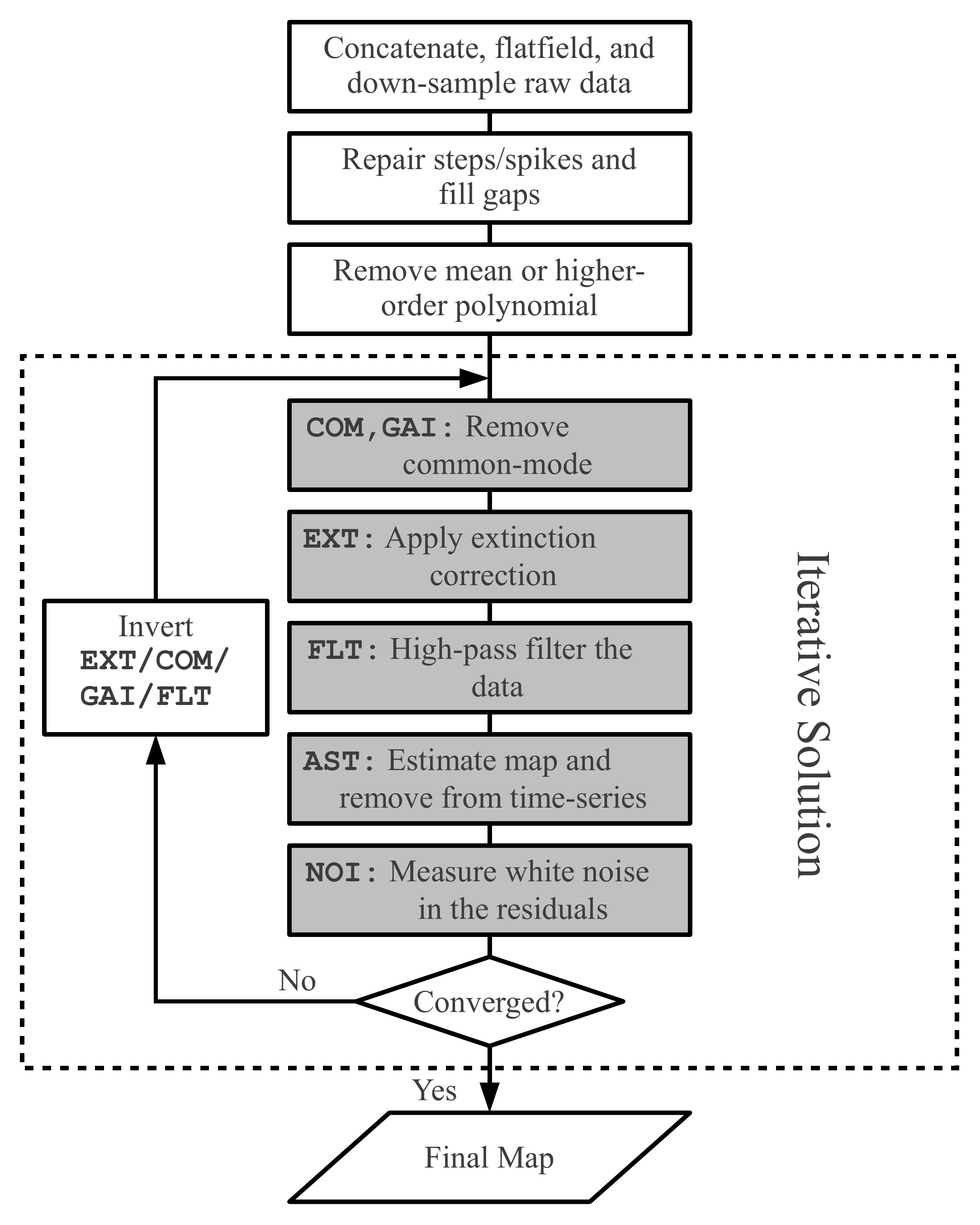}
\caption{Typical map-making algorithm. Raw data (stored in multiple
  files) are read and concatenated into continuous time-series, and
  flatfieded. Based on the scan speed of the telescope, the data are
  down-sampled to match the selected output map pixel size (usually 2
  and 4\,arcsec at 450 and 850\,\micron, respectively, to ensure
  adequate sampling of the \scuba\ PSFs). A cleaning stage repairs
  steps and spikes, and fills gaps to ensure continuity. Finally, the
  mean of each bolometer time-series is removed (higher-order
  polynomials may also be used). The iterative portion then begins
  (dashed box): estimating and removing the common-mode signal (the
  combined \model{COM} and \model{GAI} models); applying the
  extinction correction (\model{EXT}); high-pass filtering to remove
  residual independent low-frequency noise (\model{FLT}); estimating
  the map by re-gridding the data, and then removing its projection
  from the time-series (\model{AST}); and finally measuring the noise
  in the residual time-series (\model{NOI}). If the solution has
  converged, the map is written to disk. Otherwise any multiplicative
  factors that may have been applied to the data are inverted (i.e.,
  the extinction correction, \model{EXT}), and then the models for the
  low-frequency noise; the common-mode (\model{COM,GAI}) and high-pass
  filter (\model{FLT}). Each model is then re-estimated in turn until
  \model{AST}, at which point the previous estimate of the
  astronomical signal is added back into the data prior to its
  re-calculation.}
\label{fig:dimm}
\end{figure}

As we have seen in Sections~\ref{sec:bolosignal}--\ref{sec:pca},
\scuba\ data are dominated at low frequencies ($\lsim$2\,Hz) by highly
correlated signals. While it can be significantly reduced using simple
common-mode removal (subtracting the average signal from all
bolometers at each time step), the more complicated correlated
residuals at these low frequencies are difficult to model. Ultimately,
a very simple approach has been taken to handle both components, for
which the main steps in a typical reduction are shown in
Fig.~\ref{fig:dimm}.

First, the raw data are passed through a pre-processing stage which
corrects some of the more significant glitches, applies flat-field
corrections etc. Next, the iterative process begins (dashed box). Most
of the low-frequency noise is removed using common-mode subtraction
(\model{COM,GAI}, Section~\ref{sec:comgai}). Then, the extinction
correction (\model{EXT}, Section~\ref{sec:ext}) is applied. At this
point the data resemble the grey bolometer traces in
Fig.~\ref{fig:bolos_mix}, with some residual baseline drifts still
visible. These drifts are removed using a high-pass filter
(\model{FLT}, Section~\ref{sec:flt}) implemented with FFTs. The
residual signal is then re-gridded to estimate the map. Finally, the
map is used to estimate and remove the astronomical signal from the
bolometer data (\model{AST}, Section~\ref{sec:ast}), leaving a data
set that is appropriate for measuring the white noise level of each
bolometer (\model{NOI}, Section~\ref{sec:noi}). Since the common-mode
and filtering stages will have introduced ringing in the map in the
vicinity of bright sources, the entire process is iterated. Each
component in the dashed box of Fig.~\ref{fig:dimm} is re-calculated in
sequence. In this way, the second time the common-mode is calculated,
for example, most of the bright astronomical sources in the data have
already been removed in the previous iteration when the map was
estimated, reducing the amount of ringing in the map once it is
re-estimated. Note that the extinction correction is a multiplicative
factor, and must be inverted prior to re-calculating any of the
additive model components to preserve the data amplitude. Additive
model components from the previous iteration are generically added
back into the time-series immediately prior to their
re-calculation. However, for the special case of the common-mode and
high-pass filter stages, they are replaced simultaneously at the start
of the iteration to assist with convergence (see
Sections~\ref{sec:converge}).

In general, SMURF is highly configurable, including many options for
both the pre-processing stage and the iterative model components
(which models are used, what order they are calculated in, how they
are calculated).  In Sections~\ref{sec:dataprep} and
\ref{sec:components} we describe the data pre-processing steps and
iterative algorithm in detail. Section~\ref{sec:converge} explores
convergence tests and degeneracies between model components. Finally,
Section~\ref{sec:performance} summarises the mapping performance of
\scuba\ using SMURF.

\subsection{Data pre-processing}
\label{sec:dataprep}

Prior to executing the iterative part of the algorithm, the data must
undergo several pre-processing steps. First, the data files are read
into memory and concatenated into continuous arrays (\scuba\ data are
broken up and written to disk every $\sim$30\,s regardless of the
observation length during data acquisition). As the data are loaded,
they are multiplied by the flatfield correction \citep[see Section~2.1
in][]{dempsey2013}. Next, a series of configurable data cleaning and
filtering procedures are applied; these include the removal of large
glitches that may hinder the iterative solution from converging, or
simply tasks that do not need to be iterated. All of the data that are
flagged as bad are ignored when estimating the map.

\subsubsection{Time-series down-sampling and map pixel size}
\label{sec:downsamp}

The highest useful frequency in the nominally 200\,Hz-sampled \scuba\
data is that which corresponds to the smallest angular scale that the
instrument is sensitive to. As mentioned earlier, the usual
rule-of-thumb for a Gaussian beam is to provide at least 3 samples for
each FWHM, or roughly 2.5\,arcsec for the 7.5\,arcsec 450\,\micron\
channel, and 5\,arcsec for the 14.5\,arcsec 850\,\micron\ channel. For
a typical scan speed of 300\,arcsec\,s$^{-1}$, the maximum useful
sample rate is therefore about 120\,Hz at 450\,\micron, and 40\,Hz at
850\,\micron. In order to save execution time and memory usage (both
of which scale linearly with data volume), it is clearly advantageous
to re-sample the data to these lower rates. In practice, the default
map pixel sizes are set to 2 and 4\,arcsec at 450 and 850\,\micron,
respectively (slightly over-sampled), and down-sampling occurs as the
data are loaded to match this spatial sampling given the slew speed.

The method used by SMURF is to average together multiple samples from
the original time-series, $x_i$ to estimate the lower-frequency output
time-series, $y_j$. In general there will not be an integer number of
samples from $x_i$ in each of the $y_j$, so fractional samples from
$x_i$ are used at the $y_j$ time boundaries: $y_j = (\sum w_i
x_i)/(\sum w_i)$, where $i$ runs over all the samples from $x$ that
land within, or partially overlap, $y_j$, and the weights,
$0<w_i\le1$, indicate the fractions of each $x_i$ that land within
$y_j$. The algorithm is fast since each sample in $x_i$ need only be
visited once. From a spectral point of view, this boxcar average is
equivalent to applying a sinc-function low-pass filter. Such behaviour
is desirable since it serves as an anti-aliasing filter; most
non-white features above the Nyquist frequency in $y_j$ that could
contaminate the output data are removed.

As an alternative, we also investigated an algorithm in which the FFT
of $x_i$ is simply truncated to the target sample rate before
transforming back to the time domain (i.e., applying a hard-edged
low-pass filter). In practice the noise performance was
indistinguishable, and required slightly longer execution time
compared with the previous method, and is therefore not used.

\subsubsection{Time-domain de-spiking}
\label{sec:timedespike}

Spikes of short duration and high amplitude are often seen in the time
series data. If not removed, they can cause ringing when filtering the
data. Two alternative approaches may be used to remove these
spikes. This section describes the detection and removal of spikes
within the time-series of each bolometer, and Section~\ref{sec:ast}
describes the iterative detection and removal of spikes as part of map
estimation. In practice, map-based de-spiking usually gives superior
results, and so time-domain de-spiking is switched off by default.

Each one-dimensional bolometer time-series is processed
independently. At each time slice, the median value of the current
bolometer is found in a box centred on the time slice, and containing
a specified number of time slices (typically 50). If the residual
between the time slice value and the median value is greater than some
specified multiple (typically 10) of the local noise level, the time
slice is flagged as a spike.

If the local noise level were estimated within the same box used to
determine the median value, a spike in the box would cause the local
noise level to be over-estimated severely. For this reason, the local
noise level is taken as the standard deviation of the values within a
neighbouring box on the ``down-stream'' side of the median box (that
is, the side that has already been checked for spikes). In other
words, the high end of the noise box is just below the low end of the
median filter box. This introduces a slight asymmetry in the noise,
but this should not matter unless the noise varies significantly on a
time scale shorter than the box size.

This simple algorithm is not very good at distinguishing between spikes
and bright point sources, and so the threshold for spike detection is
usually raised when making maps of bright point sources.

\subsubsection{Step correction}
\label{sec:steps}

\begin{figure*}
\raggedright
\textbf{(a) large step} \hspace{7.3cm} \textbf{(b) small steps}\\

\vspace{0.2cm}

\includegraphics[width=\linewidth]{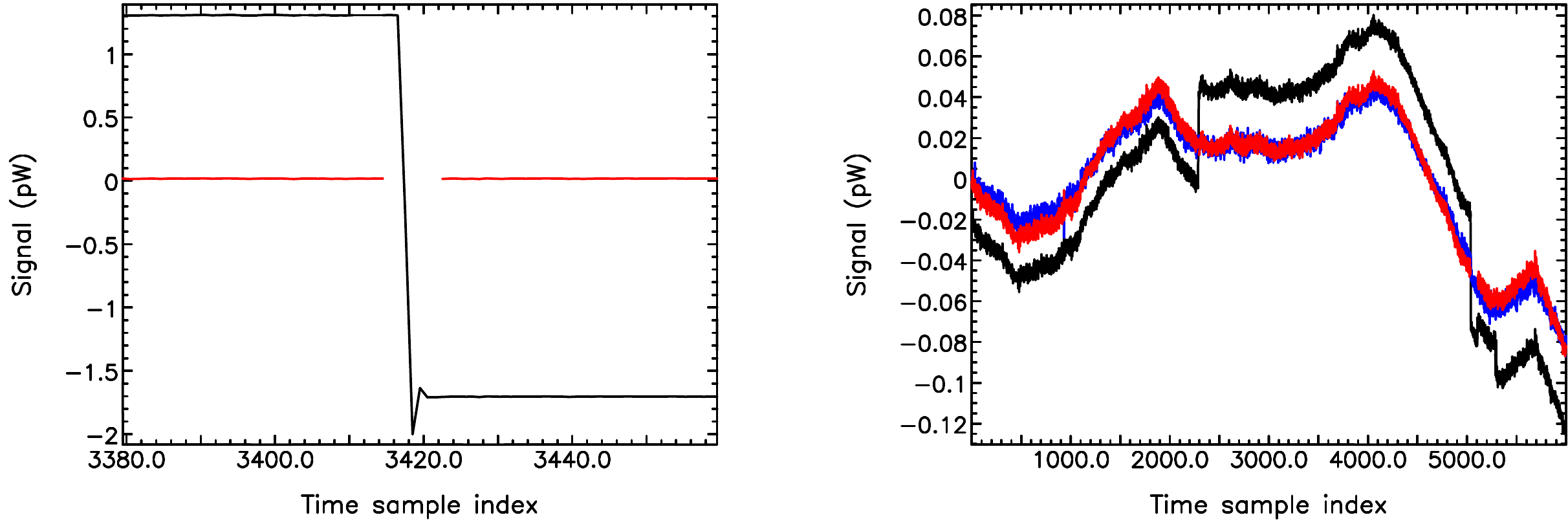}
\caption{Examples of steps in time-series data. Steps occur with a
wide range of heights from the large steps shown in (a) to the small
steps shown in (b). Large steps are often followed by a brief
``over-shoot'' as shown in (a) which is ringing caused by the
anti-aliasing filter. In both plots, the black curve is the
uncorrected time-series, and the red curve is the corrected time
series. Samples close to a step are omitted in the corrected
time-series.  In (b) the blue curve is the uncorrected time-series for
a nearby bolometer. The similarity between the red and blue curves
shows that the step correction is performing well.  }
\label{fig:steps1}
\end{figure*}

Sudden steps can occur in the time-series data from each bolometer,
with the most likely cause being cosmic ray events \citep[see
Section~3.5.3 in][]{holland2013}. The black curves in
Fig.~\ref{fig:steps1} show examples of such steps in the time-series
for two bolometers. If not removed, these steps can cause severe
ringing when filtering, and visible streaks in the final map,
corresponding to the paths of individual bolometers over the sky.

Steps occur with a wide range of heights and shapes. The ratio of step
height to noise can vary from less than 10 to several hundred. Some
steps occur over a single sample, such as the step close to sample
5000 in Fig.~\ref{fig:steps1}b, but others happen more gradually, such
as the step close to sample 5300. In addition, a step can be preceded
or followed by a short period of instability, as is visible at the
bottom of the step in Fig.~\ref{fig:steps1}a (this is probably due to
the response of the \scuba\ anti-aliasing filter; the sudden large
step occurs prior to the 200\,Hz re-sampling). Further problems are
caused by steps that occur close together in time, such as the large
downward step followed by a smaller upward step close to sample 5000
in Fig.~\ref{fig:steps1}b.

Detecting and correcting such a wide variety of steps reliably has
proved to be a challenge. In outline, the following stages are
involved in detecting steps in a single bolometer time-series:

\begin{enumerate}

\item median smooth the whole time-series;

\item find the gradient of the median smoothed time-series at each
sample;

\item smooth the gradient values to determine the local mean gradient
and subtract this local mean from the total gradient to get the
residual gradient;

\item find residual gradient values that exceed 25 times the local RMS
of the residual gradients;

\item group these high residual gradients into contiguous blocks of
samples;

\item merge blocks that are separated by less than 100 samples.
\end{enumerate}

The above process produces a list of candidate steps in each bolometer
time-series. Each candidate step is then verified, measured and corrected
using the following procedure:

\begin{enumerate}

\item the above process can misinterpret the edges of a bright source
as a step, so we ignore blocks that occur close to bright sources;

\item if the block passes the above test, a least squares linear fit
is performed to the median-filtered bolometer data just before the
block, and this fit is extrapolated to predict the data value expected
at the centre of the block;

\item a least squares linear fit is performed to the median-filtered
bolometer data just after the block, and this fit is extrapolated to
predict the data value expected at the centre of the block;

\item the difference between these two expected data values is taken
as the step height;

\item the preceding three steps are repeated several times, each time
including a different selection of samples in the two least squares
fits, with the mean and standard deviation of the corresponding set of
step heights found;

\item if the mean step height is small compared to the standard deviation
of the step heights, or compared to the noise in the bolometer data, then
the step is ignored;

\item if the above checks are passed, all subsequent bolometer samples
are corrected by the mean step height;

\item bolometer samples within the duration of the step, and a few
samples on either side, are flagged as unusable.

\end{enumerate}

Once all steps have been corrected within a bolometer time-series, a
constant value is added to all samples in the time-series to restore its
original mean value.

The results of step correction are shown by the red curves in
Fig.~\ref{fig:steps1}. For comparison, the blue curve shows the
uncorrected time-series from a nearby bolometer that does not suffer from
steps. The agreement between the red and blue curves confirms that
the step correction algorithm is working satisfactorily.

\subsubsection{Gap filling / apodisation}
\label{sec:gaps}

SMURF uses FFTs of the bolometer data extensively for filtering. Data
that have been flagged as bad for any reason (for instance, due to the
presence of spikes, steps, or unusual common-mode signal) need to be
excluded. For this reason, each contiguous block of bad data samples
is filled with artificial data before taking the FFT. A least squares
linear fit is performed to the 50 samples preceding the block, and a
similar fit is performed to the 50 samples following the block. These
are used to estimate the expected values at the start and end of the
block of bad values. The bad values are then replaced by linear
interpolation between the expected start and end values. Gaussian
noise is added with a standard deviation equal to the mean of the RMS
residuals in the two fits. These flagged and filled portions of the
data are then given a weight of zero when estimating the map.

In addition to replacing bad samples before the FFT, it is also
necessary to ensure that the data values at the start and end of the
time-series are similar. Since an FFT treats the data as a single
cycle in an infinitely repeating waveform, any large difference
between starting and ending values will effectively introduce sudden
steps at the start and end of each cycle, causing unwanted
oscillations (ringing) in the transform. Another consequence of the
cyclic nature of the FFT is that features at one end of the time
series can affect the filtered values at the other end of the time
series. Two methods are available to avoid these two problems:

\begin{enumerate}

\item Apodisation: a number of samples at the start and end of each
bolometer time-series are multiplied by a cosine function in order to
roll the data values off smoothly to zero. The default number of
samples modified at each end of the time-series is given by half the
ratio of the sampling frequency to the lowest retained frequency, the
edge frequency of the high-pass filter (Section~\ref{sec:flt}). In
addition, each end of the time-series is padded with double this
number of zeros. This method is illustrated in Fig.~\ref{fig:pad2}. It
is not used by default as it reduces the amount of data available for
the map, and can significantly hinder very short observations (e.g.,
of calibrators when focussing the telescope).

\begin{figure}
\centering
\includegraphics[width=\linewidth]{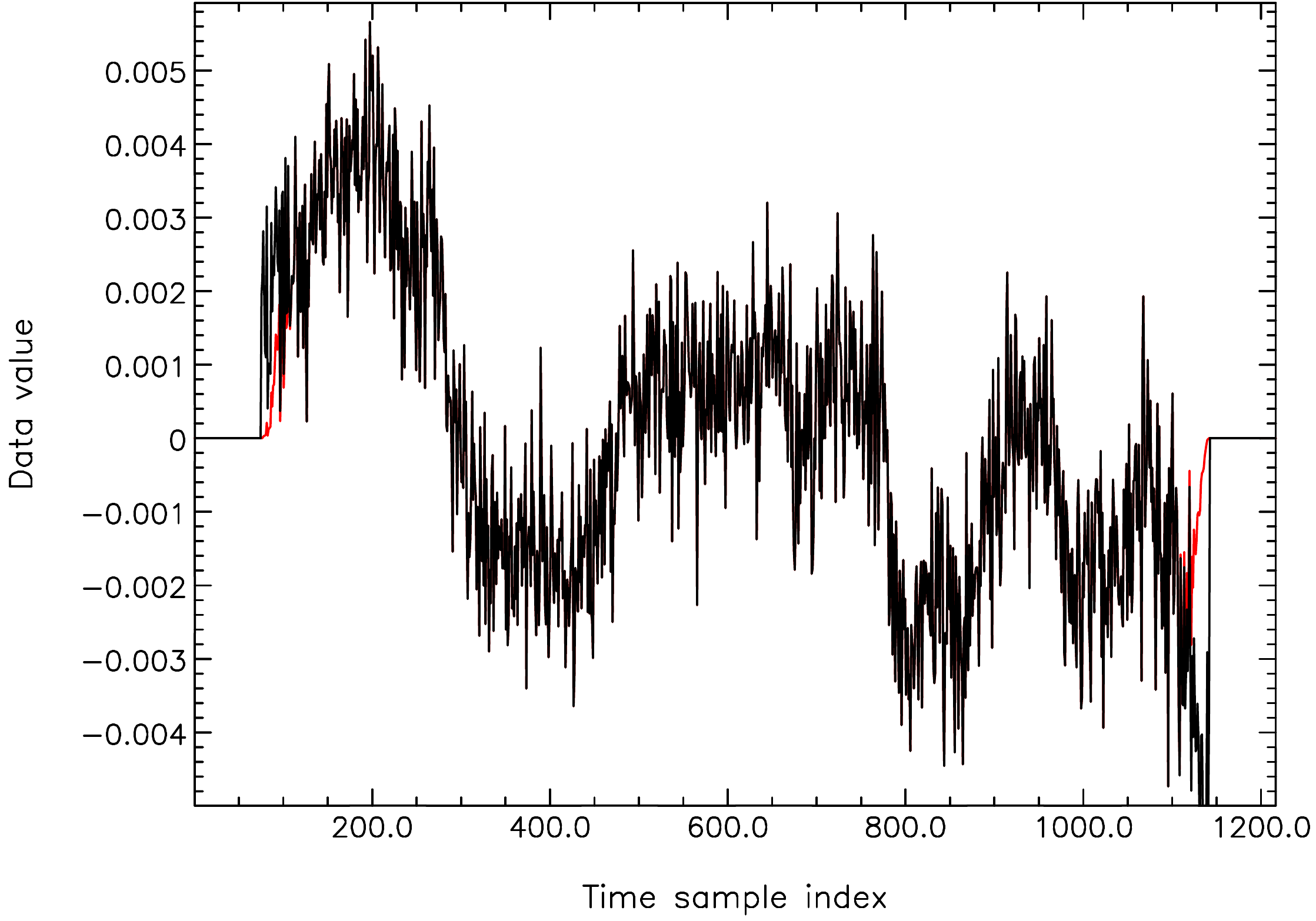}
\caption{The black curve shows a bolometer time-series, padded
with zeros. The red curve shows the time-series after apodisation.}
\label{fig:pad2}
\end{figure}

\item Padding with artificial data: instead of padding with zeros,
each time-series is padded with artificial data which connects the two
ends of the data stream smoothly and includes Gaussian noise. No
apodisation is performed. The number of samples of padding at each end
is again equal to the ratio of the sampling frequency to the lowest
retained frequency.  This is illustrated in Fig.~\ref{fig:pad1}. See
\citet{stompor2002} for a thorough discussion of this procedure within
the context of CMB map-making. This method is used by default.

\begin{figure}
\centering
\includegraphics[width=\linewidth]{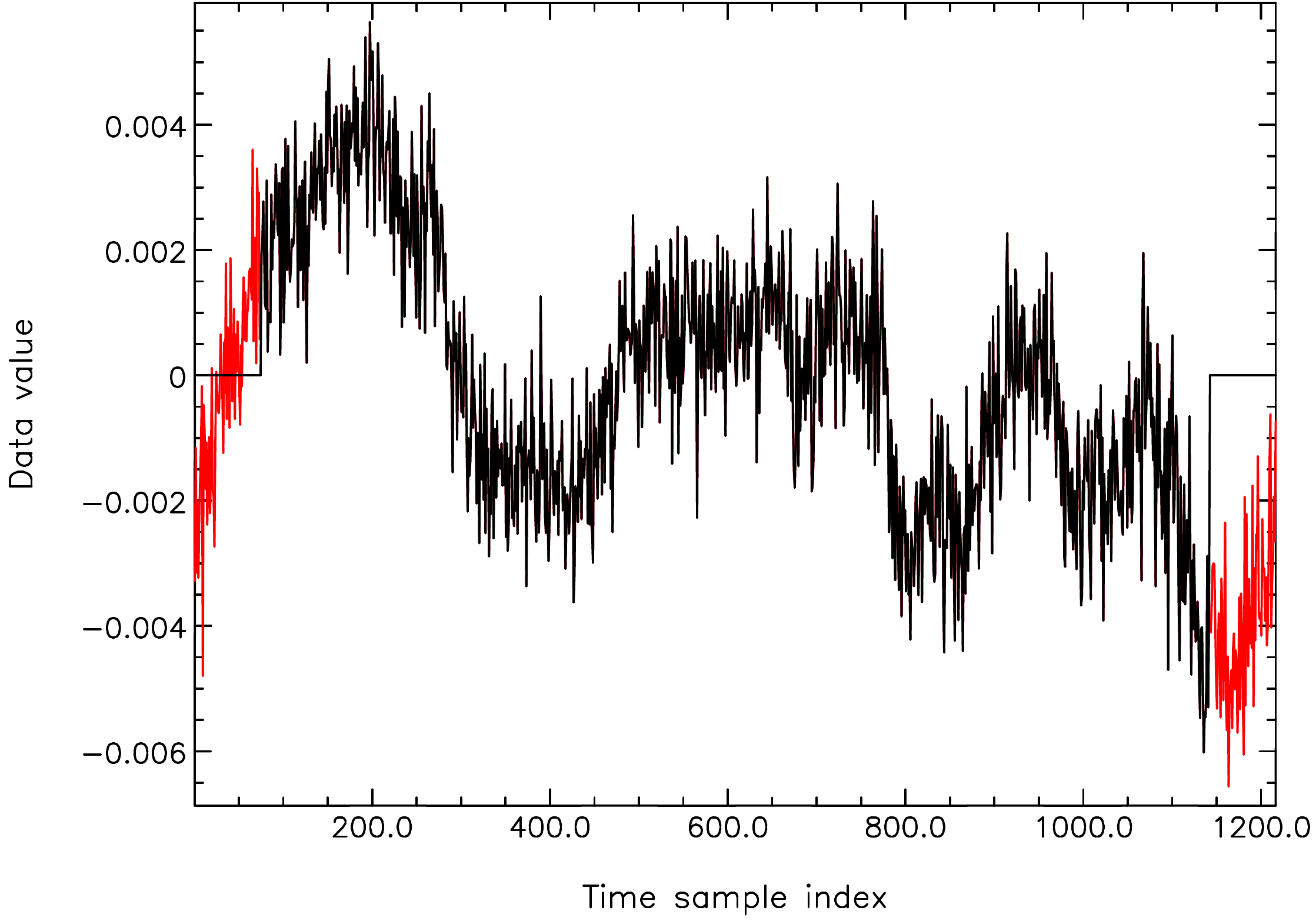}
\caption{The black curve shows the same bolometer time-series as in
Fig.~\ref{fig:pad2}, again padded with zeros. The red curve shows the
time-series after padding with artificial data.}
\label{fig:pad1}
\end{figure}

\end{enumerate}

\subsubsection{Bolometer filtering}

Despite the ability of the map-maker to iteratively remove many noise
components, under some circumstances it may be desirable to filter the
data once during the pre-processing step. Three main filtering options
are available:

\begin{enumerate}

\item The most commonly-used pre-processing filter is polynomial
subtraction. A polynomial of the requested order is fit and removed
from each bolometer time-series. At a bare minimum, the mean is
removed from all of the bolometers (order 0) in all reductions
described in this paper.

\item All of the filters available as part of the iterative Fourier
Transform Filter (Section~\ref{sec:flt}) can also be applied once
during pre-processing.

\item As an alternative to the iterative map-making procedure,
cleaning by principal component analysis (PCA) is available as an
experimental pre-processing option (Section~\ref{sec:pca}). The most
significant components in the analysis are identified, and the
projection of each bolometer time-series along their basis vectors are
removed. A single parameter specifies the threshold on the amplitude
of the components to be removed, as a number of standard deviations
away from the mean value. Subarrays are cleaned independently, once,
for the full length of each continuous chunk of data. Given the
computational expense of this method, and initial tests which showed
little improvement over simple high-pass filtering, this method has
not yet been explored in detail with \scuba\ data. However, since
systematic effects seem to come and go, it is possible that PCA could
be useful for particular data sets.

\end{enumerate}

\subsubsection{Additional data rejection}
\label{sec:flagbadbol}

Despite the cleaning operations described in the previous sections,
the data from a given bolometer may be unusable due to it being poorly
biased, having an incorrect flatfield correction applied, or having
some other particularly pathological noise contamination. Most of
these bolometers can be flagged simply by identifying outliers in the
distribution of bolometer white noise levels. By default, SMURF
measures the PSD of all bolometers between 2 and 10\,Hz, after all
other pre-processing steps have been run (identical to the measurement
in Section~\ref{sec:noi}). Despite having significant low-frequency
noise (and possibly bright astronomical source) contamination, this
higher-frequency portion of the PSD is generally quite
flat. Furthermore, even if there is contamination, the purpose of this
measurement is to identify outliers, rather than provide a meaningful
absolute measurement of a given bolometer's white noise level. Both
high and low (e.g., due to an incorrect and very small flatfield
correction being applied) outliers from the centre of the logarithm
(to reduce the impact of outliers) of the bolometer noise distribution
are flagged.

Finally, data that are taken while the telescope is either moving
extremely slowly, or extremely fast, are flagged and ignored when
estimating the map. When the telescope is moving slowly, astronomical
signals appear at low frequencies in the bolometer data that are
dominated by $1/f$ noise (in the extreme case of no motion, the
bolometers have zero sensitivity). When the telescope is moving
significantly faster than 600\,arcsec\,s$^{-1}$, both the thermal
response of the bolometers, and the anti-aliasing filter in the
readout electronics (Section~\ref{sec:bolos}) can significantly
distort signals.  The default allowable range of speed is
30--1000\,arcsec\,s$^{-1}$. These thresholds primarily reject short
periods when the telescope was stationary prior to commencing the scan
pattern (generally speaking the speed is constant once the scan
begins). The scan speed can also briefly exceed the upper limit when
approaching the maximum requested speed of 600\,arcsec as the pointing
system struggles at high elevation.

\subsubsection{Impact of lags}
\label{sec:lags}

In principle, the effects of the bolometer thermal response,
anti-aliasing filter, and data acquisition lags should be accounted
for; together they produce a net lag in the bolometer signals of about
4.5\,ms with respect to the pointing data
(Section~\ref{sec:bolos}). This can be modelled by convolving the
``pure'' bolometer signals with a single system response
function. Similarly, this effect could be removed using de-convolution
as a pre-processing step. The impact of this lag is primarily a
reduction in point-source sensitivity since scans at different
position angles will detect the source at slightly different positions
(and therefore reduce the peak signal once averaged together). This
attenuation is most significant at 450\,\micron\ given the smaller
beam size, and at the greatest scan speeds. For reference, at a speed
of 150\,arcsec\,s$^{-1}$ (typical for CV daisies of calibrators and
deep fields) the attenuation is a negligible $<$1\% at 850\,\micron,
and $\sim$1\% at 450\,\micron. However, at the maximum speed of
600\,arcsec\,s$^{-1}$ (for the largest, and typically shallowest
maps), this attenuation grows to about 5\% and 15\% at 850 and
450\,\micron, respectively. It is unlikely that this response has had
a significant impact on observations taken to date, since such scans
have been used primarily for shallow cosmology fields for which only
the 850\,\micron\ data are expected to be useful, or Galactic fields
of extended structures (which are less sensitive to this effect). In
the present Starlink \textsc{kapuahi} release there is no option to
de-convolve the system response. However, it is presently being added
and will be present in future releases.

\subsection{Iterative model calculation}
\label{sec:components}

Once the data have been cleaned, and the worst data flagged, the
iterative solution for the map and contaminating noise signals begins.

First, we describe the basic model for \scuba\ data. We express the
digitised signal observed by the $i$th bolometer as a function of
time,
\begin{equation}
\mathbf{b}_i(t) = f_i[\mathbf{e}_i(t) \mathbf{a}_i(t) + \mathbf{n}_i(t)],
\label{eq:model}
\end{equation}
where $\mathbf{a}(t)$ is the time-varying signal produced by scanning
the telescope across astronomical sources, $\mathbf{e}(t)$ is the
time-varying extinction, which is a function of the telescope
elevation and atmospheric conditions, and $\mathbf{n}_i(t)$ represents
sources of noise. The two terms in square brackets, as written, have
units of power delivered to the detectors (pW). The scale factor $f_i$
converts this effective power to the digitised units recorded on disk
-- the flatfield multiplied by a digitisation constant (applied once
during pre-processing) -- which in this formulation is assumed to be
constant in time.

We then express the noise, $\mathbf{n}_i(t)$, as the sum of several
components,
\begin{equation}
  \mathbf{n}_i(t) = \mathbf{n}^\mathrm{w}_i(t) +
  g_i\mathbf{n}^\mathrm{c}(t) + \mathbf{n}^\mathrm{f}_i(t),
\label{eq:noise}
\end{equation}
where $\mathbf{n}^\mathrm{w}_i(t)$ is uncorrelated white noise,
$\mathbf{n}^\mathrm{c}(t)$ is a correlated or common-mode signal (with
an optional scale factor $g_i$ for each bolometer), and
$\mathbf{n}^\mathrm{f}_i(t)$ is (predominantly low-frequency) noise in
excess of the white noise level, which is either uncorrelated from
bolometer-to-bolometer, or at least does not have a simple correlation
relationship that would lead to it being included in
$\mathbf{n}^\mathrm{c}(t)$.

During pre-processing, the map-maker divides by $f_i$ once. Then, in
each iteration, the map-maker models and removes
$g_i\mathbf{n}^\mathrm{c}(t)$, divides by $\mathbf{e}_i(t)$, and
applies a high-pass filter to remove $\mathbf{n}^\mathrm{f}_i(t)$ from
the bolometer time-series. This procedure isolates the astronomical
signal and white noise, $\mathbf{a}_i(t) +
\mathbf{n}^\mathrm{w}_i(t)$, for estimating the map. Finally, once the
map is estimated it is projected into the time-domain and removed from
the bolometer time-series leaving $\mathbf{n}^\mathrm{w}_i(t)$, in
which the bolometer noise can be measured. This is the most typical
sequence; in practice the user can select an arbitrary order, although
the solution converges much faster when the strongest components are
estimated first, and then subtracted, leaving a cleaner signal for
subsequent model estimates.

After the first iteration, there will almost certainly be correlated
errors between the model estimates. For example, the presence of a
bright astronomical source will contaminate $\mathbf{n}^\mathrm{c}(t)$
(which is a simple average of all of the bolometer time-series at each
instant in time), leading to an over-subtraction, and in turn,
negative bowls in the map.

Subsequent iterations, however, diminish such problems. Each additive
model component is re-estimated in the same sequence, after first
adding the previous estimate back into the time-series (but
importantly, not the other components). In this case, much of the
astronomical signal will have been identified and removed in the
previous calculation of $\mathbf{a}_i(t)$, and therefore there will be
less contamination in $\mathbf{n}^\mathrm{c}(t)$, and less negative
bowling in the map. Any multiplicative models [e.g.,
$\mathbf{e}_i(t)$] are inverted immediately prior to the start of a
new iteration to preserve the units of the data. The iterative
solution will thus converge, barring degeneracies between model
components.

The full set of model components available, and the parameters that
control them, are described in the following
sections. Table~\ref{tab:components} shows the typical order in which
they are calculated. For a discussion on convergence tests and
degeneracies, see Section~\ref{sec:converge}.

\begin{table}
  \caption{Summary of the model components that can be fit to \scuba\
    time-series data with SMURF. Only the first group of models are
    typically fit to the data (\model{COM}--\model{NOI}) in the
    indicated order. The remaining models (\model{DKS}--\model{PLN})
    are usually omitted, although they are available as options.}
  \vspace{0.2cm}
  \centering
  \begin{tabular}{c|l}
    \hline
    Model & Description \\
    \hline
    \model{COM} & remove common-mode signal \\
    \model{GAI} & common-mode scaled to each bolometer \\
    \model{EXT} & extinction correction \\
    \model{FLT} & Fourier Transform filter \\
    \model{AST} & map estimate of astronomical signal \\
    \model{NOI} & noise estimation \\
    \hline
    \model{DKS} & dark squid cleaning along columns \\
    \model{PLN} & 2-dimensional time-varying plane removal \\
    \model{SMO} & time-domain smoothing filter \\
    \model{TMP} & pointing as baseline template \\
    \hline
    \end{tabular}
  \label{tab:components}
\end{table}

\subsubsection{\model{COM,GAI}: common-mode estimation}
\label{sec:comgai}

\begin{figure}
\centering
\includegraphics[width=\linewidth]{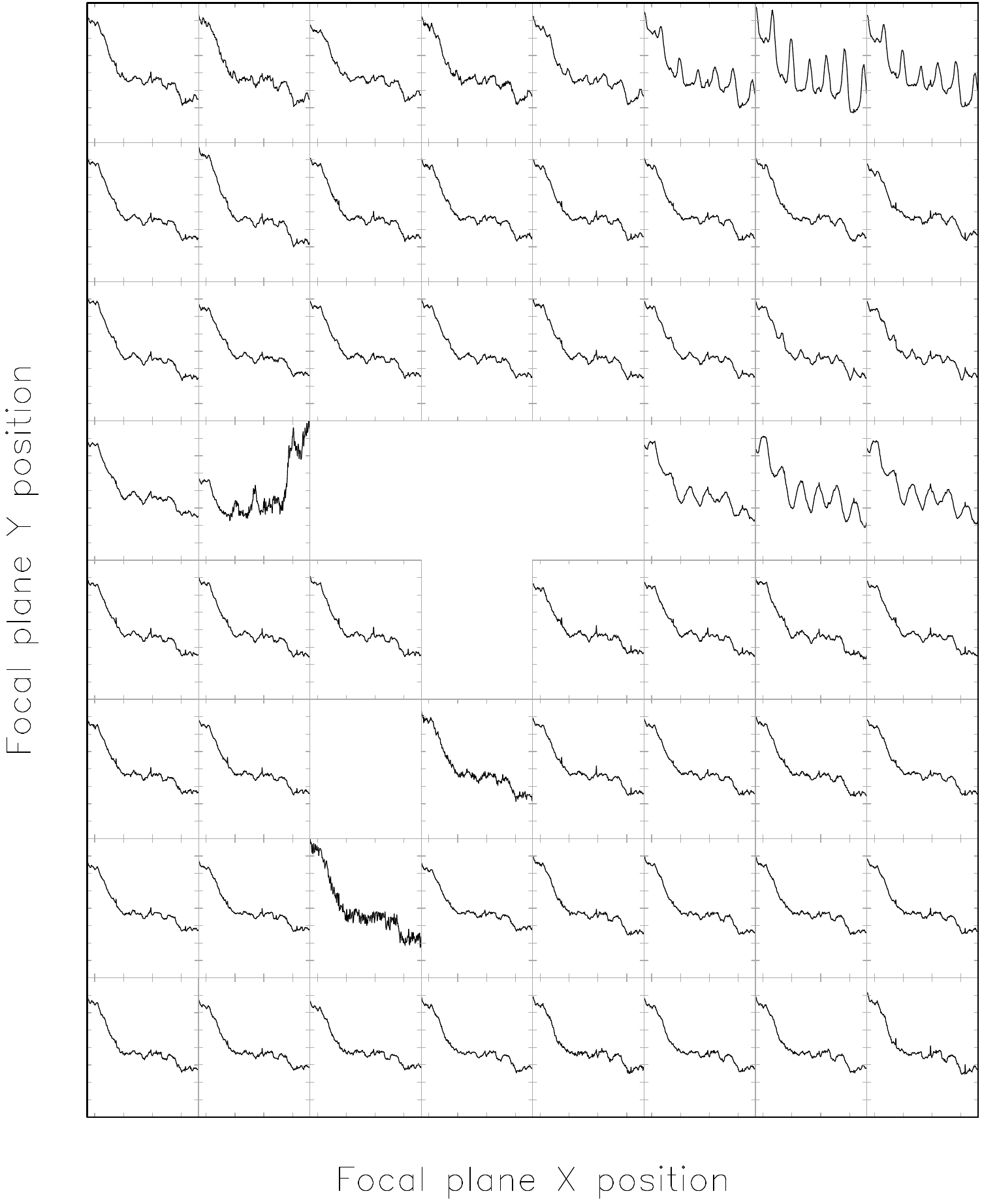}
\caption{A selection of bolometer time-series from a representative
region of the focal plane after flatfielding and removal of a constant
baseline. Each sub-plot shows the power from a single bolometer over a
30\,s interval.  The lower and upper vertical limits of each plot are
$-0.02$\,pW and $+0.02$\,pW, respectively. It can be seen that most
bolometers exhibit a common time variation overlaid with other
features. Empty squares indicate the locations of broken bolometers.}
\label{fig:com}
\end{figure}

﻿Fig.~\ref{fig:com} shows the time-series from a selection of typical
bolometers.\footnote{The data have been flat-fielded and each time
series has been adjusted to a mean value of zero.} The similarity
between most bolometers is evident, and forms the common-mode signal
-- assumed to be a consequence of variations in the atmospheric
emission and fridge temperature. This common-mode usually dominates
the astronomical signal for all but the brightest sources, and swamps
faint extended structure.  The purpose of the \model{COM} and
\model{GAI} models is to remove this common-mode signal.

The \model{COM} model is the common-mode signal itself
[$\mathbf{n}^\mathrm{c}(t)$ in Eq.~\ref{eq:noise}]. It is a single
time-series that is estimated by finding the mean of all bolometers at
each time step. Bolometer values that have been flagged as unusable
are excluded from the mean.

Even after flat-fielding, bolometers may have slightly varying
sensitivities and so the amplitude of the common-mode variations will
also vary from bolometer to bolometer. Comparing each bolometer
time-series with the common-mode signal allows an estimate of the
relative bolometer sensitivity to be obtained. In practice, a least
squares linear fit is performed between the bolometer time-series and
the common-mode to determine a gain ($g_i$ in Eq.~\ref{eq:noise}) and
offset for each bolometer.  The gain and offset for each bolometer is
known as the \model{GAI} model. Each bolometer value can then
optionally be scaled and shifted using these values so that all
bolometers share a common (but as yet unknown) calibration. This
provides an alternative (or additional) flat-fielding strategy to that
described in Section~\ref{sec:bolos}.

An option exists to cater for time-varying sensitivities. In
principle, the gain of the bolometers should be constant in
time. However, there is evidence that there are slight variations, and
this option does tend to slightly improve the noise in maps.  If used,
the least squares fits described above are performed on short blocks
of contiguous time slices, providing multiple gain and offset values
for each bolometer (one pair for each block of time slices). The gain
and offset at any required time slice can then be found by
interpolation between these values.

It can be seen from Fig.~\ref{fig:com} that some bolometers depart
radically from the common-mode, indicating some problem with the
bolometer. Such bolometers are identified by calculating the Pearson
correlation coefficient between each bolometer time-series and the
common mode. Bolometers for which the correlation coefficient is below
a specified limit, or which have unusually high or low gains compared
to the other bolometers, are flagged as bad in order to omit them from
the final map. If the option described above for handling time-varying
sensitivities is used, then a correlation coefficient can be
determined for each individual block of time slices.  This allows
individual bad blocks to be rejected from a bolometer time-series,
rather than rejecting the whole bolometer. As a warning, rejecting
data based on outlier gain values can be misleading in cases where the
data are dominated by magnetic field pickup. For example, the
common-mode signal for the data in Fig.~\ref{fig:magpickup} would
resemble the azimuthal scan pattern, but clearly the gains will have
opposite signs (such that one or the other bolometer would be
erroneously flagged). Experimental models that were investigated for
magnetic field pickup removal are described in
Section~\ref{sec:experimental}.

The common-mode value at each time slice is calculated as the
unweighted mean of the values of all bolometers that have not
previously been flagged as bad for some other reason. Bolometer values
that were flagged as bad simply because they were poorly correlated
with the common-mode on the previous iteration are, however, included
in the new common-mode estimate. If such samples are excluded, there
is a strong possibility of discontinuities appearing in the
\model{COM} model at block boundaries.  These in turn can lead to
ringing when filtering, and instabilities in the convergence
process.

Any astronomical sources that are smaller than the array size will
contribute signal to some bolometers but not other bolometers, thus
biasing the simple mean used to estimate the common mode. However, on
each iteration of the map-making algorithm illustrated in
Fig.~\ref{fig:dimm}, a large fraction of the remaining astronomical
signal is extracted from the bolometer time-series and transferred to
the output map, resulting in subsequent estimates of the common-mode
being more accurate. It is also for this reason that the previously
described step of (re-)flagging outlier portions of the data is
generally performed iteratively, rather than once as a pre-processing
step, since bright/compact astronomical sources can be the cause of
rejection in early iterations.

Any extended astronomical emission on a scale comparable to or larger
than the spatial extent of the area used to estimate the common-mode
will contribute a similar signal to all bolometers. Therefore such
extended emission is indistinguishable from the other sources of
common-mode signal (e.g., atmosphere variations) and will be removed
by the \model{COM} model. This places a limit on the spatial extent of
astronomical structure that can be recovered.

For this reason, the usual practice is to estimate a single
\model{COM} model by examining data from all four subarrays in each
waveband, since this allows spatial structure on the scale of the
whole focal plane to be recovered. However, sometimes there is
evidence that the common-mode differs from one array to another, and
so an option exists to estimate a separate \model{COM} model for each
individual subarray, with a consequent lowering in the scale of
spatial structure that can be recovered.

\subsubsection{\model{EXT}: extinction correction}
\label{sec:ext}

The extinction correction is a multiplicative factor that is normally
derived using the WVM, and is not considered to be a free parameter in
the solution [$\mathbf{e}_i(t)$ in Eq.~\ref{eq:model}]. However, it is
applied as part of the iterative solution, rather than a
pre-processing step, since any small errors will be amplified by the
large low-frequency drifts in the raw bolometer time-series. For
example, if the bolometer drift is 1000 times greater than an
astronomical source of interest, a 1 per cent error in the flatfield
will produce stripes of order 10 times the astronomical signal
amplitude in the final map! If \model{EXT} is applied after the bulk
of the low-frequency noise has been removed (e.g., \model{COM,GAI}),
then there is little potential for such small errors to affect the
final map.  For details on how it is calculated see
\citet{dempsey2013}. Note that its numerical value is calculated only
once, and simply applied as a scale factor in the iterative
solution. Unlike the additive model components, it is inverted at the
start of each iteration to preserve the amplitude of the data before
the re-calculation of other model components.

\subsubsection{\model{FLT}: Fourier Transform filter}
\label{sec:flt}

This model takes the FFT of the bolometer time-series data, and can
apply both high- and low-pass filters, as well as notch filters, at
hard frequency edges specified by the user. Alternatively, the
frequency edges of the filters may be defined in terms of an angular
scale, but converted into a frequency through knowledge of the mean
telescope slew speed. The time-series are generally gap-filled
(Section~\ref{sec:gaps}) before the transform to avoid ringing
(primarily caused by wrap-around discontinuities at the ends of the
time-series). Finally, a whitening filter may also be applied in which
a simple form, $1/f^\alpha + \mathrm{constant}$, is fit to the power
spectrum of each bolometer, and then the bolometer FFT is multiplied
by the square root of its inverse (though normalised to preserve the
white noise level).  The signals that are removed from the time-series
by this process are stored in the \model{FLT} container
array. Typically this model is used purely as a high-pass filter to
remove most of the residual $1/f$ noise following common-mode removal
[$\mathbf{n}^\mathrm{f}_i(t)$ in Eq.~\ref{eq:noise}]. Low-pass
filtering is redundant for two reasons: (i) SMURF already low-pass
filters and re-samples to a lower sample rate, as described in
Section~\ref{sec:downsamp}; and (ii) the act of re-gridding the data
to produce map estimates effectively low-pass filters the data below a
frequency that corresponds to the inverse of the crossing time of a
single map pixel. Notch filters have not been proven to be useful with
\scuba\ data, particularly since line features tend to move around
(i.e., a dynamic line-detection system would have to be developed, and
care would have to be taken that astronomical sources are not
suppressed).

\subsubsection{\model{AST}: map estimation}
\label{sec:ast}

Map estimation is accomplished using a nearest-neighbour resampling of
the data onto a pre-defined map grid. For the $i$th map pixel,
$\mathbf{m}(x_i,y_i)$, the brightness is estimated as the weighted
average of the bolometer data samples $\mathbf{b}_j$ that land within
that pixel (from any bolometer or point in time, provided that they
are not flagged as bad or gap-filled, Section~\ref{sec:gaps}),
\begin{equation}
  \mathbf{m}(x_i,y_i) = \frac{\sum_j \mathbf{w}_j \mathbf{b}_j }
                             { \sum_j \mathbf{w}_j } .
\end{equation}
For the initial iteration the weights $\mathbf{w}_j$ are set to 1, but
subsequently they are set to $1/\sigma_j^{\mathbf{w}2}$, the estimated
inverse variance expected from the bolometer white noise levels, as
discussed in Section~\ref{sec:noi}. This weighting scheme is sensible
provided that the bolometer data have no correlated (e.g.,
low-frequency) noise.

In addition to the signal map, a variance map $\mathbf{v}(x_j,y_j)$ is
estimated. The default procedure is to estimate this quantity given
the scatter in the weighted samples. This is accomplished by dividing
the biased weighted sample variance by the number of samples that went
into the average (akin to the formula for standard error on the mean,
but accounting for weights),
\begin{equation}
\label{eq:varmap}
\mathbf{v}(x_j,y_j) = \frac{\sum_j \mathbf{w}_j
                            \sum_j \mathbf{w}_j \mathbf{b}_j -
                            \left( \sum_j \mathbf{w}_j \mathbf{b}_j \right)^2 }
                           { N \left( \sum_j \mathbf{w}_j \right)^2 },
\end{equation}
where $N_j$ is the total number of bolometer samples that land in the
pixel. As written, this algorithm is numerically unstable if the two
terms in the numerator are large, and of nearly the same value:
floating point precision errors can cause the difference to be
significantly incorrect. In practice we use the superior ``weighted
incremental algorithm'' that calculates incremental differences, as
described in \citet{west1979}.

We decided not to use an unbiased estimator (e.g., the extension of
the common $(1/N-1)\sum_j (\mathbf{b}_j - \bar{\mathbf{b}})^2$
estimator using weights), since in practice it would require
accumulating an additional array of values at every map pixel, and
only results in a small difference where there are less than $\sim10$
samples per pixel (a situation that is almost never encountered in a
\scuba\ map, except in the edge pixels).

Finally, once the map estimation is complete, the map is projected
into the time domain (the signal that would be produced in each
bolometer by the signal represented by the map, $\mathbf{a}_i(t)$ in
Eq.~\ref{eq:model}) and removed.

In addition to map estimation, the \model{AST} model can also be used
to perform map-based despiking of the time-series. Unlike the
time-domain despiker (Section~\ref{sec:timedespike}), this calculation
utilises the scatter in the population of samples that land in a map
pixel (from different times and bolometers) to reject outliers. This
method is more robust against false-positive detections of
bright/compact astronomical sources since transient features in the
time-series are unlikely to occur by chance whenever a bolometer
crosses a specific location on the sky, whereas real astronomical
sources are at fixed spatial locations.

The estimated variance, $\mathbf{v_p}(x_i,y_i)$, of the normalised
weighted samples that land in the $i$th map pixel is simply the biased
weighted sample variance (i.e., the variance map value multiplied by
the number of samples):
\begin{equation}
  \mathbf{v_p}(x_i,y_i) = N_i \mathbf{v}(x_i,y_i).
\end{equation}
In order to compare the weighted differences between the samples and
the map values, [$\mathbf{b}_j - \mathbf{m}(x_i,y_i)$] to
$\mathbf{v_p}$, they must be scaled appropriately. We define a
normalised difference, $\mathbf{d}_j$, in such a way that the variance
of this new variable gives the weighted sample variance of the
underlying data points:
\begin{eqnarray}
  \frac{\sum_j \mathbf{d}_j^2}{N} &=&
  \frac{\sum_j \mathbf{w}_j [\mathbf{b}_j - \mathbf{m}(x_i,y_i)]^2}
       {\sum_k \mathbf{w}_k} \\
   \Rightarrow \mathbf{d}_j^2 &=& \frac{N \mathbf{w}_j [\mathbf{b}_j -
       \mathbf{m}(x_i,y_i)]^2}{ \sum_k \mathbf{w}_k} .
\end{eqnarray}
The map-based despiker flags those $\mathbf{d}_j$ that are further
than some threshold number of standard deviations
$\sqrt{\mathbf{v_p}}$ away from zero, so that they are not used in
subsequent iterations.

A final option available as part of the \model{AST} model is to apply
constraints to the map to improve convergence, which presently include
setting user-specified or low-\snr\ regions to a value of zero for
all but the final iteration. See the discussion in
Section~\ref{sec:converge} and examples in
Sections~\ref{sec:examples}.

\subsubsection{\model{NOI}: noise estimation}
\label{sec:noi}

The primary purpose of the noise component, \model{NOI}, is to measure
the white noise levels of each bolometer, which is approximated with a
single (non-time varying) variance, $\sigma_i^{\mathbf{w}2}$. This
value is then used to calculate weights for map estimation,
$\mathbf{w}_i = 1/\sigma_i^{\mathbf{w}2}$. The measurement generally
occurs once all of the other models have been fit and removed [i.e.,
once $\mathbf{n}^\mathrm{w}_i(t)$ from Eq.~\ref{eq:noise} is
isolated].

First, the bolometer PSDs are calculated as in Eq.~\ref{eq:psd}. An
average white noise level is then measured from 2 to 10\,Hz, a clean
region of the PSD that tends to lie above the $1/f$ knee, but below
the high-frequency line features for typical bolometer data
(Fig.~\ref{fig:pspec}). Taking this constant level for $\mathbf{P}(f)$
we then calculate the expected variance of the time-series (in
approximately 200\,Hz samples) using Eq.~\ref{eq:psd}. If the
bolometer noise were produced purely by uncorrelated sources (i.e., no
other long time-scale drifts), with no high-frequency line features,
and without the attenuation at even higher frequencies by the
anti-aliasing filter, this is the theoretical noise limit of the
detectors. These measured variances are stored, and then used in
subsequent iterations to weight the data points when calculating the
map estimate (Section~\ref{sec:ast}). They are also used for
convergence tests (Section~\ref{sec:converge}).

Since noise estimation is usually calculated as the final step in the
iteration, the data at this stage have had most of the astronomical
and other large, low-frequency noise signals removed. For this reason,
\model{NOI} may optionally perform some cleaning options, such as the
step fixer (Section~\ref{sec:steps}) and spike detection
(Section~\ref{sec:timedespike}), which may work better with these
cleaner time-series.

Finally, it should be noted that the default procedure is to calculate
the white noise levels once within \model{NOI}, after the second
iteration. The reason for fixing these values is to prevent any
potential divergence in the weight estimates with iterations, and also
to provide a fixed reference for the convergence tests
(Section~\ref{sec:converge}). Note that the absolute values of the
noise, and therefore weights calculated by the model, are irrelevant
(only their relative values matter). The reason is that the final
noise in the map is measured empirically from the scatter of the
weighted data points that land in each pixel (Section~\ref{sec:ast}).

\subsubsection{Experimental models for the removal of magnetic field
  pickup and the atmosphere}
\label{sec:experimental}

Additional models exist as options, although they are not generally
used: \model{DKS}, the use of dark squids as a template for removing
magnetic field pickup; \model{TMP} which uses the azimuth of the
telescope also as a template for magnetic field pickup; \model{SMO},
which provides a time-domain smoothing alternative to the FFT-based
filter model \model{FLT}; and \model{PLN}, which fits a plane to the
signal distribution across the focal plane at each time slice in an
attempt to remove possible coherent atmospheric sky-noise structure.

As described in Section~\ref{sec:magpickup}, wide scans can produce
significant magnetic field pickup that tracks the azimuthal motion of
the telescope. The \model{DKS} model uses the dark squid signals for
each column as a template that is simply scaled (gain and offset) to
each bolometer time-series before removal. Unfortunately the dark
squids do not work for every column (several cases in each subarray),
meaning that those columns are usually discarded when producing
maps. One possible remedy is to identify dead bolometers (with
otherwise working TES readouts) or intentionally disconnect working
bolometers, to create replacement dark readouts in those columns. This
solution has not been pursued due to the limited success at removing
pickup using the presently working dark squids. Another alternative
model is \model{TMP} in which the azimuth of the telescope itself is
used as the template. For both \model{DKS} and \model{TMP}, there is
some success at visibly decreasing the low-frequency scan-synchronous
noise in certain data sets, although often it does not (and there are
also examples in which the noise is increased). In the case of
\model{DKS}, there may simply be components of the signal seen by
working detectors that are not apparent in the dark
squids. Furthermore, in the case of \model{TMP}, the relationship
between the projection of the Earth's magnetic field on the instrument
and the magnitude of the pickup may not be linearly correlated as we
have assumed (verified in some cases by a comparison of the azimuthal
motion with working dark squid signals). In no case was the use of
these models able to reduce the $1/f$ knee substantially,
necessitating the continued use of \model{FLT} to remove the remaining
low-frequency noise. Neither \model{DKS} nor \model{TMP} are typically
used.

The \model{SMO} model uses a rolling mean or median boxcar filter to
calculate the low-frequency component of the bolometer signals, which
are then removed. In other words, this is an alternative to the
high-pass filtering for which \model{FLT} is generally used. The
primary reason for developing this model was to make it more robust
against ringing near the ends of the time-series, or residual spikes
(for which the median filtering is particularly useful). However, the
de-spiking and gap-filling algorithms that we have employed
(Section~\ref{sec:gaps}) successfully mitigate these problems, and the
\model{FLT} model is substantially faster.

Finally, we experimented with an alternative to \model{COM} for
removing correlated atmospheric noise. Rather than subtracting the
average signal at each time slice, \model{PLN} fits a plane to the
observed signal at each instant. We found that there was no obvious
improvement (either in terms of reducing the $1/f$ knee or reducing
the noise in final maps). This result is basically consistent with
those from earlier instruments \citep[e.g., the prediction for \scuba\
based on SCUBA data described in][]{chapin2002}, and suggests that the
angular scale of the emitting regions in the atmosphere are
un-resolved at the \scuba\ focal plane. \citet{sayers2010}, in
particular, examined several different ways of modelling and removing
the correlated atmospheric noise from Bolocam data at 143 and 268\,GHz
(also atop Mauna Kea), finding only marginal improvements by fitting a
plane, or even higher-order polynomials (see their figure~10, as well
as the discussion and references to earlier work in their
section~4.3). In \citet{aguirre2011} iterative PCA cleaning was used
to remove the atmosphere since the simpler methods mentioned left
substantial residuals in their data (similar to the \scuba\ data
described here, although PCA cleaning is prohibitively slow in our
case).

\subsection{Convergence tests and model degeneracies}
\label{sec:converge}

The map-maker will halt after a user-specified number of iterations,
or once some convergence criterion has been achieved. Presently two
numerical quantities are tracked after each iteration: the change in
reduced chi-squared, $\chi^2_\mathrm{r}$; and the change in the
map. While convergence in $\chi^2_\mathrm{r}$ is not, ultimately, a
good indicator of when the solution should be stopped, we discuss its
calculation and properties first as an introduction to the problem of
model degeneracies.

The bolometer residual variances, $\sigma^2_i$, are measured once the
modelled signal components have been removed. In the early iterations,
the residuals contain both white noise, and other long-timescale
features. However, when the solution has converged, this signal should
look white. Taking the component of bolometer variances due to white
noise, $\sigma_i^{\mathbf{w}2}$ (measured from the bolometer PSDs in
\model{NOI}, Section~\ref{sec:noi}), $\chi^2_\mathrm{r}$ is calculated
as $(1/N) \sum_i \sigma^2_i / \sigma_i^{\mathbf{w}2}$, where $i$ runs
over the $N$ bolometers. In other words, it is the average ratio
between the measured time-series bolometer variances and their
white-noise levels measured between 2 and 10\,Hz, and should tend to a
value of $\sim$1 if the low-frequency noise (and any bright
astronomical signals) have been successfully removed. However, this
expression does not account for the degrees of freedom in the
model. Clearly there are a large number of parameters that will
``fit-out'' some of the uncorrelated white noise, and bias this
estimate of $\chi^2_\mathrm{r}$ low. Indeed, for a converged map,
values in the range $\sim$0.85--0.95 are common, depending on the
precise configuration parameters. Regardless of the normalisation,
once the model parameterisation has been chosen, this quantity should
converge to some fixed value.

Practically speaking, $\chi^2_\mathrm{r}$ tends to converge before the
map due to model degeneracies. One simple example is the degeneracy
between large-scale astronomical structures (larger than the array
footprint), and the common-mode rejection step (\model{COM}), which
will be illustrated in Section~\ref{sec:point}. In this case, there is
a significant space of maps with spurious large-scale structures that
are anti-correlated with features in the common-mode, and the solution
is free to wander this space while producing flat residuals.

Instead, we typically use a map-based convergence statistic,
$M_\mathrm{c}$, the average absolute change in the value of map pixels
between subsequent iterations, normalised by the map pixel
uncertainties (square root of Eq.~\ref{eq:varmap}), or
\begin{equation}
M^j_\mathrm{c} = \frac{1}{N} \sum_i \frac{| \mathbf{m}(x_j,y_j) -
  \mathbf{m}(x_{j-1},y_{j-1}) |} {\sqrt{\mathbf{v}(x_j,y_j)}} ,
\end{equation}
where $i$ runs over the $N$ map pixels, and $j$ enumerates the
iterations. Experimentally we found that a change in this quantity
$<$0.05 (on average, map pixels change by $<$5\% of the estimated map
RMS in subsequent iterations) is a good point to stop, providing
correspondence with what we would choose ``by eye''. Letting the
solver run for many more iterations in several test cases yields
insignificant differences.

Another major source of divergence is correlation between \model{COM}
and \model{FLT}. Since \model{FLT} usually consists of a high-pass
filter following the application of \model{COM}, \model{COM} is
completely free to grow any large-scale structure at frequencies below
the chosen filter edge. While such structure does not appear in the
map (as it is removed by \model{FLT}), we found that the solution
could be made to converge significantly faster by ``re-mixing''
\model{COM} and \model{FLT}. Immediately prior to the calculation of
\model{COM}, the values of \model{COM} and \model{FLT} from the
previous iteration are added back into the residual simultaneously. In
this way, truly common-mode signals, even at low-frequencies, do not
leak into \model{FLT}.

In order to control the degeneracy between low-frequency signal that
is removed, and large-scale structures in the map, we have developed a
simple system for constraining regions of the map devoid of sources to
a value of zero for all but the final iteration in the \model{AST}
model (Section~\ref{sec:ast}). Such regions are either user-defined in
advance, or can be determined from the data using a cut on \snr. This
technique is explored considerably in the examples in
Sections~\ref{sec:point} and \ref{sec:extended}.

\subsection{Instrument and map-making performance}
\label{sec:performance}

From a scientific perspective, the most important goal of map-making
is to efficiently use all of the available data to achieve the
greatest sensitivity to astrononomical sources that the instrument is
capable of. Generally speaking, only a small percentage of the data
are flagged as unusable during map-making beyond the $\sim$70\%
detector yield.  As an example, for the 450\,\micron\ scan of Uranus
that will be presented in Section~\ref{sec:point}, during
pre-processing: 31\% of the data (1587/5120 bolometers) are flagged
due to data acquisition problems (dead or deactivated bolometers, or
non-linear flatfields); 0.12\% of the data are flagged due to steps
(including a small region about the steps, Section~\ref{sec:steps});
and 0.83\% of the data are flagged because the telescope was
stationary (Section~\ref{sec:flagbadbol}). Once the iterative solution
begins: 55 spikes are detected using the map-based despiker
($1.2\times10^{-6}$ \% of the data, Section~\ref{sec:ast}); and 0.72\%
of the data are flagged as outliers compared to the common-mode
(Section~\ref{sec:comgai}).

The white noise levels of \scuba\ bolometers are measured between 2
and 10\,Hz in the PSDs (as described in Section~\ref{sec:noi}),
leading to noise equivalent powers (NEPs, with units W\,s$^{1/2}$) --
the \rms\ time-series noise in a 1\,s integration \citep[see section
3.5.2 in][]{holland2013}. The NEPs are then converted into more useful
noise equivalent flux densities (NEFDs, with units Jy\,s$^{1/2}$)
through multiplication by the extinction correction and the FCF. For
an ideal scan pattern in which every bolometer spends an equal amount
of time observing each point on the sky, the expected noise in a map
pixel is $\mathrm{NEFD_{eff}}/t^{1/2}$, where $t$ is the integration
time, and the effective NEFD, $\mathrm{NEFD_{eff}} = (1/\sum
\mathrm{NEFD}_i^{-2})^{1/2}$, is the sensitivity of the entire array
including all bolometers $i$. Experimentally it has been found that
the map noise based on the weighted scatter of the data
(Eq.~\ref{eq:varmap}), and the scatter of map pixels in small
apertures, both correlate well with this expected limiting noise
performance for maps that are larger than the array footprint (for
smaller maps the assumption of uniform coverage is incorrect). As an
example of observing under favourable conditions with line-of-sight
opacities of 0.87 and 0.34 at 450 and 850\,\micron\ \citep[zenith sky
opacities at 225\,GHz of 0.04 and 0.065 at 450 and 850\,\micron,
respectively, at an average airmass of 1.2, from table 3
in][]{holland2013}, the mapping speed is approximately
$6\times10^{-5}$ and $3\times10^{-3}$\,deg$^2$\,hr$^{-1}$\,mJy$^{-2}$,
at 450 and 850\micron, for the default pixel sizes of 2 and 4\,arcsec,
respectively. If maps are filtered and cross-correlated with the PSF
for the optimal extraction of point sources (see
Section~\ref{sec:cosmo}), the mapping speeds increase to approximately
$5\times10^{-4}$ and
$1.5\times10^{-2}$\,deg$^2$\,hr$^{-1}$\,mJy$^{-2}$ in this example,
independently of the chosen map pixel sizes. See the on-line \scuba\
integration time calculator for mapping speeds under arbitrary
observing conditions and scan
modes\footnote{\url{http://www.jach.hawaii.edu/jac-bin/propscuba2itc.pl}}.

\section{Examples}
\label{sec:examples}

While most of the steps described in Section~\ref{sec:algorithm} for
reducing data with SMURF are applicable to all data sets, there is no
``default'' reduction that provides good results for a wide range of
source \snr\ and angular scales. However, SMURF does have a small
number of configurations with variations on these baseline parameters
that are applicable to most common types of data. In this section we
illustrate the differences between these configurations with the
following examples: a bright point source, Uranus
(Section~\ref{sec:point}); a blind survey of high-redshift galaxies in
the Lockman Hole (Section~\ref{sec:cosmo}); and a map of a bright
extended star-forming region in our Galaxy, M17
(Section~\ref{sec:extended}).

\subsection{Known point source}
\label{sec:point}

\begin{figure*}
\centering
\includegraphics[width=\linewidth]{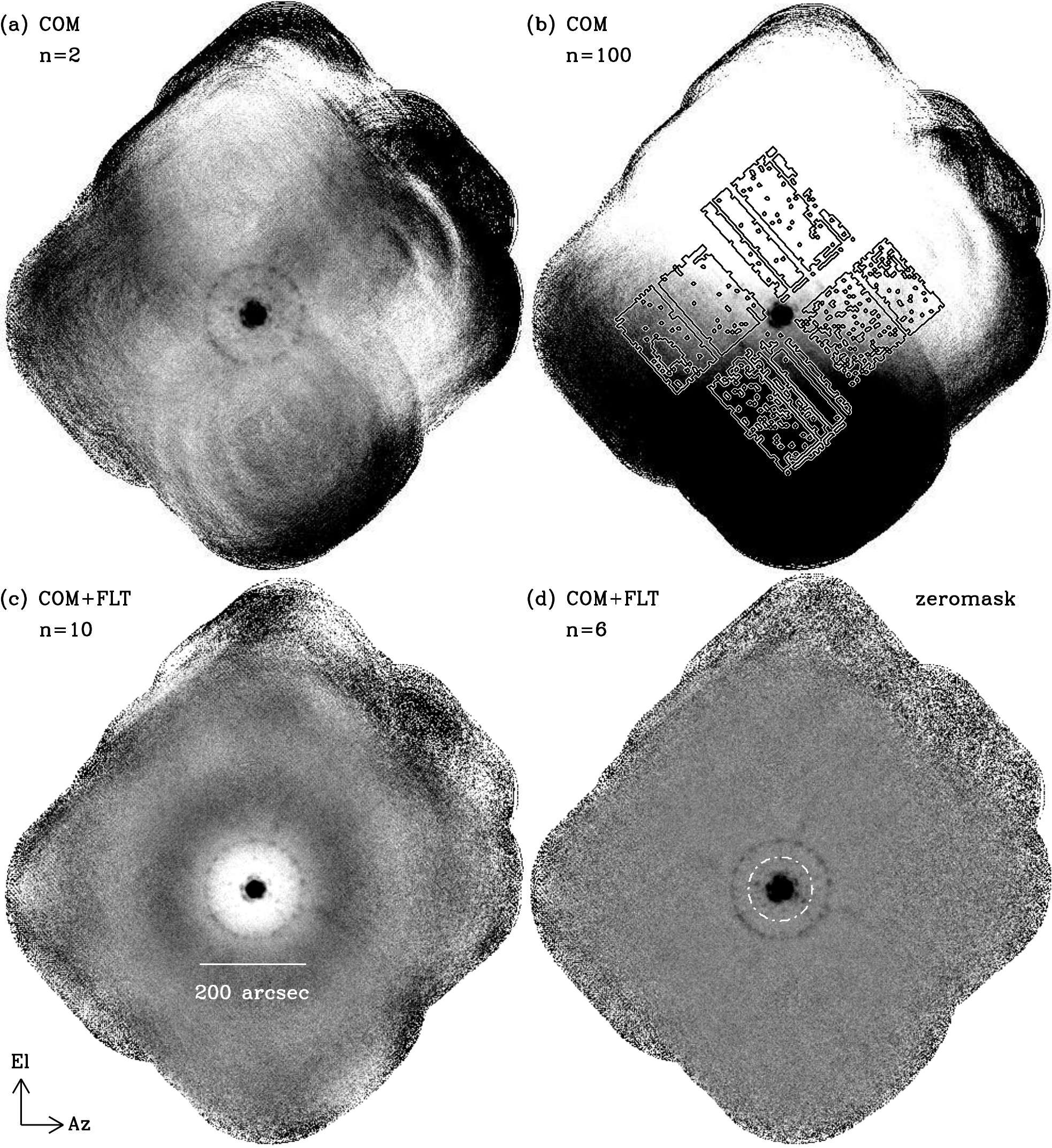}
\caption{Multiple reductions of a 450\,\micron\ CV daisy scan of
  Uranus, all scaled between $-$0.002\,pW (white) and +0.002\,pW
  (black). In all cases extinction correction has been applied
  immediately after common-mode removal. (a) Reduction in which only
  common-mode subtraction is used to suppress low-frequency noise, and
  the reduction is forced to use 2 iterations (after the first
  iteration an estimate of the source flux is removed, and the noise
  is measured in the residuals to obtain appropriate weighting for the
  second and final iteration). Circular streaks are caused by
  independent low-frequency noise that is not removed by the
  common-mode (Uranus peak 0.358\,pW). (b) Same as (a) but using 100
  iterations, illustrating the degeneracy between large-scale
  structure and the common-mode (the footprint of working bolometers
  is also shown for reference, Uranus peak 0.358\,pW). (c) Reduction
  in which high-pass filtering above 0.775\,Hz (corresponding to a
  spatial scale of 200\,arcsec, as indicated) is applied after
  common-mode removal, but before the map estimate. The map-based
  convergence test is activated and the solution halts after 10
  iterations, but leaving large-scale ringing due to the filter
  (Uranus peak 0.353\,pW). (d) Same as (c), but now the region of the
  map beyond the white dot-dashed circle is constrained to a value of
  zero until all but the final iteration.  The map is now extremely
  flat, there is no attenuation of the source flux compared with the
  first two reductions, and the diffraction pattern is clearly seen
  (Uranus peak 0.358\,pW).  }
\label{fig:pointmaps}
\end{figure*}

The accurate measurement of positions and brightnesses of known point
sources are necessary in real-time to establish telescope pointing
offsets and focus. They are also necessary to measure the FCF
(absolute calibration), and hence noise performance of the instrument
in astronomically-useful units.  In this example we reduce a
450\,\micron\ map of Uranus (observation 26 on 2011 October 17), which
is a nearly point-like source for \scuba\ that is commonly used as a
primary flux calibrator. The CV daisy pattern was used, with a scan
speed of 155 arcsec\,sec$^{-1}$. We perform several different
reductions of the data to illustrate the purpose of various model
components and the convergence properties of the solution
(Fig.~\ref{fig:pointmaps}). In all cases the maps are produced on a
grid of azimuth (horizontal) and elevation (vertical) offsets from the
position of Uranus (the origin), using 2\,arcsec pixels.

The first, simplest reduction of the data uses only the \model{COM}
model to estimate and remove the common-mode signal in order to
suppress low-frequency noise in the data. After \model{COM}, the
extinction correction is applied (\model{EXT}), and an initial map is
estimated using equal weighting for all of the detectors. This
estimate of \model{AST} is then removed from the data, and the noise
is measured in the residuals to estimate weights for the subsequent
and final iteration. The resulting map after these two iterations is
shown in Fig.~\ref{fig:pointmaps}a. While the peak \snr\ of Uranus is
clearly large ($\sim$160), enabling us to see the faint sidelobes
(circle and cross pattern), the map also has obvious circular streaks
and other large-scale ripples. The circular streaks are due to the
fact that \model{COM} does not account for all of the low-frequency
(i.e., uncorrelated) noise (see Fig.~\ref{fig:pspec}), and therefore
each bolometer leaves a trace of the circular scan pattern in the map,
as their baselines slowly drift independently. A significant
contribution to the larger-scale ripples in the map, however, can be
made by degeneracies in the map solution, as we discuss below.

\begin{figure}
\centering
\includegraphics[width=\linewidth]{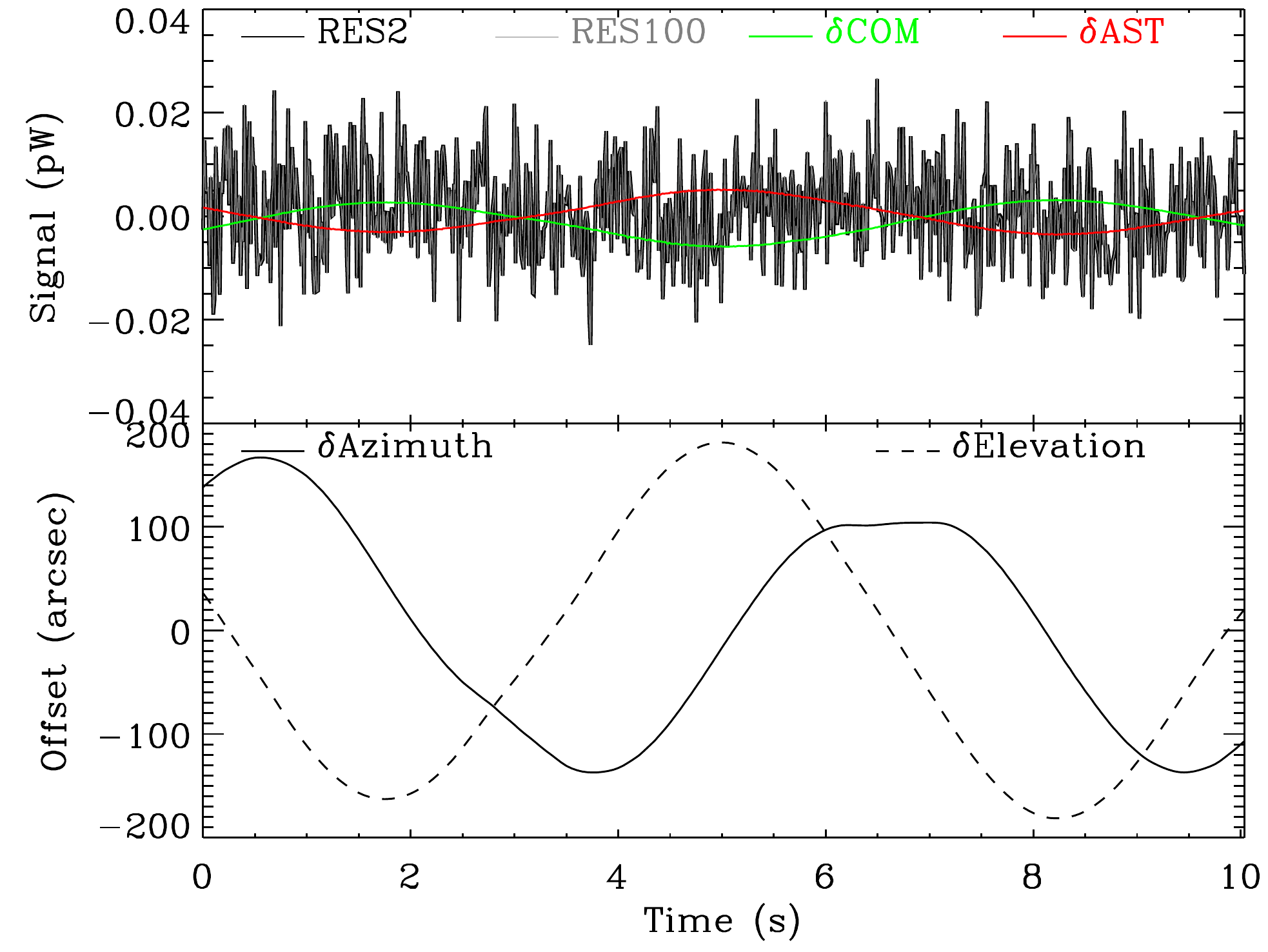}
\caption{Demonstration of the degeneracy between large-scale
  structures in the map and common-mode removal, corresponding to
  panels (a) and (b) in Fig.~\ref{fig:pointmaps}. The black and grey
  lines in the top panel show the residual signal for a single
  bolometer after 2 and 100 iterations, respectively (they are nearly
  identical; the black line lies beneath the grey line). The green and
  red lines show the difference between the \model{COM} and
  \model{AST} (the signal produced by the current map estimate for a
  given bolometer) model components for that bolometer between
  iterations 2 and 100, respectively. This shows that a strong signal
  has grown over time, and it has equal but opposite signs in the two
  model components, so that they cancel one another. The bottom panel
  shows the scan pattern of the telescope over the same period;
  clearly the \model{COM}/\model{AST} model degeneracy is correlated
  with the elevation offset, and referring to
  Fig.~\ref{fig:pointmaps}, panel (b), this corresponds to the strong
  vertical gradient that has appeared}.
\label{fig:degeneracy}
\end{figure}

To illustrate how large-scale ripples can form (and grow), the same
map solution is run for 100 iterations and shown in
Fig.~\ref{fig:pointmaps}b, now exhibiting a strong vertical
gradient. The degeneracy is easy to understand if the time-domain
behaviour of each model component is considered. The top panel of
Fig.~\ref{fig:degeneracy} shows the residual signals for a single
bolometer after 2 (black) and 100 (grey) iterations, which are nearly
identical, yet the change in the estimated
\model{COM}\footnote{\model{COM} has been multiplied by the
time-varying extinction correction to enable direct comparison with
\model{AST}.} (green) and \model{AST} (red) signals between 2 and 100
iterations are large However, it is also clear that the estimated
\model{COM} and \model{AST} signals are complementary. In other words,
the large change in the \model{AST} signal is cancelled by freedom in
the \model{COM} signal to grow with opposite sign. For comparison, the
bottom panel of Fig.~\ref{fig:degeneracy} shows the telescope pointing
for this section of data, and the shapes of the \model{AST} and
\model{COM} signals match the elevation component, which is aligned
with the gradient in Fig.~\ref{fig:pointmaps}b. Generically, the
calculation of \model{COM} will remove any information on angular
scales that are larger than the array footprint (outline shown in
Fig.~\ref{fig:pointmaps}b for reference), meaning that the map
solution is unconstrained on such large scales.

Attacking the problem of streaks, a simple method of removing residual
(un-correlated) sources of low-frequency noise is to apply a high-pass
filter after the common-mode removal. A third reduction of the data
uses the baseline map-making parameters, as described in
Section~\ref{sec:algorithm}, which adds the \model{FLT} model to
accomplish this task immediately prior to map estimation. We set the
filter edge based on an angular scale of 200\,arcsec, which, given the
scan speed of 155\,arcsec\,sec$^{-1}$, corresponds to a frequency of
0.78\,Hz (note in Fig.~\ref{fig:pspec} that this frequency is slightly
above the $1/f$ knee at 450\,\micron\ after common-mode removal). Now
using the automatic map-based convergence test
(Section~\ref{sec:converge}), the solution converges after 10
iterations (Fig.~\ref{fig:pointmaps}c). Both the circular streaks and
the large-scale gradient are now removed, but they have been replaced
by an obvious, circularly-symmetric ringing pattern about Uranus. The
reason for this ringing is that the hard-edged high-pass filter in
frequency space is equivalent to a $\sinc$ function-like response in
map space. Since the scan pattern is fairly isotropic (scans at all
position angles), and there is a bright point-like source at the
centre, the result is an azimuthally-symmetric $\sinc$ function-like
pattern in the map. The peak flux has also been attenuated to
0.353\,pW in this reduction, compared with 0.358\,pW in the first two
reductions. The gradient resulting from the degeneracy between the
common-mode and map has been removed because the filter scale of
200\,arcsec (3.3\,arcmin) is smaller than the array field-of-view
($\sim$5\,arcmin). If it were greater, degeneracies would again begin
to appear on those larger scales.

This example illustrates the need for constraints in the map solution
in many situations. For calibrators (and other previously known
bright, compact sources), a good, simple prior is to constrain the map
to a value of zero away from the known locations of emission (part of
the \model{AST} model calculation, Section~\ref{sec:ast}).  In
Fig.~\ref{fig:pointmaps}d, a solution is produced in an identical
manner to Fig.~\ref{fig:pointmaps}c, but now setting the map
explicitly to zero beyond a radius of 60\,arcsec from the location of
Uranus (much larger than the FWHM of the main lobe), for all but the
final iteration. In this case, the map converges after 6 iterations,
and the ringing has been effectively removed. The attenuation from the
previous reduction is now removed, and Uranus again has a peak value
of 0.358\,pW. For reference, the Uranus model distributed with
Starlink predicts a peak brightness of 176\,Jy on this date, yielding
an FCF of 491 Jy\,pW$^{-1}$, which is well within the typical range
for \scuba\ of $(492 \pm 67)$\,Jy\,pW$^{-1}$ \citep[see][for
details]{dempsey2013}. Since the map is now flat away from the source,
and constrained to a value of zero, it is appropriate for performing
aperture photometry directly, with no need for an additional reference
annulus.

The way this prior works can be understood from the point of view of
differential measurements. Bolometer data contain information up to
scales corresponding to the filter edge (or the scale of the array
footprint, whichever is smallest). In this example, the relevant scale
is 200\,arcsec. Since the map is constrained to zero within 60\,arcsec
of the peak of Uranus (well within 200\,arcsec), the solution is able
to accurately reconstruct the differential peak intensity of Uranus
with respect to this constrained background. This approach to
map-making is similar to that employed for poorly cross-linked scans
of compact (though resolved) sources by \citet{wiebe2009} using BLAST
data.

\subsection{Deep point source survey}
\label{sec:cosmo}

\scuba\ surveys designed to detect extremely faint point-sources
(e.g., high-redshift star-forming galaxies, and features in debris
disks) are ideally limited by the white-noise performance of the
instrument. The approach described here for maximising the \snr\ of
point-sources involves three major steps: (i) generating a map that
removes most large-scale noise sources with approximately linear
response, without prior knowledge of the location of sources; (ii)
apply a Fourier-space ``whitening filter'' to suppress residual
large-scale noise; and (iii) detecting point sources using a ``matched
filter''. Note that variations on this general procedure have been
used extensively in the submm cosmology community using previous
instruments
\citep[e.g.,][]{scott2002,borys2003,laurent2005,coppin2006,scott2008,perera2008,devlin2009}.
In this section we reduce scans of the Lockman Hole taken during S2SRO
as a pilot project for the SCUBA-2 Cosmology Legacy Survey. It
consists of $\sim8.5$\,hours of data taken in 36 separate scans
(average 15\,min. each) spread over February and March 2010. Each scan
is a 360\,arcsec rotating PONG pattern, with a scan speed of
240\,arcsec\,s$^{-1}$, covering an area of about 50\,arcmin$^2$. The
full list of dates and observation numbers includes: 2010 February 18,
63, 64, 65, 70, 71, 72, 90, 91, 92, 97, 98, 99; 2010 February 20, 111,
112, 113, 118, 128, 129, 130; 2010 March 3, 61, 62, 64, 69, 70, 72,
73, 74, 75; 2010 March 9, 87, 88, 90, 91; and 2010 March 11, 59, 64,
65, 72. These observations represent about 80 per cent of the total
data taken as part of the project; the remaining observations were
dropped due to problems in keeping the arrays properly tuned during
S2SRO, and were easily identified by their highly variable and erratic
bolometer time-series (which resulted in maps full of large streaks).

\begin{figure*}
\centering
\includegraphics[width=\linewidth]{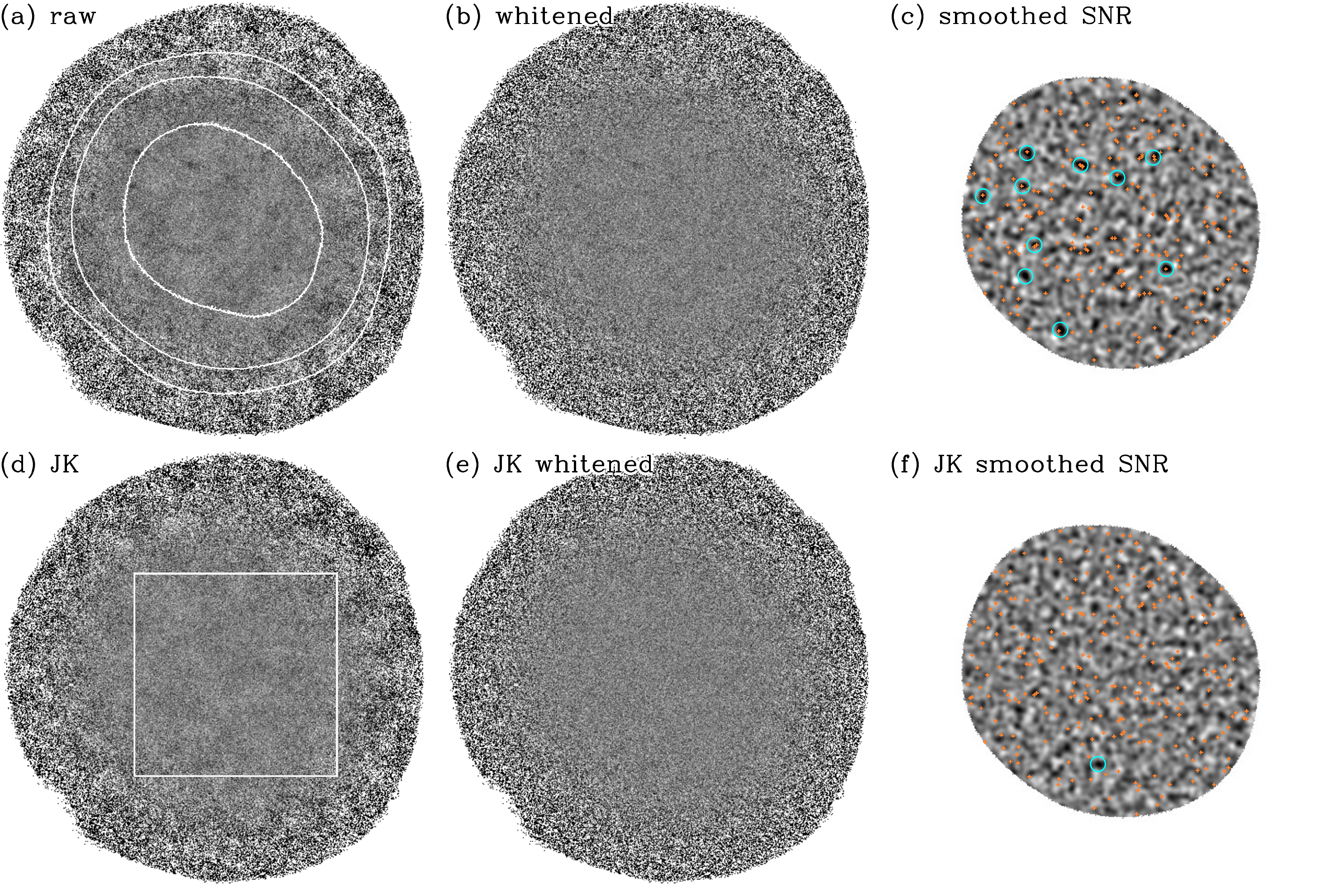}
\caption{Reduction of a blank-field survey: the Lockman Hole. (a) Raw
  output of SMURF using high-pass filtering as a pre-processing step,
  followed by 4 iterations using only the \model{COM} model to remove
  residual correlated noise. White contours correspond to estimated
  noise levels of 1.25, 2.5 and 5.0 times the minimum noise at the
  centre of the map. (b) The whitened map using a filter based on the
  angular power spectrum of the jackknife map. (c) The whitened map
  cross-correlated with the whitened PSF to identify point sources
  [restricted to a lower-noise region, within the area denoted by the
  second contour of panel (a)]. The image shows the \snr\, with
  3.8-$\sigma$ peaks indicated by blue circles (radius 8\,arcsec). The
  orange ``$+$'' signs show the locations of 1.4\,GHz radio sources
  from \citet{owen2008} with $S\gsim15$\,$\mu$Jy. Of the 10 submm
  peaks, 9 are within a search radius of 8\,arcsec of at least one
  radio source. (d) Jackknife map produced from the difference of two
  maps, using the even and odd scan numbers, respectively. Provided
  that all noise sources are statistically uncorrelated between the
  two halves of the data, the map is a plausible realisation of the
  noise without contamination from astronomical sources. (e) The
  jackknife map whitened using the same filter as that in panel
  (b). The whitened jackknife map cross-correlated with the same
  whitened PSF as in panel (c). Again, orange ``$+$'' and blue circles
  indicate radio sources and 3.8-$\sigma$ peaks, respectively. Unlike
  panel (c), there is only a single (apparently) significant peak, and
  it is not in the vicinity of a radio source.}
\label{fig:lockman_maps}
\end{figure*}

\begin{figure}
\centering
\includegraphics[width=\linewidth]{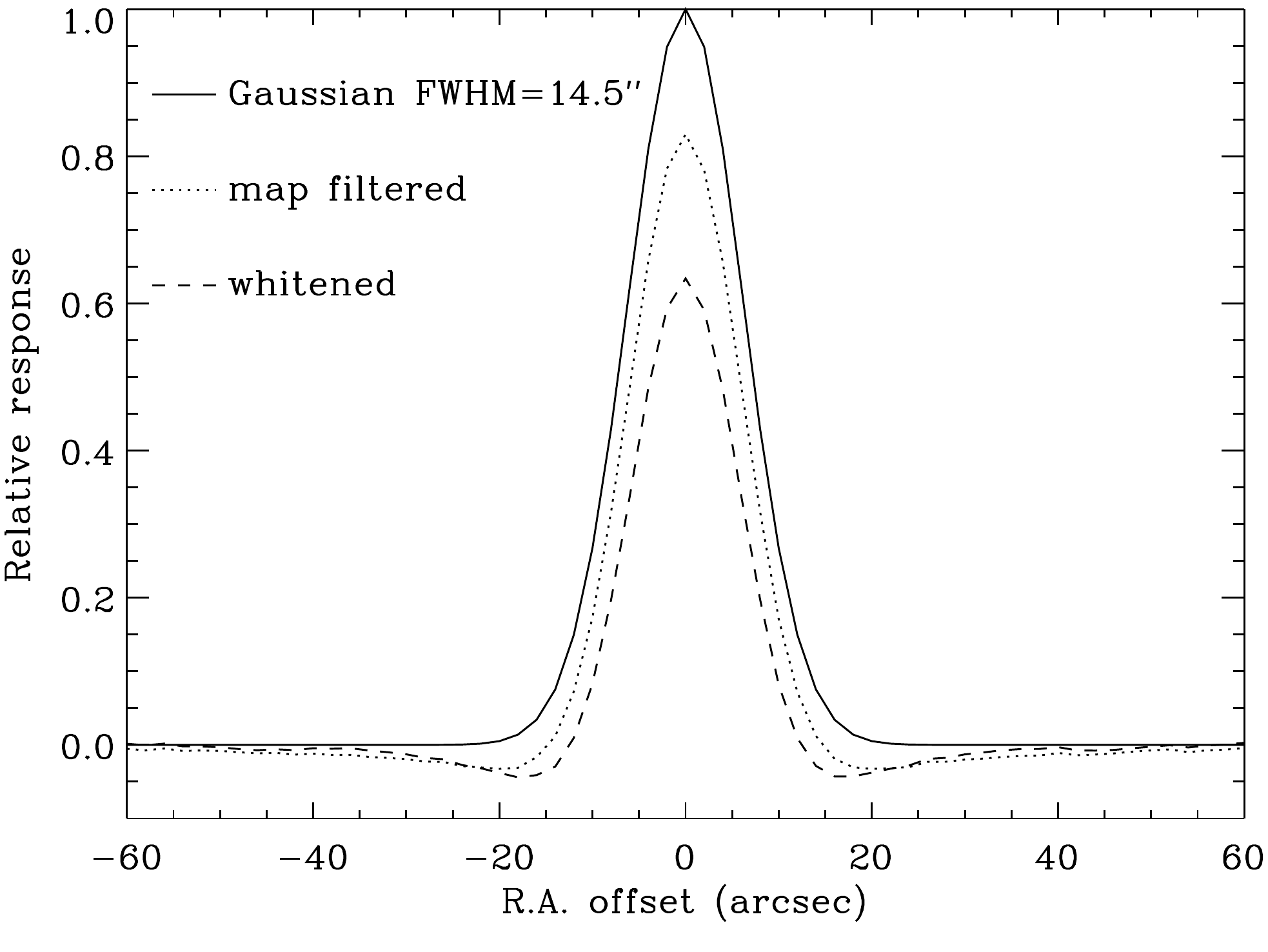}
\caption{Slice through the angular response to an ideal Gaussian point
  source (solid line) along the R.A. axis following map-making using
  the blank-field processing configuration with SMURF (dotted line),
  and upon application of the whitening filter (dashed line). The
  ``map filtered'' response is produced by adding the ideal Gaussian
  to the real data near the centre of the Lockman Hole map, and
  re-reducing the data. The resulting dotted line gives the expected
  shape of a source in panel (a) from
  Fig.~\ref{fig:lockman_maps}. Applying the whitening filter to the
  dotted line (dividing the Fourier Transform of the map filtered
  response by the square root of the solid orange line in
  Fig.~\ref{fig:lockman_pspec}, and then transforming back to real
  space) gives the whitened line (dashed), which is the expected shape
  of a source in panel (b) from Fig.~\ref{fig:lockman_maps}. Finally,
  cross-correlating the whitened map with this whitened PSF is an
  effective ``matched filter'' for identifying point-like sources, and
  this smoothed map is shown in panel (c) of
  Fig.~\ref{fig:lockman_maps}.}
\label{fig:lockman_psf}
\end{figure}

The first step, map generation, is different from that described in
Section~\ref{sec:point} in two key ways. Since the locations of
sources are unknown \emph{a priori}, a map constraint is not
employed. Large-scale diverging structures in the map must be
mitigated, and the method used in this example (the baseline
processing in SMURF) is to apply a high-pass filter to the data once,
as a pre-processing step. The iterative solution is then run using
only \model{COM}, \model{EXT}, \model{AST}, and \model{NOI}. In other
words, there is no information in the bolometer signals below some
cutoff frequency, and residual correlated high-frequency noise above
the cutoff is only removed through iterative common-mode
subtraction. In this case, the filter edge has been chosen to remove
scales larger than 200\,arcsec, or a high-pass filter above 1.2\,Hz
given the scan speed. Since the data are high-pass filtered prior to
the iterative solution, \model{GAI} (fitting an independent amplitude
of \model{COM} to each bolometer) has been de-activated, since there
is very little structure in the common-mode with which to fit an
accurate gain. The map is shown in Fig.~\ref{fig:lockman_maps}a. The
map-maker has been tested in two ways: (i) large numbers of iterations
are used to verify that the maps converge without the growth of large
structure; and (ii) adding synthetic sources to the real time-series
data (a built-in feature of SMURF) at a range of brightnesses verify
that the map-maker response to them is linear (i.e., the relative
shape and amplitude compared to the input source is independent of
brightness). The response to a synthetic point source (solid line)
after map-making (dotted line) is shown in
Fig.~\ref{fig:lockman_psf}. Clearly the use of a high-pass filter as a
pre-processing step, and having no other map-constraints, has the
down-side of introducing sidelobes around the main peak. Furthermore,
the details of this shape depends on the high-pass filter edge that
has been chosen (the higher the frequency, the greater the attenuation
of the central peak, and the larger the negative side-lobes).
However, the way this filter affects point-sources is measurable
(using the synthetic source injection facility of SMURF), and linear.

Even though the map looks quite flat, there is a mixture of faint
astronomical sources, and what is probably residual low-frequency
noise, causing faint patchiness visible to the naked eye. Since the
mixture of the two components is unknown, the first step is to
suppress the low-frequency noise, under the assumption that such
contaminants occur randomly in time, while astronomical sources are
(usually) constant.

First, the angular power spectrum of noise is estimated from a
``jackknife map'': maps are produced from two independent halves of
the total data set, and the jackknife signal in a map pixel,
$S_\mathrm{JK}$, and its variance, $\sigma^2_\mathrm{JK}$ are
estimated from the two input map fluxes, $S_1$, $S_2$, and the
corresponding variances, $\sigma^2_1$, and $\sigma^2_2$ as,
\begin{eqnarray}
S_\mathrm{JK} = \frac{S_1 - S_2}{2}, \\
\sigma^2_\mathrm{JK} = \frac{\sigma^2_1 + \sigma^2_2}{4}.
\end{eqnarray}
Provided that the noise in one half of the data is uncorrelated with
that from the other half, the signal in the jackknife map should
resemble noise drawn from the same parent distribution as that of the
real map. The astronomical signal, however, should be cleanly removed
(provided that there are no strong time-varying signals, and also
assuming that errors due to calibration between the two halves are
insignificant). The approach we have taken to minimize systematics is
to produce the two maps using odd and even scan numbers (i.e., each
map will contain a nearly uniformly-spaced sampling of data across the
full data set).

\begin{figure}
\centering
\includegraphics[width=\linewidth]{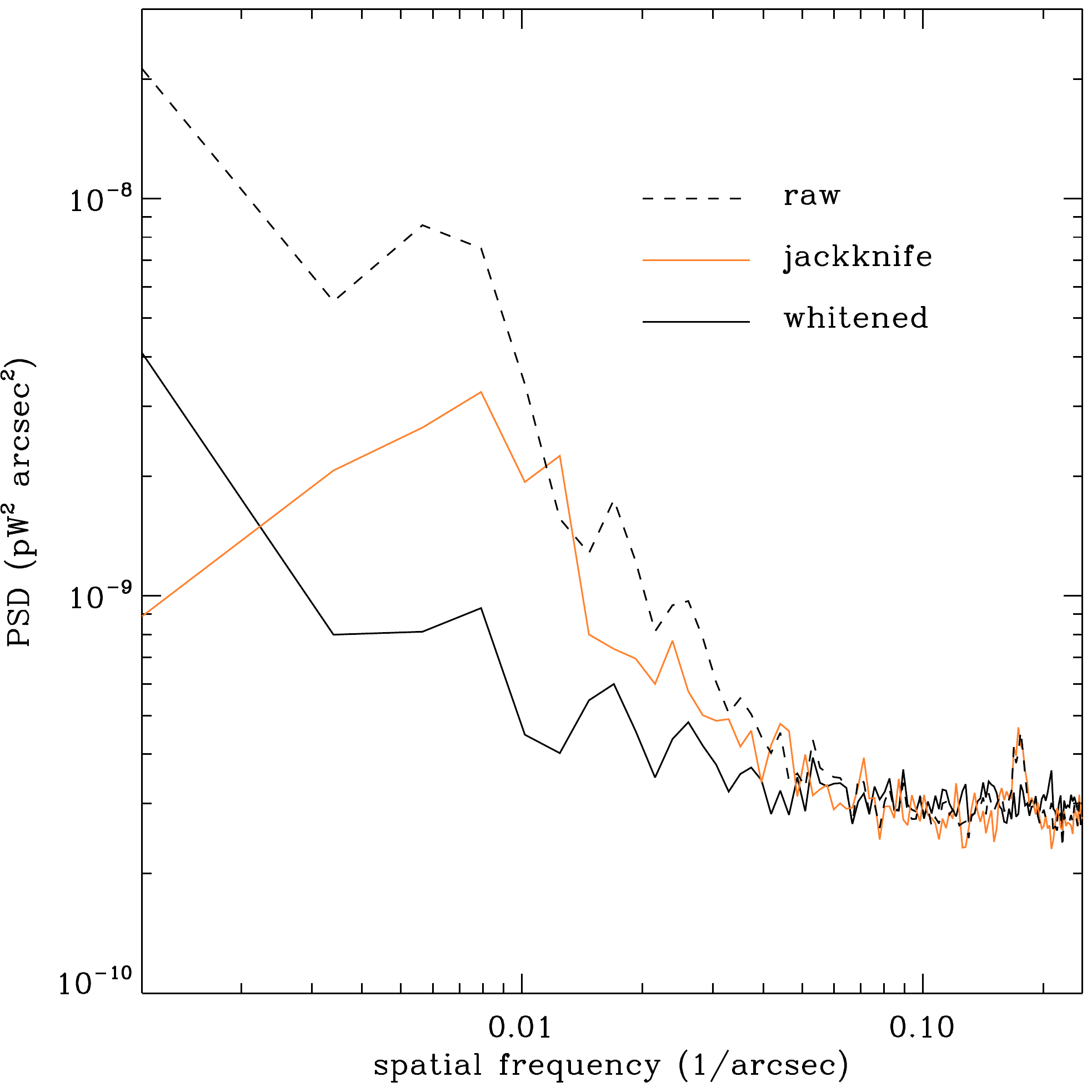}
\caption{Radial (azimuthally-averaged) angular power spectral
  densities for the raw map output by SMURF (dashed black line), the
  jackknife (solid orange line), and whitened (solid black line) maps
  (panels (a), (d), and (b) from Fig.~\ref{fig:lockman_maps},
  respectively), considering only the square region indicated in
  Fig.~\ref{fig:lockman_maps}d. Since the raw map contains spatially
  correlated signals on large scales, both due to noise and
  astronomical signals (low-spatial frequencies), the jackknife map
  (difference of two approximately equal-length subsets of the data)
  is used to generate a plausible realisation of pure noise. Assuming
  that the noise properties are isotropic, a whitening filter is
  estimated from the reciprocal of the jackknife power spectrum with
  only a radial dependence. The power spectrum of the resulting signal
  map still has residual power on large scales, which is presumably
  due to astronomical sources.}
\label{fig:lockman_pspec}
\end{figure}

Since the \scuba\ scan strategy is usually isotropic (all position
angles scanned with roughly equal weights), we make the simplifying
assumption that the angular noise power spectrum is azimuthally
symmetric. For these data, there are no obvious anisotropic structures
in the 2-dimensional FFT. The radial (azimuthally-averaged) angular
power spectrum therefore encodes all of the useful information about
the noise properties. These power spectra for the raw output of SMURF,
and the jackknife map (transforming only the approximately uniform
region indicated by the square in Fig.~\ref{fig:lockman_maps}d in each
case) are shown by the dashed black, and solid orange lines in
Fig.~\ref{fig:lockman_pspec}, respectively. Both power spectra are
approximately flat at spatial frequencies $\gsim 0.06$\,arcsec$^{-1}$
(scales $\lsim16$\,arcsec), with the exception of a spike at
$\sim0.175$\,arcsec$^{-1}$ (a scale of $\sim5.7$\,arcsec). One
possibility for this feature is that it is related to the
inter-bolometer spacing in the focal plane (which happens to be this
size): small relative drifts in the baselines of adjacent bolometers
may produce faint parallel stripes in the map along the scan direction
(the superposition of many scans at different angles then results in
an isotropic noise pattern). It is not likely that this signal is due
to astronomical sources because it appears with a nearly equal
amplitude in both the real map and the jackknife.  At lower spatial
frequencies, both the real map and the jackknife power spectra grow
significantly, as a result of the more obvious large-scale patchiness
in Figs.~\ref{fig:lockman_maps}(a) and (d).

To suppress noise in the map, we construct a whitening filter whereby
the Fourier Transform of the map is divided by the square root of the
jackknife power spectrum (orange line in
Fig.~\ref{fig:lockman_pspec}), normalised by the white-noise level
estimated from the RMS power at angular frequencies $>
0.1$\,arcsec$^{-1}$, and transforming back into real space. The
whitened map is shown in Fig.~\ref{fig:lockman_maps}b, and for
comparison, the jackknife map has also been whitened in
Fig.~\ref{fig:lockman_maps}e. In both cases, the maps are visibly
flatter than the non-whitened cases.

The angular power spectrum of the whitened signal map is shown with a
solid line in Fig.~\ref{fig:lockman_pspec}. At low angular frequencies
($\lsim 0.5$\,arcsec$^{-1}$) there is significantly less power than
for the raw map. However, it also clearly has some power in excess of
the white noise level. In theory, this residual signal is produced by
astronomical sources, although its origin cannot be determined from
this plot alone; nor are astronomical sources readily visible in the
map (Fig.~\ref{fig:lockman_maps}b). To identify sources, we next apply
a ``matched filter'' to the whitened maps.

For blind, high-redshift surveys, individual sources are expected to
be un-resolved by the \scuba\ 7.5--14.5\,arcsec FWHM beams. Under this
assumption, and also assuming that the map noise is white,
cross-correlation between the map and the known PSF, or matched
filtering, yields the maximum-likelihood flux density of supposed
point-sources centred over every location in the resulting map
\citep[an extremely well-known result throughout astronomy,
see][]{stetson1987}. Peak identification in such smoothed maps have
been used extensively in the submillimetre community, as both an
efficient source-detection and photometric measurement strategy.  For
the case at hand, we may use this cross-correlation technique since
the map has been whitened. However, we must first establish the
effective shape of point sources in this map due both to map-making
itself, and the whitening filter. We determine the effect of
map-making by adding a synthetic (and high-\snr) point source to the
real data, and measure its resulting shape in the map (solid and
dotted lines in Fig.~\ref{fig:lockman_psf}, respectively). Next, the
Fourier Transform of the map filtered PSF is divided by the square
root of the jackknife noise power spectrum to calculate the final
whitened PSF (dashed line in Fig.~\ref{fig:lockman_psf}).  Both the
whitened signal and jackknife maps are smoothed by this shape and
shown in Figs.~\ref{fig:lockman_maps}c and f, respectively. Note that
these images are plotted in \snr\ units, where the smoothed noise maps
have been calculated by propagating the original noise maps output by
SMURF through both the whitening and matched filters (each of which is
a linear operations). In terms of the angular power spectra, this
complete process can be thought of as an optimal band-pass filter that
has both suppressed low-frequency noise, and information on scales
that are smaller (higher frequencies) than the beam.

Have real astronomical sources been detected using the matched filter?
For both the smoothed signal and jackknife maps, blue circles denote
3.8-$\sigma$ peaks. While not justified here, this threshold is fairly
typical for other ground-based submillimetre surveys in recent years
\citep[e.g.,][]{coppin2006,perera2008,2009ApJ...707.1201W} leading to
estimated false-identification rates of order $\sim$5\%, and is chosen
as a convenient reference. In the former, 10 peaks are found, whereas
there is only 1 in the latter. However, this test does not preclude
the possibility that some correlated noise made it into the jackknife
map, in which case the estimated \snr s are misleading.

\begin{figure}
\centering
\includegraphics[width=\linewidth]{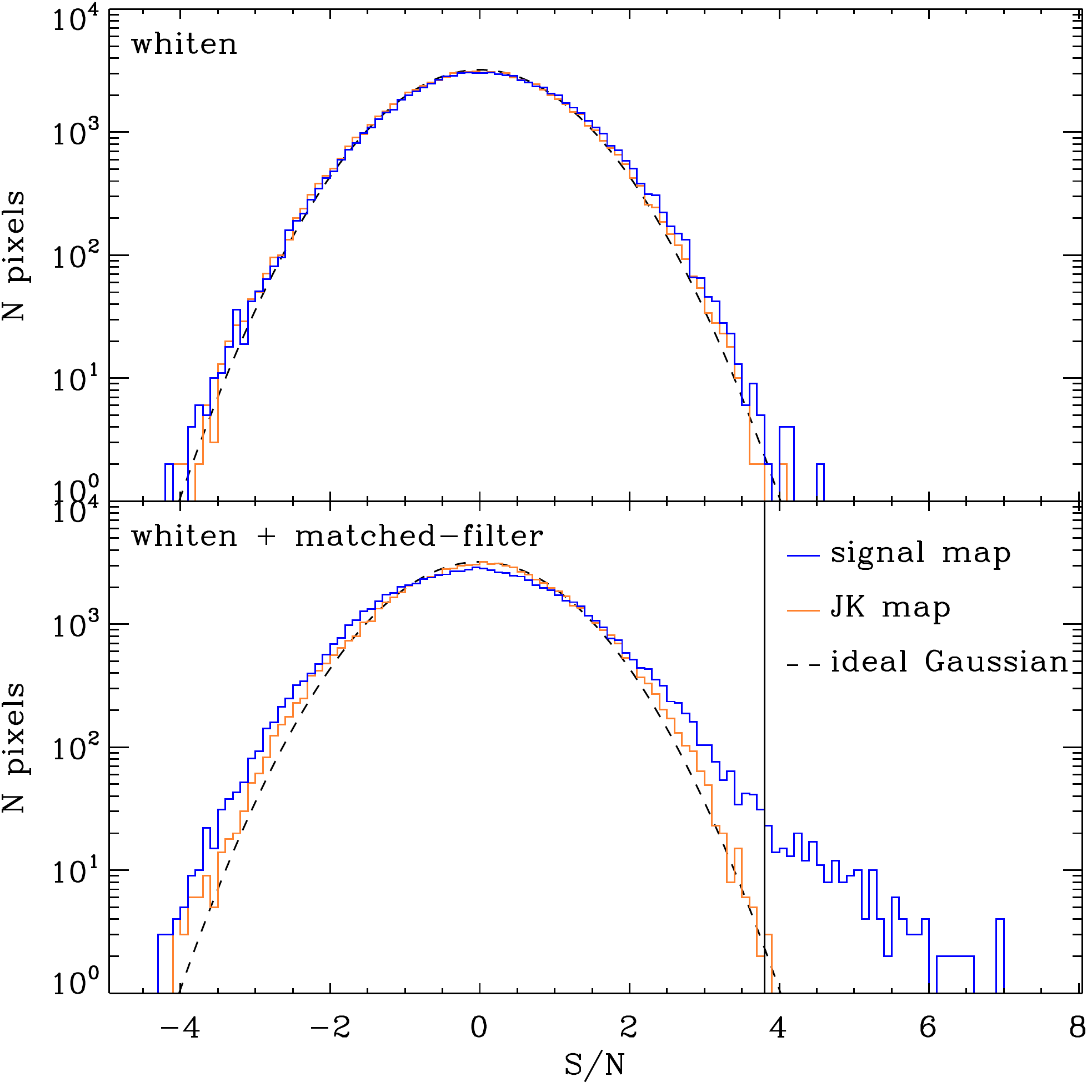}
\caption{Histograms of \snr\ using pixels from the central region of
  the Lockman Hole (second contour) in
  Fig.~\ref{fig:lockman_maps}a. The top panel shows the histograms for
  pixels from the whitened signal and jackknife maps
  (Figs.~\ref{fig:lockman_maps}b,e), compared to a Gaussian
  distribution with mean zero, standard deviation one, and an area
  normalised to the total number of pixels -- the expected
  distribution for a map of spatially-uncorrelated Gaussian noise. The
  good agreement indicates that these maps are indeed dominated by
  white noise. The lower panel shows the results for matched filtered
  signal and jackknife maps (Figs.~\ref{fig:lockman_maps}c,f). The
  filtered jackknife map distribution is still close to the
  expectation (Gaussian), but now the matched filter has picked out
  significant signal, leading to the large positive tail. The vertical
  solid line shows the chosen 3.8-$\sigma$ source-detection
  threshold.}
\label{fig:lockman_hist}
\end{figure}

One simple test of the calculated noise properties is to compare the
signal and jackknife \snr\ distributions with ideal Gaussians. The top
panel of Fig.~\ref{fig:lockman_hist} shows the whitened (but not
match-filtered) signal (blue) and jackknife (histograms), along with a
Gaussian (mean 0, standard deviation 1, and area normalised to the
number of map pixels) as a dashed line. In this case, it is clear that
the \snr\ distributions for both maps are nearly indistinguishable
from the theoretical distribution of white noise. This result shows us
that: (i) the whitening filter appears to have removed correlated
large-scale noise, since the jackknife map histogram is consistent
with white noise; and (ii) any potential astronomical signals are
small compared to the typical white noise in most map pixels
(unsurprising given the appearance of
Fig.~\ref{fig:lockman_maps}b). Next, we examine the \snr\ histograms
for maps processed with the matched filter in the bottom panel of
Fig.~\ref{fig:lockman_hist}. Again, the histogram of the jackknife
\snr\ data appears consistent with pure noise. However, the signal map
now deviates significantly from a Gaussian distribution, with a clear
positive tail (as one would expect for emitting sources). In fact,
integrating the positive tails beyond our 3.8-$\sigma$
source-detection threshold (vertical solid line) yields 229 map pixels
in the signal map, compared with 3 in the jackknife map (out of 80603
pixels in the entire region). Recall that given the small pixel size
of the map, several pixels generally contribute to each peak. This
procedure gives only a flavour of the analysis that is usually
required to produce a robust source lists. Additional tests, along
with a careful consideration of completeness and false-positive rates,
usually require a series of Monte Carlo simulations that are beyond
the scope of this work. We direct the interested reader to a selection
of papers on the subject:
\citet{scott2002,coppin2006,perera2008,2009ApJ...707.1201W} and
\citet{chapin2011}.

As an additional external check, we have over-plotted orange ``$+$''
signs at the locations of 1.4\,GHz radio sources from \citet{owen2008}
with $S\gsim15$\,$\mu$Jy. Such radio catalogues have historically
proven invaluable for the precise identifications of high-redshift
submillimetre galaxies due to their low surface densities (compared
with optical catalogues, for example), and a strong correlation
between the radio and submillimetre emission mechanisms
\citep[e.g.,][]{smail2000,pope2006,ivison2007,chapin2009b}. Taking a
representative search radius of 8\,arcsec from these studies with
similar \snr\ sources and source sizes (the same size as the blue
circles), 9 out of 10 peaks in the smoothed signal map have potential
radio counterparts, whereas the single peak in the smoothed jackknife
map does not lie near any radio source. Again, a proper
cross-identification analysis must inevitably include simulations to
establish completeness, false-positive rates, as well as the effects
of point source clustering and confusion. See the aforementioned
papers and references therein for examples.

\begin{figure*}
\centering
\includegraphics[width=\linewidth]{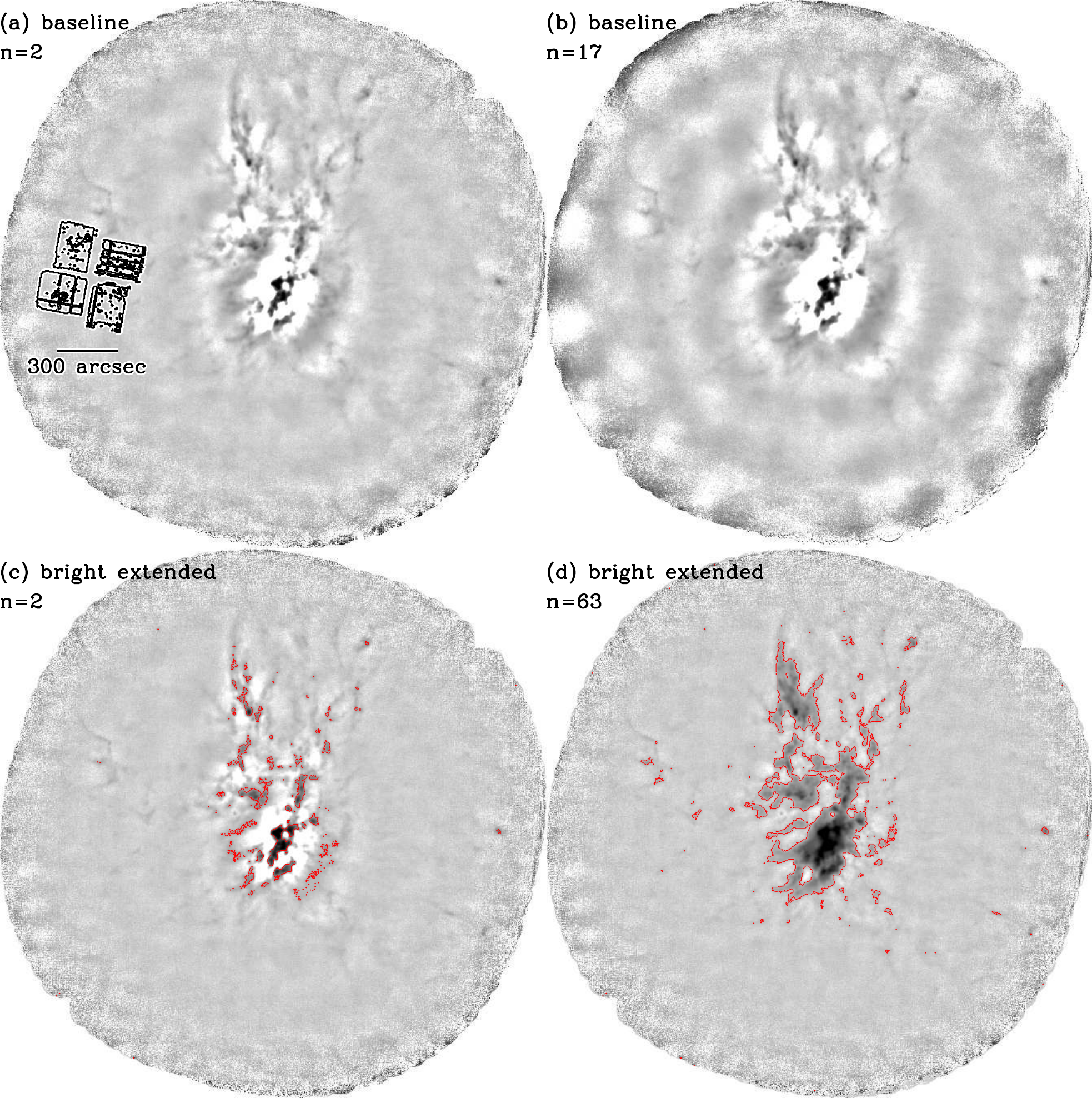}
\caption{An 850\,\micron\ rotating PONG map of M17. Intensity is
logarithmically scaled between $-$0.0003\,pW (white) and +0.01\,pW
(black). Iteration numbers are given in the corner of each
panel. Panels (a) and (b) show the results for a reduction using the
baseline parameters (the solution halted after reaching the map-based
convergence criterion in 17 iterations). Panel (a) also depicts the
array footprint (position angle indicative of the start of the
observation), and a 300\,arcsec line shows the spatial scale
corresponding to the \model{FLT} high-pass filter. Similar to
Fig.~\ref{fig:pointmaps}c, the high-pass filtering introduces ringing
around bright sources. Panels (c) and (d) show the ``bright extended''
reduction, in which a zero-mask is created iteratively from all of the
pixels that lie below a \snr\ of 5. While this region (outside the red
contour) only avoids the brightest peaks early in the solution, in the
final iteration it skirts most of the bright, extended emission, and
significantly helps with negative ringing.}
\label{fig:m17}
\end{figure*}

For future, significantly deeper \scuba\ maps, in which the RMS in a
PSF-smoothed map is dominated by point-source confusion, rather than
instrumental noise, a modified matched filter will offer improved
results. See Appendix~A in \citet{chapin2011}, which shows how to
include confusion (when known \emph{a priori}) explicitly as a noise
term in the calculation of such filters.

\subsection{Bright extended emission}
\label{sec:extended}

In this final example, we analyse a map of M17 which contains bright,
extended emission. The data are from observation 11 on 2011 May 31
using the 850\,\micron\ array. It is a rotating PONG scan covering a
diameter of 0.375\,deg, with a scan speed of 300\,arcsec\,s$^{-1}$,
and a transverse spacing of 180\,arcsec, taking 37.5\,min to complete.

The baseline reduction of these data is shown in the top panels of
Fig.~\ref{fig:m17}, after 2 iterations (the first map estimated after
the noise weights have been measured) and 17 iterations (when the map
has converged), which uses iterative common-mode subtraction and
high-pass filtering. The first panel also depicts the array footprint,
and the angular scale (300\,arcsec) corresponding to the high-pass
filter edge (0.6\,Hz). Much like the reduction of a point source
without any prior constraints (Fig.~\ref{fig:pointmaps}c), there are
ripples around bright sources due to the filtering.

Unlike the case of a known point-source (Section~\ref{sec:point}), it
may not be possible for the observer to define, in advance, a mask of
regions containing blank sky. Indeed, for this map, much of the field
clearly contains extended structure. Furthermore, the goal of such
maps may be to detect previously unknown cool, dense regions of the
ISM that may not have appeared at other wavelengths (e.g., the first
optically-thick cloud-collapse stages of star-formation). While the
option does exist for the user to supply their own mask, a facility
has been added to SMURF to generate one automatically by flagging
pixels below some \snr\ threshold to be set to zero after each
iteration as part of the ``bright extended'' configuration.

The results of this automatic masking are shown in the bottom panels
of Fig.~\ref{fig:m17}. After the second iteration, everything but the
brightest peaks are set to zero (outside the red contours). However,
as the solution progresses, the negative bowls around the bright
sources are slowly reduced and the mask ``grows'' out from the
brightest areas. By the final iteration, most of the obvious
structures in the data are excluded in the mask, negative bowling is
significantly reduced, and the brightest regions are more extended.

\begin{figure*}
\centering
\includegraphics[width=\linewidth]{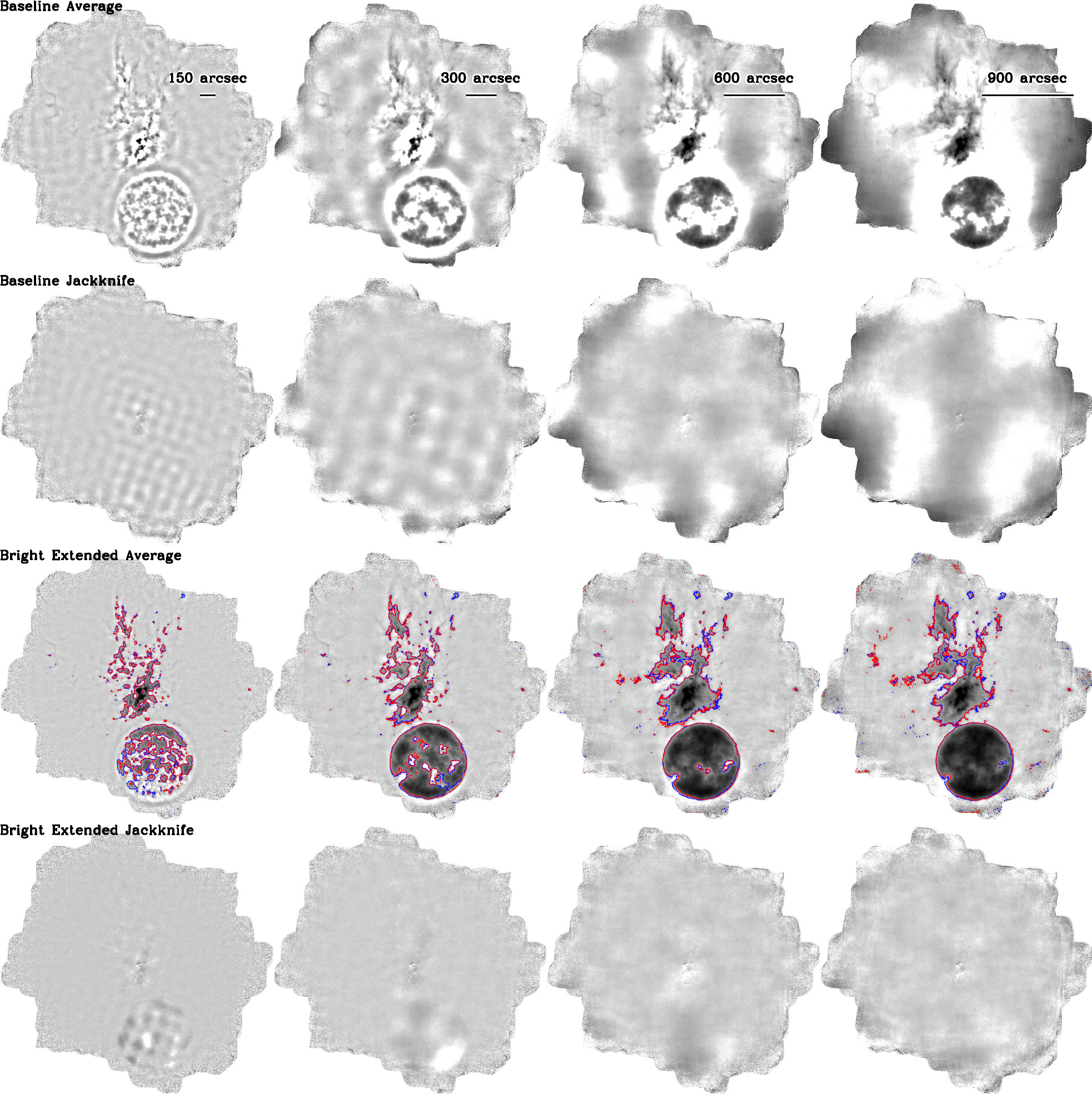}
\caption{Maps of M17 used to characterise the noise properties and
transfer function of SMURF (same intensity scale as
Fig.~\ref{fig:m17}). For each column, a different high-pass filter
edge scale was adopted (indicated in the top panels). First (top) row:
average of two halves of the data analysed independently, using the
baseline configuration. The data have had added to them synthetic
signal within a 600\,arcsec diameter circle (south of M17) created as
a realisation of noise from a $P(k) \propto k^{-3}$ angular power
spectrum multiplied by the PSF, subtracting the minimum to make it
positive, scaling it to a similar signal range as M17 itself, and
rolling-off the edges smoothly using half a period of a radial
(1+cosine)/2 function across 100\,arcsec beyond the edge of the
600\,arcsec region. Second row: jackknife maps produced from the
differences of the maps of each half of the data that went into the
first row. Third row: average of the two halves of the data using the
bright extended reduction. The regions outside the blue and red
contours are constrained to zero for each half of the data (note for
the 300\,arcsec case that these regions about M17 closely match the
mask in Fig.~\ref{fig:m17}d using a full reduction). Clearly as the
filter scale is increased, due to the high \snr\ of the data, larger
emission regions are (correctly) detected and reproduced in the
map. Fourth (bottom) row: Jackknife maps for the bright extended
reductions.}
\label{fig:m17_jk}
\end{figure*}

\begin{figure*}
\centering
\begin{minipage}[h]{0.495\linewidth}
\textbf{(a) baseline reduction PSD} \\

\includegraphics[width=\linewidth]{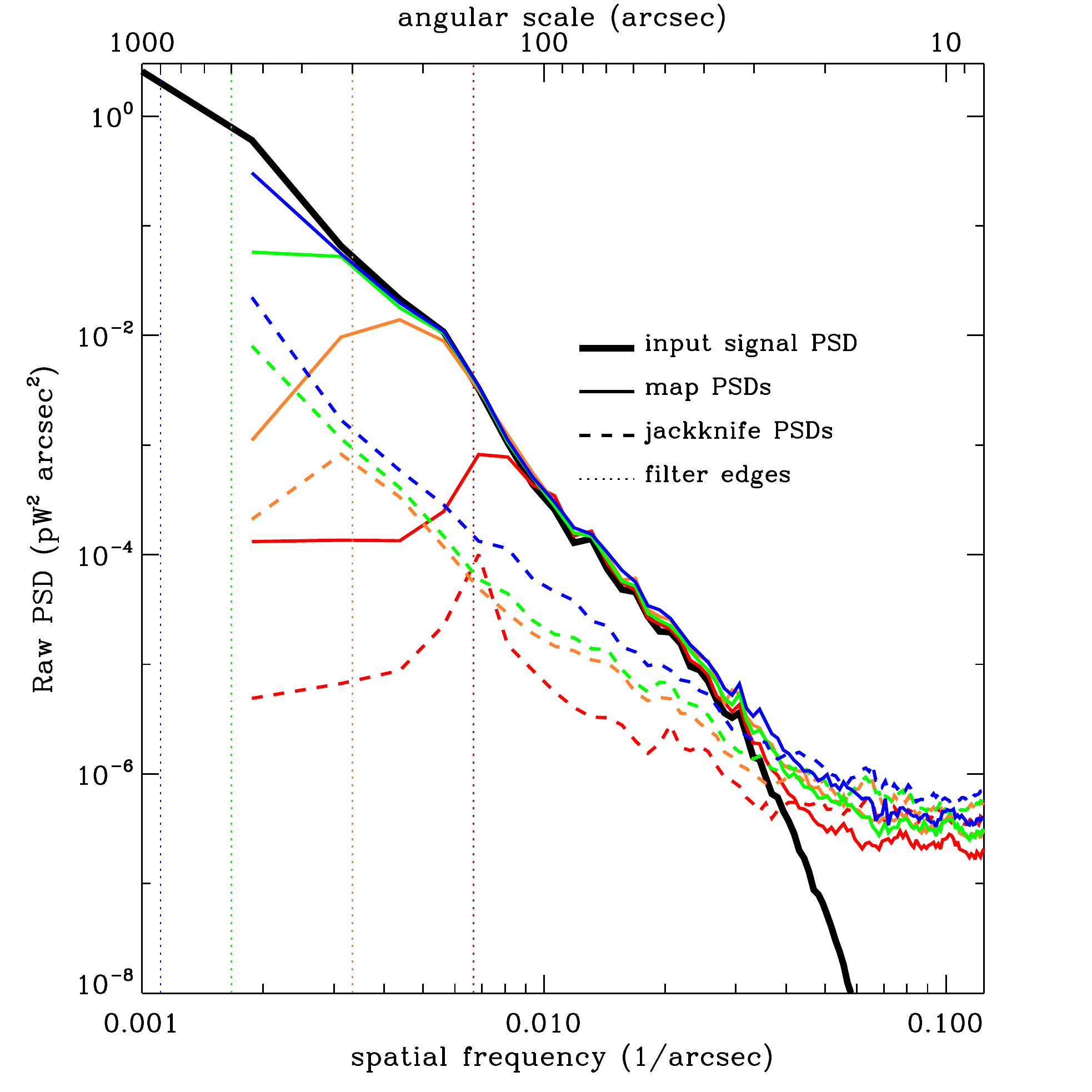}
\end{minipage}
\begin{minipage}[h]{0.495\linewidth}
\textbf{(b) baseline reduction transfer function} \\

\includegraphics[width=\linewidth]{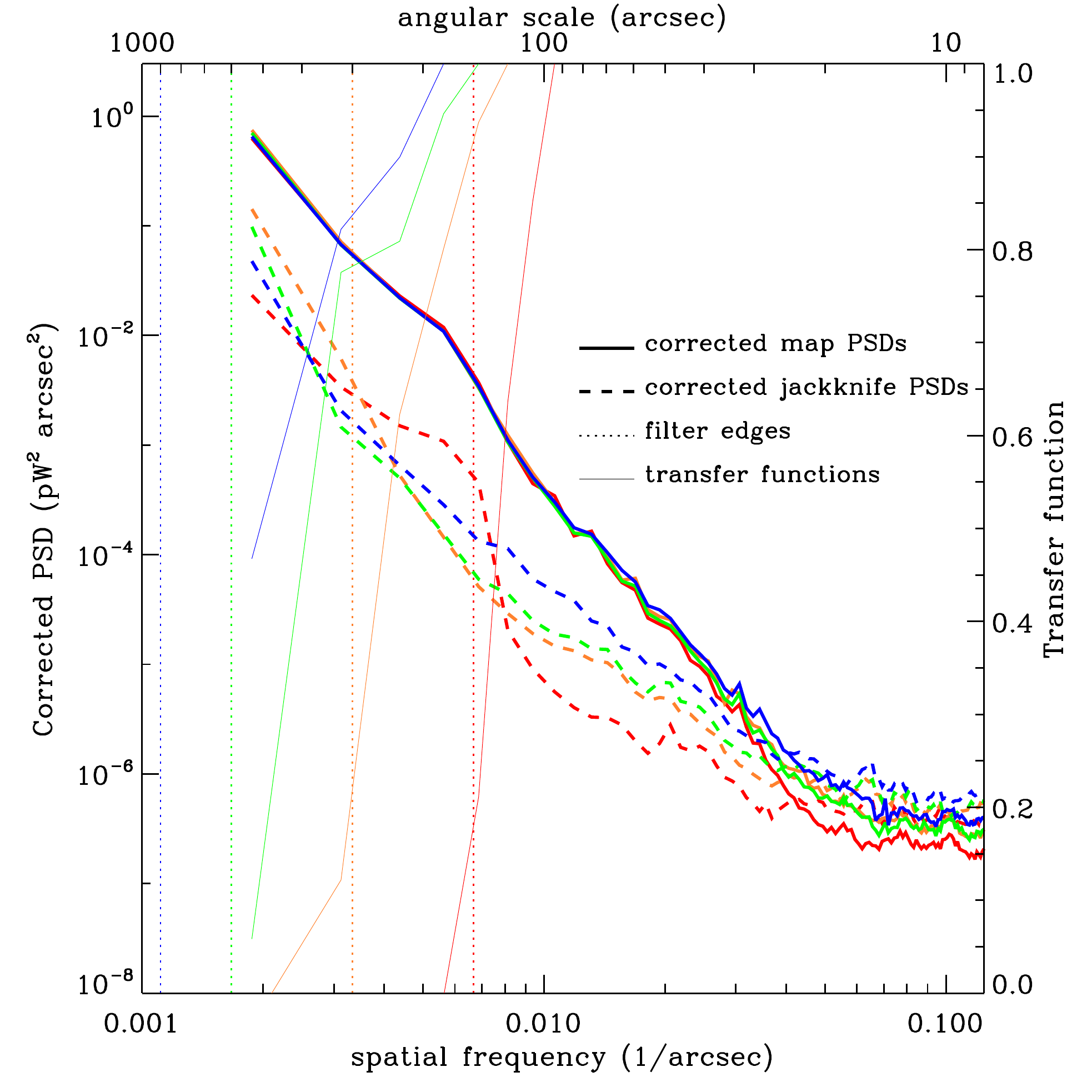}
\end{minipage}
\caption{Angular PSDs for the region of the M17 map in
Fig.~\ref{fig:m17_jk} containing synthetic data, using the baseline
configuration. (a) Raw PSDs for the input (noiseless) simulated data
(thick black line), the signal (average of each half) PSDs (thin solid
lines), and noise PSDs estimated from the jackknives (dashed
lines). Vertical dotted lines indicate the high-pass filter scales:
150\,arcsec (red); 300\,arcsec (orange); 600\,arcsec (green); and
900\,arcsec (blue). PSDs are also colour-coded by filter scale. (b)
Since the input PSD is known, it is possible to measure the transfer
function of the map-maker as the ratio between the difference of the
output map signal PSDs and jackknife PSDs, and the input PSDs, giving
the thin coloured lines (linear vertical axis shown on right of
plot). The remaining lines are as in (a), but now corrected by the
transfer function. This plot shows that increasing the filter scale
improves the \snr\ at intermediate scales ($\sim$200--600\,arcsec),
although the \snr\ worsens at smaller scales ($\lsim$200\,arcsec).}
\label{fig:m17_def_ps}
\end{figure*}

\begin{figure*}
\centering
\begin{minipage}[h]{0.495\linewidth}
\textbf{(a) bright extended reduction PSD} \\

\includegraphics[width=\linewidth]{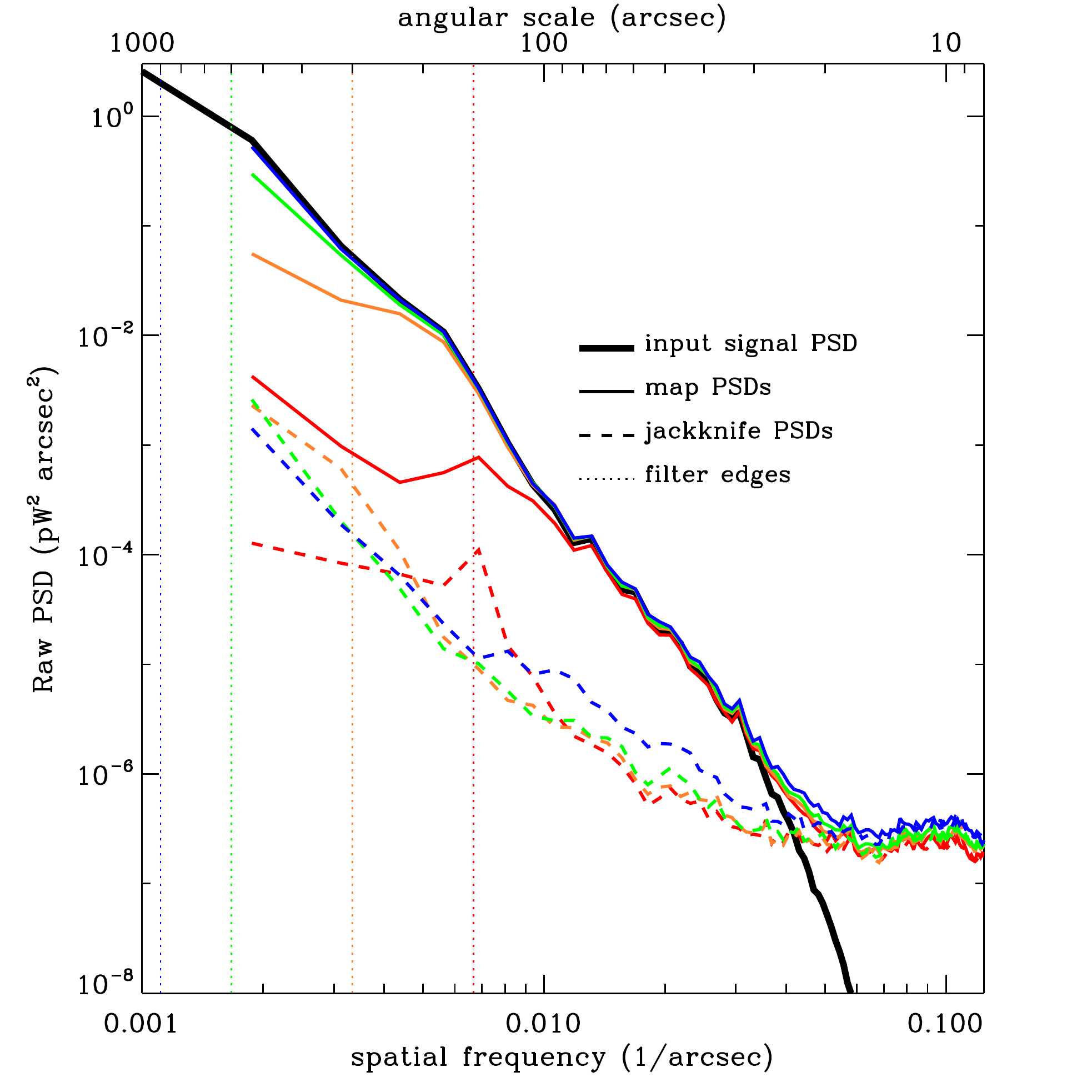}
\end{minipage}
\begin{minipage}[h]{0.495\linewidth}
\textbf{(b) bright extended reduction transfer function} \\

\includegraphics[width=\linewidth]{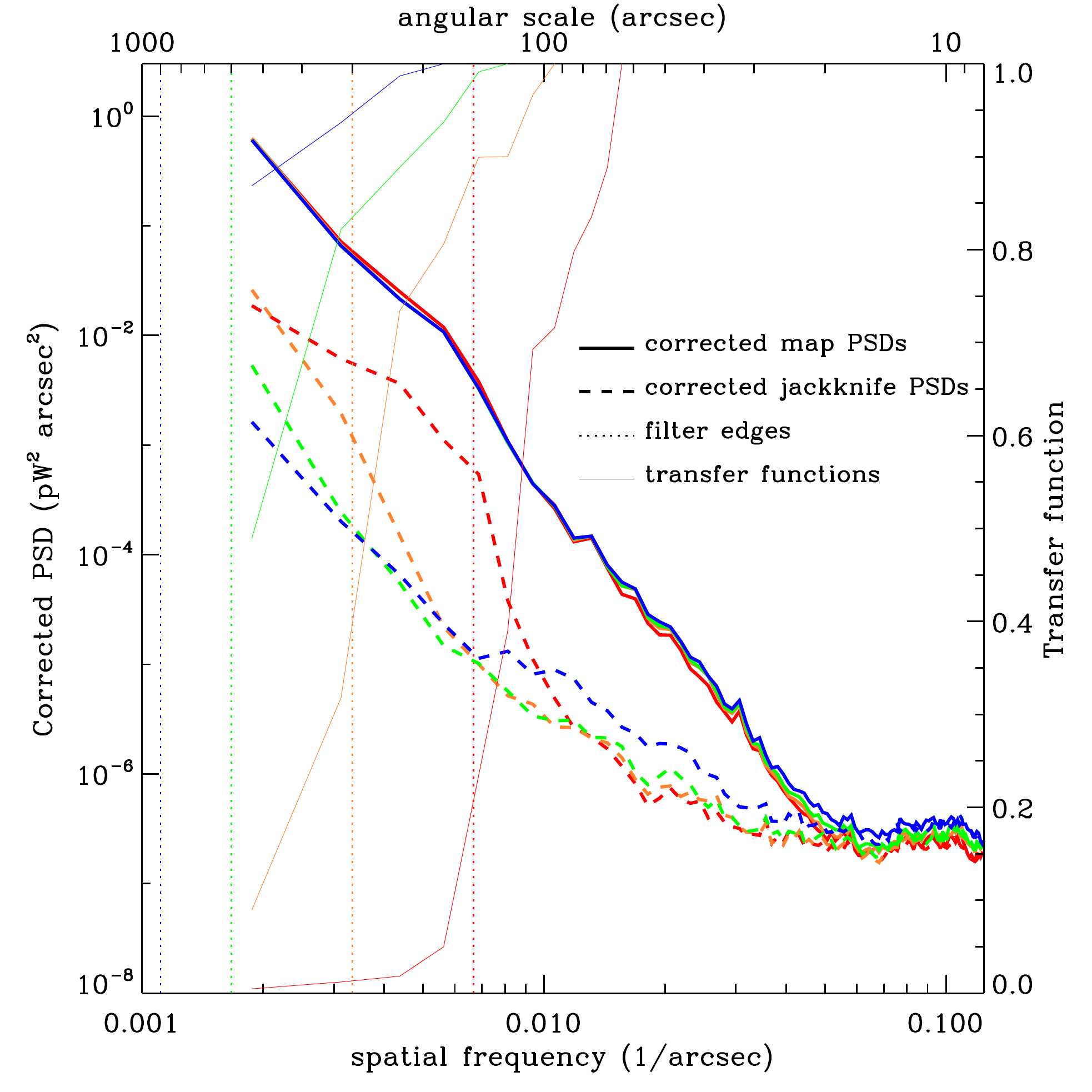}
\end{minipage}
\caption{Lines have same meaning as in Fig.~\ref{fig:m17_def_ps},
except now using the bright extended configuration. This configuration
has resulted in an improved transfer function and \snr\ at large
angular scales. Also note that increasing the filter scale does not
have as large an impact on the small-scale noise as in the baseline
configuration.}
\label{fig:m17_be_ps}
\end{figure*}

While the reduction in the bottom panels of Fig.~\ref{fig:m17} is (in
a cosmetic sense) superior to those in the top panels, it is important
to quantify both the noise properties of the maps, and the response to
real structures (the transfer function). We would also like to know
how each are affected by our choice of filter scale.  Similar to the
previous section, we will use a jackknife test to estimate the noise,
as well as injecting known sources into the data to observe how they
are attenuated.

Maps are produced using the first and second continuous halves of the
data in Fig.~\ref{fig:m17_jk}. This is not an ideal situation, since
the noise properties may evolve with time (e.g., due to changing sky
conditions), leading to a biased estimate of the parent noise
distribution in the complete map from the jackknife. Also, since the
zero-masking depends on the \snr\ of the map, it will be restricted to
regions approximately $\sqrt{2}$ times shallower in these maps than
for the full data set. Finally, the cross-linking (position angles
sampled) is similar, though not identical in the two halves. Ideally
one would have many full maps, as in the case of the Lockman Hole data
in the previous section, from which alternating maps could be
combined.

Since our goal in this section is to measure the response of the
map-maker to extended structures, we inject a simulated signal with
power at a range of scales into a relatively empty region of the map.
It is created by drawing a realisation of noise from an angular power
spectrum $P(k) \propto k^{-3}$, which is appropriate for Galactic
cirrus clouds \citep[e.g.,][]{gautier1992}, within an 800\,arcsec
on-a-side box. It is then filtered again with a 14.5\,arcsec Gaussian
to model the effect of the \scuba\ optical response. The RMS of these
fluctuations is then normalised to 0.002 pW so that they are
comparable to the dynamic range of M17 itself. The minimum is then
subtracted so that the signal is positive. Finally, multiplication by
half a period of a (1+cosine)/2 function is used to roll-off the
signal to zero between radii of 300 and 400 arcsec.

The first row of Fig.~\ref{fig:m17_jk} shows the total signal image
averaging the maps made of each independent half of the data, for the
baseline configuration (inverse-variance weighting has been used). The
columns show reductions using 150, 300, 600, and 900\,arcsec filter
edges. The synthetic data are clearly seen as the circular region
south of M17. As the filter scale is increased, the size of the
ripples increases accordingly. While larger astronomical structures do
seem to appear, negative bowls are a major problem without any other
map constraints. Since the largest scale that is completely inscribed
by the array footprint is about 400\,arcsec, and the diagonal of the
array is about 600\,arcsec, scales ranging from $\sim$400--600\,arcsec
up to the scale of the filter will be unconstrained due to the
degeneracy between the common-mode and map (see discussion in
Section~\ref{sec:point}). This degeneracy therefore causes some of the
large-scale patchiness in the 600 and 900\,arcsec filtered maps that
lie away from the central bright sources.  The second row of
Fig.~\ref{fig:m17_jk} shows the jackknife maps. The astronomical
emission is almost perfectly removed, except for a slight impression
of M17 near the centre of the map, which is probably due to some
combination of detector gain and pointing drifts. Otherwise the
jackknife appears to be a plausible realisation of noise from the same
parent distribution as for the averaged maps.

The third and fourth rows in Fig.~\ref{fig:m17_jk} repeat this
exercise using the bright extended configuration, in which the \snr\
threshold of 5 is again used to identify low-significance pixels that
are set to zero after each iteration. As the filter scale is
increased, more of the extended structure in M17 is reproduced in the
map, as evidenced by the blue and red contours (masks generated from
the first and second halves of the data, respectively). The masking
does a generally good job of suppressing the largest-scale ripples
that are produced by the baseline reduction. However, the noise away
from regions of bright emission does increase noticeably (mottled
appearance) --- due to residual $1/f$ noise following common-mode
removal that has a knee at a frequency above the filter edge. With
filter scales $\gsim600$\,arcsec, virtually the entire region of
synthetic emission is correctly identified by the \snr\ mask and
allowed to vary freely in the solution.

Next, we analyse the angular power spectral densities (PSDs) of the
maps to understand the signal and noise properties of the map-maker in
the region of synthetic sources, as a function of filter scale. In
Fig.~\ref{fig:m17_def_ps}a we show the raw PSDs for the input
synthetic signal (thick black line), the output map signals (thin
solid lines), and the jackknife maps (dashed lines). Colours encode
the filter scales used: 150\,arcsec (red); 300\,arcsec (orange);
600\,arcsec (green); and 900\,arcsec (blue). Note that, with the
exception of the synthetic data, we only plot the PSDs down to the
second-lowest spatial frequency bin of
$1.875\times10^{-3}$\,arcsec$^{-1}$, or 533\,arcsec, since the lowest
is very poorly sampled, and therefore noisy. At small scales
($\lsim$20\,arcsec) the map and jackknife PSDs flatten. However,
unlike Fig.~\ref{fig:lockman_pspec}, these white noise levels are
slightly larger in the jackknife maps, suggesting that the differences
have not completely removed the astronomical signals. A strong
possibility is that the two half-maps that go into the differences
have not converged equally, probably due to the lack of prior
constraints.  As the filter edge is increased, more power is measured
in the map PSDs at larger scales. However, much of this power is
clearly produced by noise which appears in the jackknife PSDs. We
estimate the map-maker transfer function as the ratio between the
portion of the signal PSDs not produced by noise, to the input PSD, or
\begin{equation}
G(k) = \frac{P_\mathrm{M}(k) - P_\mathrm{JK}(k)}{P_\mathrm{S}(k)},
\end{equation}
where $k$ is the spatial frequency, and the subscripts ``M'', ``JK'',
and ``S'' refer to the signal map, jackknife map, and synthetic map,
respectively.

The transfer functions $G(k)$ are plotted as thin solid lines in
Fig.~\ref{fig:m17_def_ps}b. This formula produces extremely noisy
values at large frequencies (small scales), and we therefore set it to
a value of 1 above 0.015\,arcsec$^{-1}$, as well as any point in the
curve where $P_\mathrm{M}(k)$ exceeds $P_\mathrm{S}(k)$ [i.e., $G(k)$
is assumed to be $\le$1]. As expected, the larger the scale of the
filter, the lower the spatial frequency at which the map transfer
function falls. We then correct the map and jackknife PSDs by dividing
by $G(k)$ to produce the thick solid, and dashed lines in
Fig.~\ref{fig:m17_def_ps}b. This shows us that, even though the raw
noise in Fig.~\ref{fig:m17_def_ps}a is lower at small scales when a
smaller-scale filter is used (e.g., the red dashed line), once we
correct for the transfer function, we actually win in a \snr\ sense at
large scales using the larger-scale filter (the corrected noise is
lower), as would be expected.

These tests are then repeated using the bright extended reduction, in
Fig.~\ref{fig:m17_be_ps}. The most obvious improvement with this
reduction over the baseline reduction is that the transfer functions
fall more slowly at large angular scales, accompanied by a slower
increase in noise; in other words, there is greatly improved \snr\ at
large angular scales (an obvious conclusion given the appearance of
the maps in Fig.~\ref{fig:m17_jk}). In fact, using the 900\,arcsec
filter edge, the map response is still above 80 per cent right out to
the largest scale accurately measured in the PSDs, 533\,arcsec, which
is about the largest scale that should be recoverable, given the size
of the array footprint and the fact that we use common-mode rejection.
Another interesting feature of these reductions is that the increase
in small-scale noise as the filter edge is increased is not as drastic
as in the baseline reduction. Finally, note that both the map and
jackknife white noise levels (at scales $\lsim$20\,arcsec) are in
excellent agreement, unlike the previous example.

One case in which the \snr\ is worse using the bright extended
reduction is when using a 150\,arcsec filter. Here the noise is
considerably larger in the bright extended reduction, as evidenced by
the ``kink'' near 150\,arcsec. Referring to the mask contours in the
left panel of the third row in Fig,~\ref{fig:m17_jk}, it is clear that
the map-maker has failed to identify much of the bright, extended
emission in the region of the synthetic source. Each area that is not
within the contours is constrained to zero throughout the solution,
therefore suppressing power (and lowering the transfer function), and
subsequently reducing the \snr\ of the final result. This measurement
serves as a warning: the map-maker response is non-linear when using
\snr\ masking. Harsh filtering can provide misleading results, as in
this example. Maps of faint extended emission will also suffer
considerably, as the structures of interest may lie below the \snr\
threshold for the mask.

Note that alternatives to the zero-masking approach do exist for other
iterative map-makers. For example, \citet{kovacs2008} typically
restricts the solution to a small fixed number of iterations
($\lsim$10), so that there is probably little opportunity for
degeneracies between the map and common-mode to appear in the map.
Experimentation with the model order is also advocated to gain an
impression of the convergence properties. This approach is perhaps
more relevant to their SHARC-2 data for which a more complicated model
is developed; the degeneracies we have observed for \scuba\ maps using
our more brute-force approach with a high-pass filter tend to look
similar regardless of the order (only the convergence time is
affected).  For Bolocam data of the Galactic Plane,
\citet{aguirre2011} used a maximum-entropy filtering step to suppress
large-scales. Regardless of the method used, simulations are always
required to establish the transfer function and noise properties as a
function of angular scale.

\section{Conclusions}
\label{sec:conclusions}

This paper has described the Submillimetre User Reduction Facility
(SMURF), which was designed to produce maps from the rapidly sampled
$\sim10^4$ bolometers of the \scuba\ instrument. While similar to
other algorithms in the literature that iteratively model and remove
correlated noise components from the data, successively improving the
map \citep[e.g.,][]{kovacs2008,aguirre2011,schuller2012}, we have
shown that a fairly simple approach provides good results for \scuba\
data, with reasonable computational requirements. This conclusion is
encouraging for the development of future large-format bolometer
arrays, for which our specific approach should be useful when even
larger volumes of data are involved.

A major obstacle to making maps of \scuba\ data is low-frequency
correlated noise (probably a mixture of atmospheric signals and
magnetic field pickup), which occurs at predominantly $\lsim$2\,Hz.
Much of this signal can be removed using common-mode rejection,
although principal component analysis (PCA) identifies a number of
other less-significant correlated noise patterns at low
frequencies. Unfortunately these patterns are difficult to model, and
their character varies from observation to observation. One could use
the PCA directly to remove them, but this technique is prohibitively
slow for existing high-end desktop computers, given the volume of
\scuba\ data. However, we have demonstrated that high-pass filtering
can remove the bulk of these residual correlated noise components
using significantly more efficient Fast Fourier Transforms (FFTs). The
basic algorithm therefore consists of iterative common-mode rejection
and high-pass filtering along with estimates of the map, which allows
us to approach the white-noise limit of the instrument.

We have found that the iterative solution tends to diverge on large
angular scales due to the degeneracy between the map, and the
low-frequency signal components that are removed (namely the
common-mode). In addition, the high-pass filtering produces
significant ringing around bright sources. A simple strategy of
constraining empty regions of the map to zero (using either a
user-supplied mask for known sources, or an iterative determination of
signal below some \snr\ threshold) provides good constraints for both
compact objects, and bright/extended structures. Particularly in the
latter case, using a combination of synthetic sources and an empirical
measurement of the map noise from jackknife tests (differences of
independent portions of the data), we have demonstrated that we can
effectively recover angular scales up to the order of the array
footprint (approximately 5\,arcmin).

For maps of faint point-sources, a single (non-iterative) high-pass
filter at the start of the reduction produces maps that are nearly
white-noise limited and linear (i.e., the response does not depend on
\snr). Residual large-scale noise can be removed with a whitening
filter (also established from jackknife estimates of the noise) based
on the Fourier Transform of the maps, and sources detected using a
matched-filter (smoothing with the effective filtered point spread
function).

The iterative solution is stopped once convergence in the map itself
is achieved. This enables SMURF to run in a pipeline setting without
user interaction for a wide variety of observations. Furthermore, the
execution times are typically shorter than the observation lengths,
and memory requirements for even the longest \scuba\ observations are
within the capabilities of single, high-end desktop computers. SMURF
can therefore provide real-time feedback at the telescope to
observers.

One regime in which SMURF does not presently perform well is in maps
of faint extended structures, since the zero-masking technique we have
adopted cannot be used. Since SMURF is both highly configurable and
extensible, it may be possible to develop an improved data model
and/or map constraint to assist in these situations, as more
experience with the instrument is gained. However, provided sufficient
computing power is available, the best solution in the long-term will
be a maximum-likelihood algorithm, such as SANEPIC
\citep{patanchon2008}. Even in this case, the existing iterative
solution from SMURF will probably be used as an initial step, since it
can quickly clean the bolometer time-series, as well as perform
map-based despiking (a necessarily iterative procedure).

\section{Acknowledgements}

The James Clerk Maxwell Telescope is operated by the Joint Astronomy
Centre on behalf of the Science and Technology Facilities Council of
the United Kingdom, the Netherlands Organisation for Scientific
Research, and the National Research Council of Canada. Additional
funds for the construction of \scuba\ were provided by the Canada
Foundation for Innovation. This research used the facilities of the
Canadian Astronomy Data Centre operated by the National Research
Council of Canada with the support of the Canadian Space Agency.  This
research was supported in part by the Natural Sciences and Engineering
Research Council of Canada.  EC thanks CANARIE/CANFAR for additional
funding.  The authors thank the members of the SCUBA-2 commissioning
team for testing the map-maker and reporting anomalies; in particular
Antonio Chrysostomou and Jessica Dempsey.  We also thank Mark Halpern,
Matthew Hasselfield, and Gaelen Marsden for many useful discussions;
observers who provided helpful feedback, especially David Nutter and
Todd McKenzie; and Mandana Amiri and Dan Bintley for consultations
regarding the SCUBA-2 Multi-Channel Electronics.  We acknowledge the
contributions of Dennis Kelly, Alex van Engelen and Jennifer Balfour
for early investigations related to SMURF; and Mark Thompson, Craig
Walther and S\'{e}verin Gaudet for being on the Critical Design Review
panel. We thank Per Friberg and Gary Davis for their helpful comments
on the manuscript. Finally, we thank the anonymous referee for their
thorough review that helped to clarify several key areas of the paper.

\bibliographystyle{mn2e}
\bibliography{mn-jour,refs}

\end{document}